\documentclass[12pt]{article}

\usepackage{ graphicx, setspace, amsmath, amssymb, url, multirow, booktabs, verbatim, bm,threeparttable, amsthm, color}%,xcolor}
\usepackage{indentfirst}
\usepackage{bm}
\usepackage{multirow} 
\usepackage{natbib}
\usepackage{appendix}
\usepackage{lscape}
\usepackage{caption}
\usepackage{enumerate}
\usepackage{graphicx}
\usepackage{algorithm}  
\usepackage{algorithmic}
\usepackage{subfig}
\usepackage{mathrsfs}

\newtheorem{thm}{Theorem}
\newtheorem{lem}{Lemma}
\newtheorem{cor}{Corollary}

\newtheorem{rem}{Remark}

\newcommand{\blind}{0}

\begin{document}

\def\spacingset#1{\renewcommand{\baselinestretch}%
{#1}\small\normalsize} \spacingset{1}

%%%%%%%%%%%%%%%%%%%%%%%%%%%%%%%%%%%%%%%%%%%%%%%%%%%%%%%%%%%%%%%%%%%%%%%%%%%%%%

\if0\blind
{
  \title{\bf Heterogeneity-aware Clustered Distributed Learning for Multi-source Data Analysis}
  \author{Yuanxing Chen\\
    Department of Statistics, School of Economics, Xiamen University \\
    Qingzhao Zhang \\
    MOE Key Laboratory of Econometrics, Department of Statistics, \\School of Economics, Wang Yanan Institute for Studies in Economics, \\
    and Fujian Key Lab of Statistics, Xiamen University\\
    Shuangge Ma \\
    Department of Biostatistics, Yale University \\
    Kuangnan Fang \\
    Department of Statistics, School of Economics, Xiamen University\\}
  \maketitle
} \fi

\bigskip
\begin{abstract}
In diverse fields ranging from finance to omics, it is increasingly common that data is distributed and with multiple individual sources (referred to as ``clients'' in some studies). Integrating raw data, although powerful, is often not feasible, for example, when there are considerations on privacy protection. Distributed learning techniques have been developed to integrate summary statistics as opposed to raw data. In many of the existing distributed learning studies, it is stringently assumed that all the clients have the same model. To accommodate data heterogeneity, some federated learning methods allow for client-specific models. In this article, we consider the scenario that clients form clusters, those in the same cluster have the same model, and different clusters have different models. Further considering the clustering structure can lead to a better understanding of the ``interconnections'' among clients and reduce the number of parameters. To this end, we develop a novel penalization approach. Specifically, group penalization is imposed for regularized estimation and selection of important variables, and fusion penalization is imposed to automatically cluster clients. An effective ADMM algorithm is developed, and the estimation, selection, and clustering consistency properties are established under mild conditions. Simulation and data analysis further demonstrate the practical utility and superiority of the proposed approach. 
\end{abstract}

\noindent%
\vfill

\newpage
\spacingset{1.9} % DON'T change the spacing!
\section{Introduction}
\label{sec:intro}

In diverse fields, it is increasingly common that data is distributed and with multiple individual sources (referred to as ``clients'' in this article and some published studies). For example, in financial studies, it is common that data is with, for example, multiple individual bank branches. In omics studies, it is common that multiple independent studies have generated their own data and address the same scientific question. The power of integrating data from multiple sources has been well identified \citep{liu2014integrative}. A family of studies/methods integrate raw data  \citep{tang2016fused,huang2017promoting}. Such integrative analysis methods, although effective, are not always feasible. For example, with financial and omics data, there are often strong considerations on data privacy protection \citep{Abbe2012,cai2021individual}. Additionally, when data is large, sending raw data from individual clients to a central server and constructing a statistical model with the pooled data may lead to considerable computational cost \citep{bhowmick2018protection}.

To avoid sharing raw data, distributed learning techniques have been developed, with which all clients share summary statistics (which have no privacy concerns and much smaller sizes) as opposed to raw data. It is noted that this is also closely related to the divide-and-conquer (DC) strategy, which has been designed to reduce cost and improve feasibility and performance in the analysis of big data \citep{lee2017communication,Battey2018}. Some distributed learning methods demand multiple communications between the local clients and central server \citep{Jordan2019}. There are also one-shot methods \citep{dobriban2020wonder}, with which the local clients only need to communicate once with the central server. Such methods may enjoy lower cost and higher convenience. In most of the existing studies, it has been assumed that all the clients have the same data generation model. This assumption, although convenient, may be overly stringent. In raw data-based integrative analysis \citep{tang2016fused,huang2017promoting}, it has been well established that data may be heterogeneous and demand different models. Here, heterogeneity can be caused by differences in sample characteristics, data collection techniques, and multiple other factors \citep{ghosh2020efficient}.

To accommodate data heterogeneity, some federated learning methods allow client-specific models \citep{smith2017federated}. Of the most relevance to this study is clustered federated learning, under which clients form clusters, those in the same cluster share the same model, and different clusters have different model \citep{ghosh2020efficient,marfoq2021federated}. This type of analysis may have been more popular in computer science than statistics. Intuitively, assuming and identifying a clustering structure can lead to a better understanding of the ``interconnections'' among clients (for example, those in the same cluster are more alike and can be more closely related to each other) and a smaller number of model parameters. A common limitation shared by the existing clustered federated learning methods is that it is usually challenging to determine the number of clusters. For example, \cite{ghosh2020efficient} and \cite{marfoq2021federated} first pre-specified the number of clusters and then alternately updated the cluster membership of each client and model parameters for each cluster. Similar to classic clustering analysis, results can be sensitive to the number of clusters, and in practice, usually there is not enough information to accurately specify this number. 

In the statistical literature, there are also a few heterogeneous distributed learning methods that allow for client-specific models. For example, \cite{zhao2016partially} proposed a heterogeneous distributed learning method with a partially linear model, under which the nonparametric parameter is assumed to be shared by all clients, while the parametric parameters are allowed to be client-specific. \cite{duan2022heterogeneity} extended the surrogate likelihood function approach to allow client-specific nuisance parameters by adopting a surrogate estimating equation technique. It is noted that these two (and some other) studies are limited to low-dimensional settings. \cite{cai2021individual} further studied the high-dimensional heterogeneous setting by aggregating local summary statistics under a generalized linear model. As recognized in \cite{tang2021individualized}, allowing all clients to have individual models may lead to a large number of redundant parameters, negatively affecting estimation and inference. 

In this article, we consider the integrative analysis of multi-source data, where only summary statistics can be shared. To sufficiently accommodate data heterogeneity, clients are allowed to have different models. We focus on the scenario where those models have the same sparsity structure (set of important variables) and note that the proposed strategy can be extended to accommodate different sparsity structures. Further, motivated by the success of clustered federated learning, we consider the scenario where clients form clusters, and the models are cluster specific. For simultaneous estimation, variable selection, and clustering, we develop an integrative clustered regression (ICR) method, which may advance from the existing literature in multiple important ways. First, compared to methods that assume homogeneity \citep{lee2017communication}, it is more flexible and can effectively accommodate data heterogeneity. Second, compared to methods that allow for client-specific models \citep{zhao2016partially,duan2022heterogeneity}, it can lead to a better understanding of the similarity/differences among datasets and a smaller number of parameters (and hence improved estimation). Third, compared to the existing clustered federated learning methods \citep{ghosh2020efficient,marfoq2021federated}, it can data-dependently and conveniently determine the number of clusters, with the assistance of penalized fusion. Last but not least, it can accommodate multiple types of data/models, and our computational and theoretical developments can shed broader insights.  

The rest of the article is organized as follows. In Section 2, we introduce the data/model settings, the proposed approach, and an effective proximal ADMM algorithm. In Section 3, we rigorously establish that the proposed approach enjoys the estimation, variable selection, and clustering consistency properties. Numerical studies, including simulation in Section 4 and data analysis in Section 5, demonstrate the practical utilization and superiority of the proposed approach. Brief discussions are provided in Section 6. Proofs of the theoretical results and additional numerical results are provided in the Appendix.

\section{Methods}
\subsection{Integrative Analysis under Privacy Constraints}
Suppose that there are $K$ independent clients, and for the $k$th client, there are $n_k$  observations. The total sample size is $N=\sum_{k=1}^Kn_k$. For the $k$th client, let $y_{i}^{(k)}$ and $\bm x_i^{(k)}=(x_{i1}^{(k)}, \dots,x_{ip}^{(k)})^{\top}\in\mathbb R^p$ be the response and covariate vector of the $i$th observation, respectively, where the first element of $\bm x_i^{(k)}$ is fixed as $x_{i1}^{(k)}\equiv 1$ to accommodate intercept. Accordingly, let $\mathbf X^{(k)}=(\bm x_1^{(k)},\dots,\bm x_{n_k}^{(k)})^{\top}$ and $\mathbf Y^{(k)}=(y_1^{(k)},\dots,y_{n_k}^{(k)})^{\top}$ denote the design matrix and response vector of the $k$th client, respectively. Let $f(\cdot)$ be the pre-specified twice-differentiable loss function and $\bm\theta^{(k)}=(\theta_1^{(k)},\dots,\theta_{p}^{(k)})^{\top}$ be the $p$-dimensional coefficient vector. The empirical local and global loss functions are defined as
\[
\widehat{\mathcal L}_k(\bm\theta^{(k)})=\frac1{n_k}\sum_{i=1}^{n_k}f\big(\bm x_i^{(k)\top}\bm\theta^{(k)}, y_i^{(k)}\big),\ k\in[K] \quad\text{and}\quad\widehat{\mathcal L}(\bm\theta)=\frac1{N}\sum_{k=1}^Kn_k\widehat{\mathcal L}_k(\bm\theta^{(k)}),
\]
respectively, where $\bm\theta=(\bm\theta^{(1)},\cdots,\bm\theta^{(K)})$ is a $p\times K$ coefficient matrix with the $j$th row $\boldsymbol\theta_{j}=(\theta_{j}^{(1)},\cdots,\theta_{j}^{(K)})^{\top}$, and $[d]$ denotes the index set $\{1,\dots,d\}$ for an integer $d$. 
Assume that the indexes of $\boldsymbol\theta$ can be classified into $M$ non-overlapping subsets $\{\mathcal G^{(1)}, \dots, \mathcal G^{(M)}\}$. Additionally, for each covariate, its coefficients across the $K$ datasets can be viewed as a group \citep{cai2021individual}, leading to $p$ groups corresponding to the covariates. 

For simultaneous regularized estimation, variable selection, and identification of the clustering structure of datasets, we propose the objective function with the ideal pooling (IP) strategy
\begin{equation}\label{obj1}
\begin{aligned}
\widehat{\mathcal Q}_{\text{IP}}(\bm\theta)=&\ \widehat{\mathcal L}(\bm\theta)+\mathcal P_{\lambda_1}(\bm\theta)+\mathcal P_{\lambda_2}(\bm\theta)\\
=&\ \frac1{N}\sum_{k=1}^Kn_k\widehat{\mathcal L}_k(\bm\theta^{(k)})+\sum_{j=2}^{p}p_{\tau}(||\boldsymbol\theta_j||_2,\lambda_1)+\sum_{k<k^{\prime}}p_{\tau}(||\boldsymbol\theta^{(k)}-\boldsymbol\theta^{(k^{\prime})}||_2,\lambda_2),
\end{aligned}
\end{equation}
where penalty $\mathcal P_{\lambda_1}(\bm\theta)$ is mainly for regularized estimation and variable selection, and penalty $\mathcal P_{\lambda_2}(\bm\theta)$ is mainly for clustering. Here $p_{\tau}(,)$ is a penalty function with concavity parameter $\tau$, $||\cdot||_2$ is the $L_2$ norm, and $\lambda_1$, $\lambda_2$ are two non-negative tuning parameters.

\begin{figure}[hptb]
\centering
\includegraphics[height = 9cm, width = 16cm]{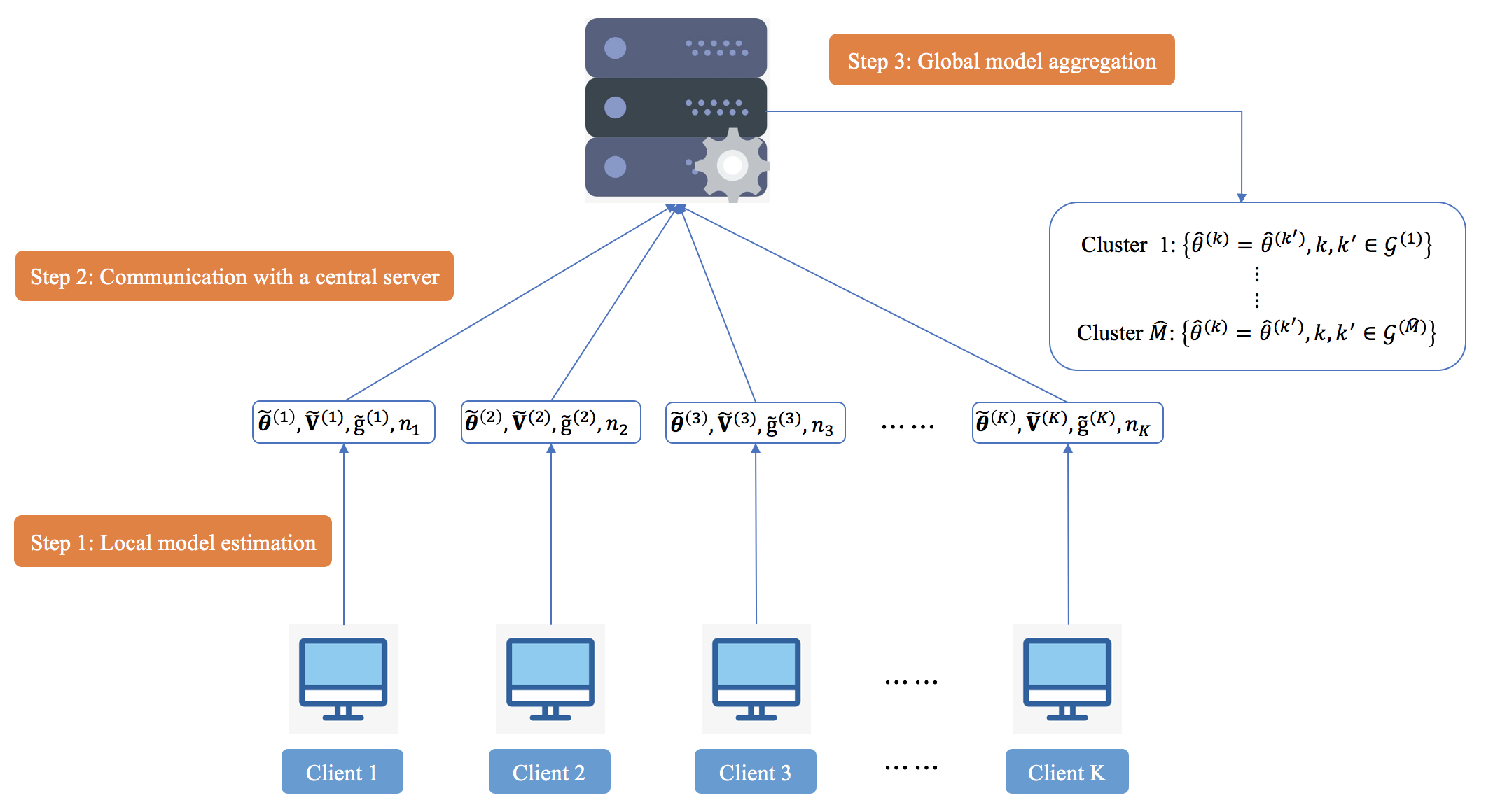}
\caption{Scheme of the proposed analysis.
}
\label{diagram}
\end{figure}

With the privacy-preservation constraints, raw data of the individual datasets is not available, and hence objective function $\widehat{\mathcal Q}_{\text{IP}}(\bm\theta)$ in (\ref{obj1}) cannot be directly implemented. To tackle this problem, we adopt the least-square approximation (LSA) of \cite{He2016} and \cite{Zhu2021}, which leads to the objective function
\begin{equation}\label{obj2}
\widehat{\mathcal Q}_1(\bm\theta)=\frac{1}{N}\sum_{k=1}^Kn_k(\boldsymbol\theta^{(k)}-\widetilde{\boldsymbol\theta}^{(k)})^{\top}\widetilde{\mathbf V}^{(k)}(\boldsymbol\theta^{(k)}-\widetilde{\boldsymbol\theta}^{(k)})+\mathcal P_{\lambda_1}(\bm\theta)+\mathcal P_{\lambda_2}(\bm\theta),\end{equation}
where $\widetilde{\bm\theta}^{(k)}$ is the local estimator of the $k$th client, and $\widetilde{\mathbf V}^{(k)}=\partial^2\widehat{\mathcal L}_k(\widetilde{\boldsymbol\theta}^{(k)})/\partial\bm\theta^{(k)}\partial\bm\theta^{(k)\top}$ is the Hessian matrix of ${\widehat{\mathcal L}_k}(\boldsymbol\theta^{(k)})$ with respect to $\bm\theta^{(k)}$ at $\widetilde{\bm\theta}^{(k)}$. \cite{He2016} recommended adopting ordinary least square (OLS) estimates as the local estimators when $p<n_k$.  Under high-dimensional settings, OLS estimates are not available, and a ``straightforward'' approach is to replace the OLS estimates with the Lasso estimates. However, in general, the Lasso estimates are biased.  Motivated by \cite{cai2021individual}, we adopt the debiased Lasso estimators \citep{vandegeer2014} and propose the ICR estimator by minimizing the following objective function
\begin{equation}\label{obj3}
\widehat{\mathcal Q}_{\text{ICR}}(\bm\theta)= \frac{1}{N}\sum_{k=1}^Kn_k\Big(\bm\theta^{(k)\top}\widetilde{\mathbf V}^{(k)}\bm\theta^{(k)}-2\bm\theta^{(k)\top}\widetilde{\bm\zeta}^{(k)}\Big)+\mathcal P_{\lambda_1}(\bm\theta)+\mathcal P_{\lambda_2}(\bm\theta),
\end{equation}
where $\widetilde{\bm\zeta}^{(k)}=\widetilde{\mathbf V}^{(k)}\widetilde{\bm\theta}^{(k)}-\widetilde{\mathbf g}^{(k)}$ and $\widetilde{\mathbf g}^{(k)}=\partial\widehat{\mathcal L}_k(\widetilde{\boldsymbol\theta}^{(k)})/\partial\bm\theta^{(k)}$ is the gradient of ${\widehat{\mathcal L}_k}(\boldsymbol\theta^{(k)})$ with respect to $\bm\theta^{(k)}$ at $\widetilde{\bm\theta}^{(k)}$. For the penalty function, viable choices include SCAD \citep{Fan2001SCAD}, MCP \citep{Zhang2010MCP}, and others. We adopt MCP in our numerical studies. The overall analysis approach is schematically presented in Figure \ref{diagram}. It consists of generating individual estimates based on raw data by individual clients, sending summary estimates from local clients to a central server, conducting the proposed estimation, and outputting the final estimators to guide downstream analysis/actions.

This approach has been motivated by the following considerations. 
In $\widehat{\mathcal Q}_{\text{ICR}}(\bm\theta)$, we only make use of four summary statistics, namely the initial local estimators $\{\widetilde{\bm\theta}^{(k)}\}_{k=1}^K$, corresponding gradient vectors $\{\widetilde{\mathbf g}^{(k)}\}_{k=1}^K$, Hessian matrices $\{\widetilde{\mathbf V}^{(k)}\}_{k=1}^K$, and local sample sizes $\{n_k\}_{k=1}^K$. That is, the proposed approach and estimate are fully based on the summary statistics as opposed to the raw data -- data privacy protection is thus achieved. 
In (\ref{obj3}), the first term measures lack-of-fit, and similar forms have been considered in the literature \citep{Zhu2021,cai2021individual}. When higher-order estimation properties are not of interest, the estimates and Hessian matrices from the local clients contain sufficient information. The first penalty determines which covariates have overall nonzero effects, under the assumptions that one covariate may have different effects/coefficients for different clients, but the effects are either all nonzero or all zero. It is possible to replace it with more complex penalties, for example those that can conduct two-level selection \citep{huang2017promoting}, to obtain ``more subtle'' information. The second is a fusion penalty \citep{ma2017concave}, with which some clients may have exactly equal estimates. Clients $k$ and $k'$ are clustered together if and only if their estimates are equal. For identifying clustering structures, fusion penalization has been recognized to have multiple unique advantages and been popular in the recent literature \citep{Yang2019,Chen2021}. For example, it translates clustering to an ``easier'' estimation problem and can more conveniently determine the number/structure of clusters (by examining the estimates).

\subsection{Computational Algorithm}
We use local linear approximation -- LLA \citep{zou2008one} to approximate the fused penalty and propose an iterative algorithm. Specifically, in the $t$th iteration, we update the coefficients by solving
\[
\mathop{\mathrm{arg\ min}}\limits_{\bm\theta\in\mathbb R^{p\times K}}\ \frac{1}{N}\sum_{k=1}^Kn_k\Big(\bm\theta^{(k)\top}\widetilde{\mathbf V}^{(k)}\bm\theta^{(k)}-2\bm\theta^{(k)\top}\widetilde{\bm\zeta}^{(k)}\Big)+\sum_{k<k^{\prime}}\omega_{kk^{\prime}}^{t-1}||\bm\theta^{(k)}-\bm\theta^{(k^{\prime})}||_2+\sum_{j=2}^{p}p_{\tau}(||\boldsymbol\theta_j||_2,\lambda_1),
\]
where $\omega_{kk^{\prime}}^{t-1}=p_{\tau}^{\prime}(||\bm\theta^{(k),t-1}-\bm\theta^{(k^{\prime}),t-1}||_2,\lambda_2)$ denotes the weight and $p_{\tau}^{\prime}(x,\lambda)$ is the derivative of $p_{\tau}(x,\lambda)$ with respect to $x$. The above minimization problem can be reformulated as a constrained minimization problem
\[
\begin{aligned}
\mathop{\mathrm{arg\ min}}\limits_{\bm\theta\in\mathbb R^{p\times K}}\Big\{  \ell(\bm\theta):=&\ g(\bm\theta)+h_1(\bm\alpha)+h_2(\bm\theta)\\
=&\ \frac{1}{N}\sum_{k=1}^Kn_k\Big(\bm\theta^{(k)\top}\widetilde{\mathbf V}^{(k)}\bm\theta^{(k)}-2\bm\theta^{(k)\top}\widetilde{\bm\zeta}^{(k)}\Big)+\sum_{k<k^{\prime}}\omega_{kk^{\prime}}^{t-1}||\bm\alpha_{kk^{\prime}}||_2+\sum_{j=2}^{p}p_{\tau}(||\boldsymbol\theta_j||_2,\lambda_1)\Big\},\\
&\ \text{subject to}\quad \boldsymbol\theta^{(k)}-\boldsymbol\theta^{(k^{\prime})}=\boldsymbol\alpha_{kk^{\prime}},\ 1\le k<k^{\prime}\le K,
\end{aligned}
\]
where $\boldsymbol\alpha=(\bm\alpha_{12},\dots,\bm\alpha_{(K-1)K})$ is a $p\times K(K-1)/2$ matrix composed of the auxiliary variables. This optimization problem is equivalent to the minimization of the augmented Lagrangian
\begin{equation}\label{aug}
\ell_{\nu}(\bm\theta,\bm\alpha,\bm\xi)= \ell(\bm\theta)+\sum_{k<k^{\prime}}\boldsymbol\xi_{kk^{\prime}}^{\top}(\boldsymbol\theta^{(k)}-\boldsymbol\theta^{(k^{\prime})}-\boldsymbol\alpha_{kk^{\prime}})+\frac{\nu}{2}\sum_{k<k^{\prime}}||\boldsymbol\theta^{(k)}-\boldsymbol\theta^{(k^{\prime})}-\boldsymbol\alpha_{kk^{\prime}}||_2^2,
\end{equation}
where $\boldsymbol\xi=(\bm\xi_{12},\dots,\bm\xi_{(K-1)K})$ is a $p\times K(K-1)/2$ matrix composed of the dual variables. $\nu$ is a small positive constant. 
Following \cite{shimmura2022converting}, we can minimize objective function $\ell_{\nu}(\bm\theta,\bm\alpha,\bm\xi)$ in (\ref{aug}) via the following iterations
\begin{equation}\label{sequence}
\begin{aligned}
&(\boldsymbol\theta^{t},\boldsymbol\alpha^{t})=\mathop{\mathrm{arg\ min}}\limits_{\bm\theta,\bm\alpha}\ \ell_{\nu}(\boldsymbol\theta,\boldsymbol\alpha,\boldsymbol\xi^{t-1}),\\
&\bm\xi_{kk^{\prime}}^{t}=\bm\xi_{kk^{\prime}}^{t-1}+\nu(\bm\theta^{(k),t}-\bm\theta^{(k^{\prime}),t}-\bm\alpha_{kk^{\prime}}^{t}),\ 1\le k<k^{\prime}\le K.\\
\end{aligned}
\end{equation}

To update $(\bm\theta,\bm\alpha)$ via (\ref{aug}), we minimize $\eta(\bm\theta)$ defined by
\begin{equation}\label{eta_iteration}
\begin{aligned}
\eta(\bm\theta):=&\ \mathop{\mathrm{arg\ min}}\limits_{\bm\alpha}\ \ell_{\nu}(\boldsymbol\theta,\boldsymbol\alpha,\boldsymbol\xi^{t-1})\\
=&\ \mathop{\mathrm{arg\ min}}\limits_{\bm\alpha}\Bigg\{\sum_{k<k^{\prime}}\Big[\omega_{kk^{\prime}}^{t-1}||\bm\alpha_{kk^{\prime}}||_2+{\boldsymbol\xi_{kk^{\prime}}^{t-1}}^{\top}(\boldsymbol\theta^{(k)}-\boldsymbol\theta^{(k^{\prime})}-\boldsymbol\alpha_{kk^{\prime}})+\frac{\nu}{2}||\boldsymbol\theta^{(k)}-\boldsymbol\theta^{(k^{\prime})}-\boldsymbol\alpha_{kk^{\prime}}||_2^2\Big]\Bigg\}\\
&\ \qquad\ +g(\bm\theta)+h_2(\bm\theta)\\
 :=&\ \mathop{\mathrm{arg\ min}}\limits_{\bm\alpha}\eta_1(\bm\theta,\bm\alpha)+g(\bm\theta)+h_2(\bm\theta) :=\eta_2(\bm\theta)+h_2(\bm\theta).\\
\end{aligned}
\end{equation}
Following \cite{chi2015splitting}, we define the proximal map with respect to $\Omega(\mathbf v)$ as
\[
\mathrm{prox}_{\sigma\Omega}(\mathbf u)=\mathop{\mathrm{arg\ min}}\limits_{\mathbf v}\bigg[\sigma\Omega(\mathbf v)+\frac12||\mathbf u-\mathbf v||_2^2\bigg].
\]
Besides, the conjugate function of $\Omega(\mathbf v)$ is defined by $\Omega^*(\mathbf u)=\mathrm{sup}_{\mathbf v}[\mathbf u^{\top}\mathbf v-\Omega(\mathbf v)]$.
Then, it is easy to show that $\eta_1(\bm\theta,\bm\alpha)$ is minimized when 
\begin{equation}\label{alpha_iteration}
\bm\alpha(\bm\theta)=\mathrm{prox}_{\nu^{-1} h_1}(\bm\theta\mathbf A+\nu^{-1}\bm\xi^{t-1}),
\end{equation}
where $\mathbf A=(\mathbf e_{12},\dots,\mathbf e_{(K-1)K})$ is a $K\times K(K-1)/2$ matrix and $\mathbf e_{kk^{\prime}}=\mathbf e_k-\mathbf e_{k^{\prime}}$, in which $\mathbf e_k$ is a $K\times 1$ vector whose $k$th element is 1 and the remaining elements are 0.
Plugging (\ref{alpha_iteration}) into $\eta_1(\bm\theta,\bm\alpha)$ in (\ref{eta_iteration}) and combining the results of Theorem 1 and Lemmas 1-2 of  \cite{shimmura2022converting}, we can show that $\eta_2(\bm\theta)$ is differentiable, and the gradient of $\eta_2(\bm\theta)$ is
\[
\frac{\partial\eta_2(\bm\theta)}{\partial\bm\theta}=\bm\theta_g+\Big[\mathrm{prox}_{\nu h_1^*}(\nu\bm\theta\mathbf A+\bm\xi^{t-1})\Big]\mathbf A^{\top},
\]
where $\bm\theta_g=2/N\big[n_1(\widetilde{\mathbf V}^{(1)}\bm\theta^{(1)}-\widetilde{\bm\zeta}^{(1)}),\dots,n_K(\widetilde{\mathbf V}^{(K)}\bm\theta^{(K)}-\widetilde{\bm\zeta}^{(K)})\big]$ is a $p\times K$ matrix and $h_1^*$ is the conjugate function of $h_1$. 
Then, we can adopt proximal gradient techniques, such as  FISTA \citep{parikh2014proximal}, to obtain the solution of the first minimization problem in (\ref{sequence}). Further, the second iterative step in (\ref{sequence}) can be reformulated as
\[
\bm\xi^{t}=\Big[\mathrm{prox}_{\nu h_1^*}(\nu\bm\theta^{t}\mathbf A+\bm\xi^{t-1})\Big].
\]
The proposed proximal ADMM algorithm is summarized as follows.
\begin{enumerate}[Step 1.]
\item Obtain the initial estimates with $(\bm\theta^0, \bm\xi^0)$.
\item At iteration $t, t=1,2,\dots$, update $\bm\theta^t$ as follows.
\begin{enumerate}[Step 2.1.]
\item Initialize $\bm u^{t-1,0}=\bm\theta^{t-1,0}=\bm\theta^{t-1}$ and $\rho_0=1$.
\item At iteration $s, s=1,2,\dots$, compute
\[
\begin{aligned}
&\omega_{kk^{\prime}}^{t-1,s}\leftarrow p_{\tau}^{\prime}(||\bm u^{(k),t-1,s-1}-\bm u^{(k^{\prime}),t-1,s-1}||_2,\lambda_2),\quad 1\le k<k^{\prime}\le K,\\
&\bm\theta^{t-1,s}\leftarrow\mathrm{prox}_{\varsigma h_2}\bigg[\bm u^{t-1,s-1}-\varsigma\frac{\partial\eta_2(\bm u^{t-1,s-1})}{\partial\bm\theta}\bigg],\\
&\rho_s\leftarrow\frac{1+\sqrt{1+4\rho_{s-1}^2}}{2},\quad\bm u^{t-1,s}\leftarrow\bm\theta^{t-1,s}+\frac{\rho_{s-1}-1}{\rho_s}(\bm\theta^{t-1,s}-\bm\theta^{t-1,s-1}).
\end{aligned}
\]
\item Repeat Step 2.2 until convergence, and set $\bm\theta^t\leftarrow\bm\theta^{t-1,s}$.
\end{enumerate}
\item For $1\le k<k^{\prime}\le K$, update $\omega_{kk^{\prime}}^{t}\leftarrow p_{\tau}^{\prime}(||\bm\theta^{(k),t}-\bm\theta^{(k^{\prime}),t}||_2,\lambda_2)$.
\item Update $\bm\xi^t\leftarrow\mathrm{prox}_{\nu h_1^*}(\nu\bm\theta^{t}\mathbf A+\bm\xi^{t-1})$.
\item Repeat Steps 2-4 until convergence, and set $\bm\alpha^t\leftarrow\mathrm{prox}_{\nu^{-1} h_1}(\bm\theta^t\mathbf A+\nu^{-1}\bm\xi^{t})$.
\end{enumerate}
In the above calculation, we conclude convergence if the absolute difference of estimates from two consecutive iterations is smaller than a predefined cutoff.

\begin{rem}\label{rem1}
There exist closed-form solutions for the proximal maps of $\nu h_1^*$, $\nu^{-1}h_1$, and $\varsigma h_2$. Specifically, the proximal map of $\nu h_1^*$ is a projection function. And the proximal maps of  $\nu^{-1}h_1$ and $\varsigma h_2$ can be easily derived as in \cite{ma2017concave} with $\tau>\varsigma$.
In Step 2.2, $\varsigma$ denotes the step size. As in  \cite{shimmura2022converting}, we can derive the Lipschitz constant of $\eta_2(\bm\theta)$, denoted by $L_{\eta}=1+2\nu\max_{k\in[K]} (\mathbf A\mathbf A^{\top})_{k,k}$, and then set $\varsigma=L_{\eta}^{-1}$. With $\nu=1$ and $\tau=3$, $\tau>\varsigma$ since $\varsigma\le 1/3$.
%By the nested proximal gradient descent algorithm in Step 2.2, the sparsity structure can be easily obtained based on $\bm\theta$. Furthermore, the clustering structure can be obtained from $\bm\alpha$. Specifically, if $\bm\alpha_{kk^{\prime}}=\bm 0$, clients $k$ and $k^{\prime}$ belong to the same cluster.
\end{rem}

\noindent{\bf  Tuning parameter selection}
Following the literature, we set $\nu=1$ and the concavity related parameter $\tau=3$. Following \cite{Yang2019}, we select $\lambda_1$ and $\lambda_2$ with the modified BIC defined as
\[
\text{mBIC}(\lambda_1,\lambda_2)=\frac{1}{N}\sum_{k=1}^Kn_k(\bm\theta^{(k)\top}\widetilde{\mathbf V}^{(k)}\bm\theta^{(k)}-2\bm\theta^{(k)\top}\widetilde{\bm\zeta}^{(k)})+C_N\frac{\log N}{N}\widehat q,
\]
where $\widehat q$ is the number of nonzero distinct coefficient vectors, and $C_N$ is a positive constant depending on $N$. Following \cite{ma2017concave}, we adopt $C_N=\log(\log(Kp))$, which can automatically adapt to a diverging number of parameters.

\section{Theoretical Properties}
Here we establish that the proposed ICR estimator has the well-desired estimation consistency, model selection consistency, and clustering consistency properties. Although sharing some similar spirit with the existing studies, with a significantly different problem and penalized estimation, our theoretical development can have unique value.

\subsection{Notations and Definitions}
For a vector $\mathbf z=(z_1,\dots,z_p)\in\mathbb R^p$, and $1\le l<\infty$, define $||\mathbf z||_l=(\sum_{j=1}^p|z_j|^l)^{1/l}$ and $||\mathbf z||_{\infty}=\max_{j\in[p]}|z_j|$. Given an index set $\mathcal S$, let $\mathbf z_{\mathcal S}$ denote the subvector of $\mathbf z$ corresponding to the elements of $\mathcal S$.
For a matrix $\mathbf Z_{s\times p}$, let $||\mathbf Z||_2=\sup_{\mathbf v\in\mathbb R^p,||\mathbf v||_2=1}||\mathbf Z\mathbf v||_2$, $||\mathbf Z||_{\infty}=\max_{1\le i\le s}\sum_{j=1}^p|Z_{ij}|$, $||\mathbf Z||_{\max}=\max_{1\le i\le s, 1\le j\le p}|Z_{ij}|$, and $||\mathbf Z||_{F}=\sqrt{\sum_{i=1}^s\sum_{j=1}^pZ_{ij}^2}$. For two index sets $\mathcal S_1$ and $\mathcal S_2$, let $\mathbf Z_{\mathcal S_1\mathcal S_2}$ denote the submatrix of $\mathbf Z$ corresponding to the rows in $\mathcal S_1$ and  columns in $\mathcal S_2$, and $\mathbf Z_{\mathcal S_1}$ denote the submatrix of $\mathbf Z$ corresponding to the rows in $\mathcal S_1$.
For a vector $\mathbf v_0\in\mathbb R^p$, let $\mathcal B_r(\mathbf v_0)=\{\mathbf v\in\mathbb R^p:||\mathbf v-\mathbf v_0||_2\le r\}$ be the $\ell_2$-ball around $\mathbf v_0$ with radius $r>0$. For a random variable $X$, its sub-Gaussian norm is defined by $||X||_{\psi_2}=\sup_{s\ge 1}s^{-1/2}(\mathbb E|X|^s)^{1/s}$. For a random vector $\mathbf z\in\mathbb R^p$, its sub-Gaussian norm is defined by $||\mathbf z||_{\psi_2}=\sup_{\mathbf v\in\mathcal B_1(\bm 0)}||\mathbf v^{\top}\mathbf z||_{\psi_2}$. For a symmetric matrix $\mathbf H$, its maximum and minimum eigenvalues are denoted by $\Lambda_{\max}(\mathbf H)$ and $\Lambda_{\min}(\mathbf H)$, respectively. For two sequences of real numbers $\{a_n\}\ge 1$ and $\{b_n\}\ge 1$, let $a_n\gg b_n$ (or $b_n\ll a_n$) denote $b_n/a_n=o(1)$, and $a_n\asymp b_n$ denote $a_n$ is of the same order as $b_n$.  Let $f^{\prime}(a,y)=\partial f(a,y)/\partial a$ and $f^{\prime\prime}(a,y)=\partial^2 f(a,y)/\partial a^2$, where $\partial f(a,y)/\partial a$ and $\partial^2 f(a,y)/\partial a^2$ denote the first and second order derivatives of $f(a,y)$ with respect to $a$, respectively. 

Let $\mathbf V^{(k)}(\bm\theta^{(k)})=\partial^2\widehat{L}_k(\bm\theta^{(k)})/\partial\bm\theta^{(k)}\partial\bm\theta^{(k)\top}$ and $\mathbf g^{(k)}(\bm\theta^{(k)})=\partial\widehat{L}_k(\bm\theta^{(k)})/\partial\bm\theta^{(k)}$. We further denote $\widetilde{\mathbf V}^{(k)}=\mathbf V^{(k)}(\widetilde{\bm\theta}^{(k)})$, $\mathbf V^{*(k)}=\mathbf V^{(k)}(\bm\theta^{*(k)})$, $\widetilde{\mathbf g}^{(k)}=\mathbf g^{(k)}(\widetilde{\bm\theta}^{(k)})$ and $\mathbf g^{*(k)}=\mathbf g^{(k)}({\bm\theta}^{*(k)})$ for simplicity. Let $\varphi^{(k)}=||\mathbb E(\mathbf V_{\mathcal A^c\mathcal A}^{*(k)})\big[\mathbb E(\mathbf V_{\mathcal A\mathcal A}^{*(k)})\big]^{-1}||_{\infty}$ for any $k\in[K]$ and  $\varphi_{\max}=\max_{k\in[K]}\varphi^{(k)}$.
Let $N_m=\sum_{k\in\mathcal G^{(m)}}n_k$, $N_{\max}=\max_{m\in[M]}N_m$, and $N_{\min}=\min_{m\in[M]}N_m$. Let $|\mathcal G^{(m)}|$ be the cardinality of index set $\mathcal G^{(m)}$ with $m\in[M]$, and denote $|\mathcal G_{\max}|=\max_{m\in[M]}|\mathcal G^{(m)}|$ and $|\mathcal G_{\min}|=\min_{m\in[M]}|\mathcal G^{(m)}|$.  

Let $\mathcal M_{\mathcal G}$ be a subspace of $\mathbb R^{p\times K}$ defined as
\[
\mathcal M_{\mathcal G}=\Big\{\bm\theta\in\mathbb R^{p\times K}: \bm\theta^{(k)}=\bm\theta^{(k^{\prime})}=\boldsymbol\psi^{(m)}, \text{for any}\ k,k^{\prime}\in\mathcal G^{(m)}, 1\le m\le M\Big\},
\]
where $\boldsymbol\psi^{(m)}$ is the distinct coefficient vector for the $m$th cluster. Further, we define the $p\times M$ common coefficient matrix $\boldsymbol\psi=(\boldsymbol\psi^{(1)},\dots,\boldsymbol\psi^{(M)})=(\bm\psi_1,\dots,\bm\psi_p)^{\top}$,where $\bm\psi^{(m)}=(\psi_1^{(m)},\dots,\psi_{p}^{(m)})^{\top}$ and $\bm\psi_j=(\psi_j^{(1)},\dots,\psi_j^{(M)})^{\top}$.

Let $\bm\theta^*$ be the true coefficient matrix corresponding to $\bm\theta$ and $\bm\psi^*$ be the true coefficient matrix corresponding to $\bm\psi$. Without loss of generality, assume the first $q$ groups of covariates have nonzero effects, and the rest $(p-q)$ have zero effects. Let $\mathcal A=\{1,\dots, q\}$ and $\mathcal A^c=\{q+1,\dots, p\}$. Further, denote $d_1=\min_{j\in \mathcal A}||\bm\psi_j^{*}||_2$ and $d_2=\min_{m,m^{\prime}\in[M], m\ne m^{\prime}}||\bm\psi_{\mathcal A}^{*(m)}-\bm\psi_{\mathcal A}^{*(m^{\prime})}||_2$.

If the underlying true cluster memberships $\mathcal G^{(1)},\dots,\mathcal G^{(M)}$ are known, we can define the cluster-oracle objective function for $\bm\theta$ by
\begin{equation}\label{cor_theta}
\mathop{\mathrm{arg\ min}}\limits_{\bm\theta\in\mathcal M_{\mathcal G}}\ \Bigg\{\mathcal L(\bm\theta)=\frac{1}{N}\sum_{k=1}^Kn_k\Big(\bm\theta^{(k)\top}\widetilde{\mathbf V}^{(k)}\bm\theta^{(k)}-2\bm\theta^{(k)\top}\widetilde{\bm\zeta}^{(k)}\Big)+\sum_{j=2}^{p}p_{\tau}(||\boldsymbol\theta_j||_2,\lambda_1)\Bigg\}.
\end{equation}
Accordingly, the cluster-oracle objective function for the common coefficient matrix $\bm\psi$ is
\begin{equation}\label{cor_psi}
\begin{aligned}
\mathcal L^{\mathcal G}(\bm\psi)=&\ \frac{1}{N}\sum_{m=1}^M\Bigg[\bm\psi^{(m)\top}\Bigg(\sum_{k\in\mathcal G^{(m)}}n_k\widetilde{\mathbf V}^{(k)}\Bigg)\bm\psi^{(m)}-2\bm\psi^{(m)\top}\Bigg(\sum_{k\in\mathcal G^{(m)}}n_k\widetilde{\bm\zeta}^{(k)}\Bigg)\Bigg]\\
&\ +\sum_{j=2}^pp_{\tau}\Bigg(\sqrt{\sum_{m=1}^M\big(|\mathcal G^{(m)}|^{1/2}{\psi}_j^{(m)}\big)^2},\lambda_1\Bigg).
\end{aligned}
\end{equation}

\subsection{Asymptotic Properties} 
Assume that $n_k\asymp N/K$ for $k\in[K]$, and we denote $n^*\asymp n_k$. We further assume the following mild conditions.

\begin{enumerate}[(C1)]
\item For each  $k\in [K]$ and $i\in[n_k]$, $\{\bm x_i^{(k)}, y_i^{(k)}\}$’s are  independent and identically distributed. There exists a constant $C_x>0$ such that $\max_{k\in[K], i\in[n_k]}||\bm x_i^{(k)}||_{\infty}\le C_x$ and $\max_{\mathbf x\in\mathcal B_1{(\bm 0)}}\mathbb E(\mathbf x^{\top}\bm x_i^{(k)})^2\le C_x$.  
\item For each  $k\in [K]$ and $i\in[n_k]$, $f^{\prime}(\bm\theta^{*(k)\top}\bm x_i^{(k)},y_i^{(k)})$’s are sub-Gaussian. That is, there exists a constant $\kappa_x>0$ such that $||f^{\prime}(\bm\theta^{*(k)\top}\bm x_i^{(k)},y_i^{(k)})||_{\psi_2}\le\kappa_x$. 

\item For each  $k\in [K]$, there exist two constants $C_{\min}$ and $C_{\max}$ such that $0<C_{\min}\le\Lambda_{\min}\big(\mathbb E({\mathbf V}_{\mathcal A\mathcal A}^{*(k)})\big)\le\Lambda_{\max}\big(\mathbb E({\mathbf V}_{\mathcal A\mathcal A}^{*(k)})\big)\le C_{\max}$.

\item For each $k\in[K]$, if $\delta=o(1)$, then there exists a constant $C_L>0$ such that
\[
|f^{\prime\prime}(\bm\theta^{(k)\top}\bm x_i^{(k)},y_i^{(k)})|\le C_L,\quad\text{for all}\ \bm\theta^{(k)}\in\mathcal B_{\delta}(\bm\theta^{*(k)}).
\]
Further, the second-order derivatives are Lipschitz continuous. That is, 
\[
\big|f^{\prime\prime}(a,y)-f^{\prime\prime}(b,y)\big|\le C_L|a-b|,\quad \text{for any}\ a,b,y\in\mathbb R.
\]
\item The local estimators satisfy
\[
\max_{k\in[K]}||\widetilde{\bm\theta}^{(k)}-\bm\theta^{*(k)}||_2\asymp \max_{k\in[K]} n_k^{-1/2}||\mathbf X^{(k)}(\widetilde{\bm\theta}^{(k)}-\bm\theta^{*(k)})||_2=O_p(\sqrt{q\log p/{n^*}}).
\]
\item The penalty function $p_{\tau}(t, \lambda)$ is non-decreasing and concave in $t$ for $t\in[0,\infty)$. For $\tau>0$, $\lambda^{-1}p_{\tau}(t, \lambda)$ is a constant for all $t\ge \tau\lambda$, and $p_{\tau}(0, \lambda)=0$. In addition, $p_{\tau}^{\prime}(t, \lambda)$ exists and is continuous
except for a finite number of $t$ values and $\lambda^{-1}p_{\tau}^{\prime}(0+, \lambda)=1$.
\item $(|\mathcal G_{\max}|/|\mathcal G_{\min}|)^2q^4\log p\ll n^*$ and $K\ll p$.
\end{enumerate}
Condition (C1) assumes that all covariates are uniformly bounded. Similar conditions have been commonly assumed in the literature, especially including \cite{cai2021individual}. It is satisfied under many practical scenarios.
Condition (C2) controls the tail behavior of $x_{ij}^{(k)}f^{\prime}(a,y)$ and bounds the random error $\mathbf g^{*(k)}$. Condition (C3) has been commonly assumed to ensure that the eigenvalues of $\mathbb E({\mathbf V}_{\mathcal A\mathcal A}^{*(k)})$ are bounded above and below. The first part of Condition (C4) assumes that the second-order derivatives of the loss function are bounded, and the second part is a Lipschitz condition to ensure that the loss function is sufficiently smooth. Condition (C5) provides the error bounds for the local estimators, and similar conditions have been assumed in \cite{cai2021individual} and \cite{Battey2018}. It is noted that such error bounds have been established in \cite{Negahban2012}. Condition (C6) is commonly assumed under high-dimensional settings, and it can be easily verified that both MCP and SCAD satisfy this condition.

\begin{thm}
Suppose that Conditions (C1)-(C7) hold. If $|\mathcal G_{\min}|^{1/2}d_1>\tau\lambda_1$, $\lambda_1\gg |\mathcal G_{\max}|^{1/2}r_{1N}+\varphi_{\max}r_{2N}$, and $r_{1N}=o(1)$, then there exists a strictly local minimizer $\widehat{\bm\psi}^{or}$ of $\mathcal L^{\mathcal G}(\bm\psi)$ in (\ref{cor_psi}) such that
\[
||\widehat{\bm\psi}_{\mathcal A}^{or}-\bm\psi_{\mathcal A}^*||_F=O_p(r_{1N}),\qquad\widehat{\bm\psi}_{\mathcal A^c}^{or}=\bm 0,
\]
with probability approaching 1 as $N\to\infty$, where
\[
\begin{aligned}
&r_{1N}=\sqrt{\frac{(K/|\mathcal G_{\min}|)q}{N_{\min}}}+\frac{|\mathcal G_{\max}|M^{1/2}q^{3/2}\log p}{N_{\min}},\\
&r_{2N}=\sqrt{\frac{(|\mathcal G_{\max}|/|\mathcal G_{\min}|)M\log p}{KN}}+\frac{(|\mathcal G_{\max}|/|\mathcal G_{\min}|^{1/2})M^{1/2}q\log p}{N}.
\end{aligned}
\]
\end{thm}

Theorem 1 establishes the estimation consistency and model selection consistency of the oracle estimator $\widehat{\bm\psi}^{or}$. Note that the second term of $r_{1N}$ is the additional error due to the aggregation of summary statistics as opposed to raw data. If $M|\mathcal G_{\max}|=o(\sqrt{N/[q^2(\log p)^2]})$, the second term in the error bound can be dominated by the first term, which means that the additional errors are asymptotically negligible.  Furthermore, if $M$ is fixed and $|\mathcal G_m|\asymp K/M$ for $m\in[M]$, then $r_{1N}$ turns to be $\sqrt{{q}/{N}}+{Kq^{3/2}\log p}/{N}$. Similarly, the additional errors are asymptotically negligible if $K=o(\sqrt{N/q^2(\log p)^2})$, and this condition is similar to that for the distributed sparse high-dimensional models \citep{Battey2018}.

\begin{cor}
Suppose that the conditions of Theorem 1 hold. Then there exists a strictly local minimizer $\widehat{\bm\theta}^{or}$ of $\mathcal L(\bm\theta)$ such that
\[
||\widehat{\bm\theta}_{\mathcal A}^{or}-\bm\theta_{\mathcal A}^*||_F=O_p(|\mathcal G_{\max}|^{1/2}r_{1N}),\qquad\widehat{\bm\theta}_{\mathcal A^c}^{or}=\bm 0,
\]
with probability approaching 1 as $N\to\infty$.
\end{cor}
Corollary 1 directly follows from Theorem 1 and can be used to derive the following theorem.

\begin{thm}
Suppose that the conditions of Theorem 1 hold. If $\lambda_1\gg \varphi_{\max}(\log p/N)^{1/2}$, $d_2>\tau\lambda_2$ and $\lambda_2\gg |\mathcal G_{\max}|^{1/2}r_{1N}$, then there exists a strictly local minimizer $\widehat{\bm\theta}$ of $\widehat{\mathcal Q}_{\mathrm{ICR}}(\bm\theta)$ that satisfies
\[
P(\widehat{\bm\theta}=\widehat{\bm\theta}^{or})\to 1.
\]
\end{thm}
Theorem 2 shows that the oracle estimator $\widehat{\bm\theta}^{or}$ is a local minimizer with a high probability. As clustering is based on estimation, the clustering consistency follows from Theorems 1 and 2.

\section{Simulation Study}
We conduct simulation to gauge performance of the proposed approach. For benchmarking, we consider: 
(a) the Local estimator obtained by minimizing a local loss function for each client separately; 
(b) the SK estimators obtained by applying the sparse K-means approach \citep{witten2010framework} to the local estimators. This is achieved using R package {\it sparcl}. We adopt two criteria, namely the Hartigan statistic \citep{hartigan1975clustering} and gap statistic \citep{tibshirani2001estimating}, to choose the number of clusters -- this is realized using R package {\it NbClust}. The two approaches are accordingly referred to as SK(har) and SK(gap), respectively;
(c) the SMA estimator \citep{He2016} obtained after executing the sure independent screening procedure \citep{Fan2008Lv} to reduce dimension to $n/(3\log n)$ as recommended by \cite{He2016}; 
and (d) the Oracle estimator obtained by minimizing objective function $\mathcal L^{or,\mathcal G}(\bm\psi)$ in (\ref{or_obj}). Note that this alternative is not realistic in practice. Here, it serves as the ideal target. 
For the alternatives, tuning parameters are chosen in a way compatible with the proposed. 

Four examples are designed. Examples 1-2 are on logistic regression and logistic loss, 
and Examples 3-4 are on linear regression and squared loss. The true number of clusters is $M=2$ in Examples 1 and 3, and $M=4$ in Examples 2 and 4. For all examples, $n_1=\cdots=n_K=n$. More specific settings are as follows.

\noindent\textbf {Example 1.} $\boldsymbol\psi^{(1)}=(0.4\times\mathbf 1_8^{\top},\mathbf 0_{p-8}^{\top})^{\top}$ and $\boldsymbol\psi^{(2)}=(-0.4\times\mathbf 1_8^{\top},\mathbf 0_{p-8}^{\top})^{\top}$. We generate $\bm x_{i,-1}^{(k)}, i\in[n_k],k\in[K]$, where $\bm x_{i,-1}^{(k)}=(x_{i2}^{(k)},\dots, x_{ip}^{(k)})^{\top}$, from a multivariate normal distribution with mean $\mathbf 0$ and $\text{cov}(X_w, X_t)=\rho^{|w-t|}$ for $w,t\in\{2,\cdots,p\}$ and $\rho=0.5$. Given $\mathbf X^{(k)}$, we generate $\mathbf Y^{(k)}$ from a logistic model. We set the number of datasets in each cluster as $|\mathcal G_1|=|\mathcal G_2|=K/2$. We further set $n=200$ and consider  $K\in\{16,32,64\}$ and $p\in\{100,500\}$.

\noindent\textbf {Example 2.} $\boldsymbol\psi^{(1)}=(0.6\times\mathbf 1_4^{\top},-0.6\times\mathbf 1_4^{\top},\mathbf 0_{p-8}^{\top})^{\top}$, $\boldsymbol\psi^{(2)}=(0.6\times\mathbf 1_2^{\top},-0.6\times\mathbf 1_2^{\top},0.6\times\mathbf 1_2^{\top},-0.6\times\mathbf 1_2^{\top},\mathbf 0_{p-8}^{\top})^{\top}$, $\boldsymbol\psi^{(3)}=(-0.6\times\mathbf 1_2^{\top},0.6\times\mathbf 1_2^{\top},-0.6\times\mathbf 1_2^{\top},0.6\times\mathbf 1_2^{\top},\mathbf 0_{p-8}^{\top})^{\top}$, and $\boldsymbol\psi^{(4)}=(-0.6\times\mathbf 1_4^{\top},0.6\times\mathbf 1_4^{\top},\mathbf 0_{p-8}^{\top})^{\top}$. We set the number of datasets in each cluster as $|\mathcal G_1|=|\mathcal G_2|=|\mathcal G_3|=|\mathcal G_4|=K/4$. $\mathbf X^{(k)}$ and $\mathbf Y^{(k)}$ are generated in a similar manner as in Example 1. We consider $K\in\{64,128\}$ and $n\in\{200,400,800\}$ and set $p=100$.

\noindent\textbf {Example 3.} The data generation is the same as in Example 1. The difference is that the response is generated from a linear regression model, where the random error has a normal distribution $\mathcal N(0,\sigma^2)$.
We consider $K\in\{16,32,64\}$ and $\sigma\in\{1,2\}$ and set $p=100$ and $n=100$.

\noindent\textbf {Example 4.} The data generation is the same as in Example 2. The difference is that the response is generated from a linear regression model, where the random error has a normal distribution $\mathcal N(0,\sigma^2)$. We consider $\sigma\in\{1,2\}$ and  $n\in\{100,200,400\}$ and set $p=100$ and $K=64$.

For each example, we generate 100 replicates. We first observe that the proposed computational algorithm has satisfactory convergence properties. With all of our simulated datasets, convergence is achieved within 100 iterations. Additionally, the proposed approach is computationally affordable. For example, the analysis of one simulated dataset under Example 1 with $K=32$, $p=100$ and 25 candidate tuning parameter values takes about 3 minutes using a desktop with standard configurations -- here we note that penalized fusion estimation is in general computationally more expensive.
For evaluation and comparison, we comprehensively consider the following measures. Denote the set of selected variables as $\widehat{\mathcal A}=\{j:{\widehat{\boldsymbol\theta}}_{j}\ne 0\}$. 
For evaluating variable selection accuracy, we consider: (1) TPR, the percentage of correctly identified important variables across the $K$ studies; (2) FPR, the percentage of falsely identified important variables across the $K$ studies; and (3) MS, the model size defined by $\text{MS}=\sum_{j=1}^p \widehat q_j$, where $\widehat q_j$ is the number of distinct nonzero coefficients of the $j$th variable. 
For evaluating clustering accuracy, we consider: (4) $\widehat M$, the number of identified clusters;
(5) Per, the percentage of fully accurate identification;
(6) RI, the Rand Index defined as $\mathrm{RI}=(\mathrm{TP+TN})/(\mathrm{TP+FP+FN+TN})$, where TP (true positive) is the number of pairs of datsets from the same cluster classified into the same cluster, and TN (true negative), FP (false positive), and FN (false negative) are defined accordingly. Since Rand index tends to be large even under random clusterings, we also adopt (7) ARI, the adjusted Rand index defined by $\mathrm{ARI=(RI-\mathbb E(RI))/(\max{(RI)}-\mathbb E(RI))}$, where $\mathrm{\mathbb E(RI)}$ and $\mathrm{\max{(RI)}}$ are the expected value and maximum value of Rand index, respectively. 
Rand index ranges from 0 to 1, adjusted Rand index ranges from -1 to 1, and higher values indicate a better agreement between the identified and true clustering structures.
For evaluating estimation, we consider: (8) RMSE, the root mean squared error of $\widehat{\boldsymbol\theta}$ defined as $\mathrm{RMSE}=\sqrt{\sum_{j\in \mathcal A}||\widehat{\bm\theta}_j-\bm\theta_j^*||_2^2/K}$.

\begin{figure}[hptb]
\centering
\includegraphics[height = 9cm, width = 18cm]{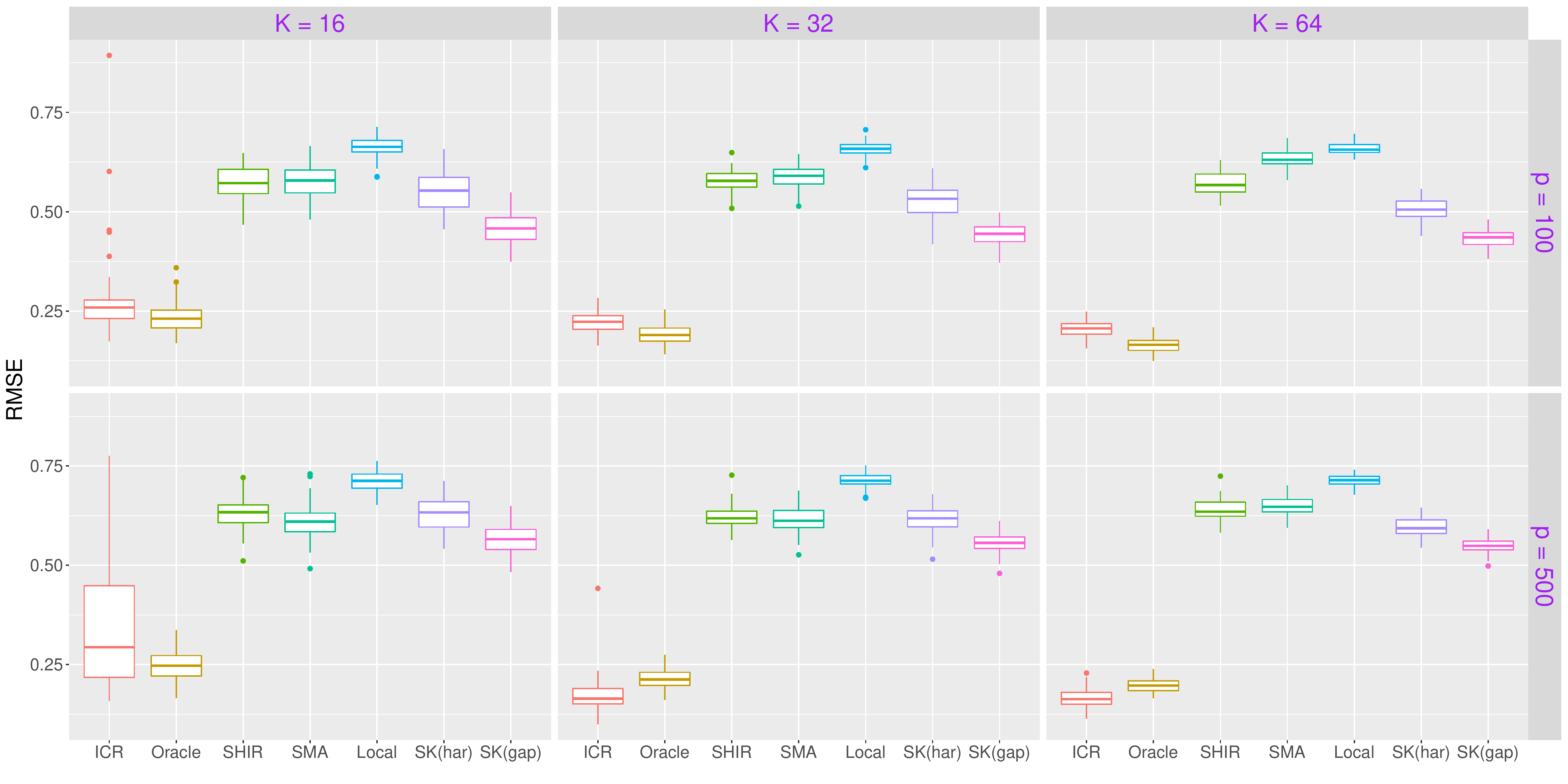}
\caption{Simulation Example 1: boxplot of RMSE. }
\label{ex1_estimation}
\end{figure}

Results for Example 1 are summarized in Table \ref{ex1_variable}, Table \ref{ex1_cluster}, and Figure \ref{ex1_estimation}. 
It is observed that the proposed approach has larger TPR and smaller FPR values, and performance improves as $K$ increases. It has the average MS values around 16, which is the true model size, while the  alternatives generate much larger models. When $K$ is sufficiently large, it outperforms the alternatives in variable selection. Compared to the integrative analysis alternatives SHIR and SMA, the Local approach has worse variable selection performance, which may be caused by the limited sample sizes. Besides, compared to the two-step alternatives SK(har) and SK(gap),it has better variable selection performance. The proposed method and SK(gap) have higher clustering accuracy, while SK(har) often overestimates the number of clusters. Figure \ref{ex1_estimation} suggests that estimation accuracy of the proposed approach is very close to that of Oracle, especially when $K$ is large.

\begin{figure}[hptb]
\centering
\includegraphics[height = 9cm, width = 18cm]{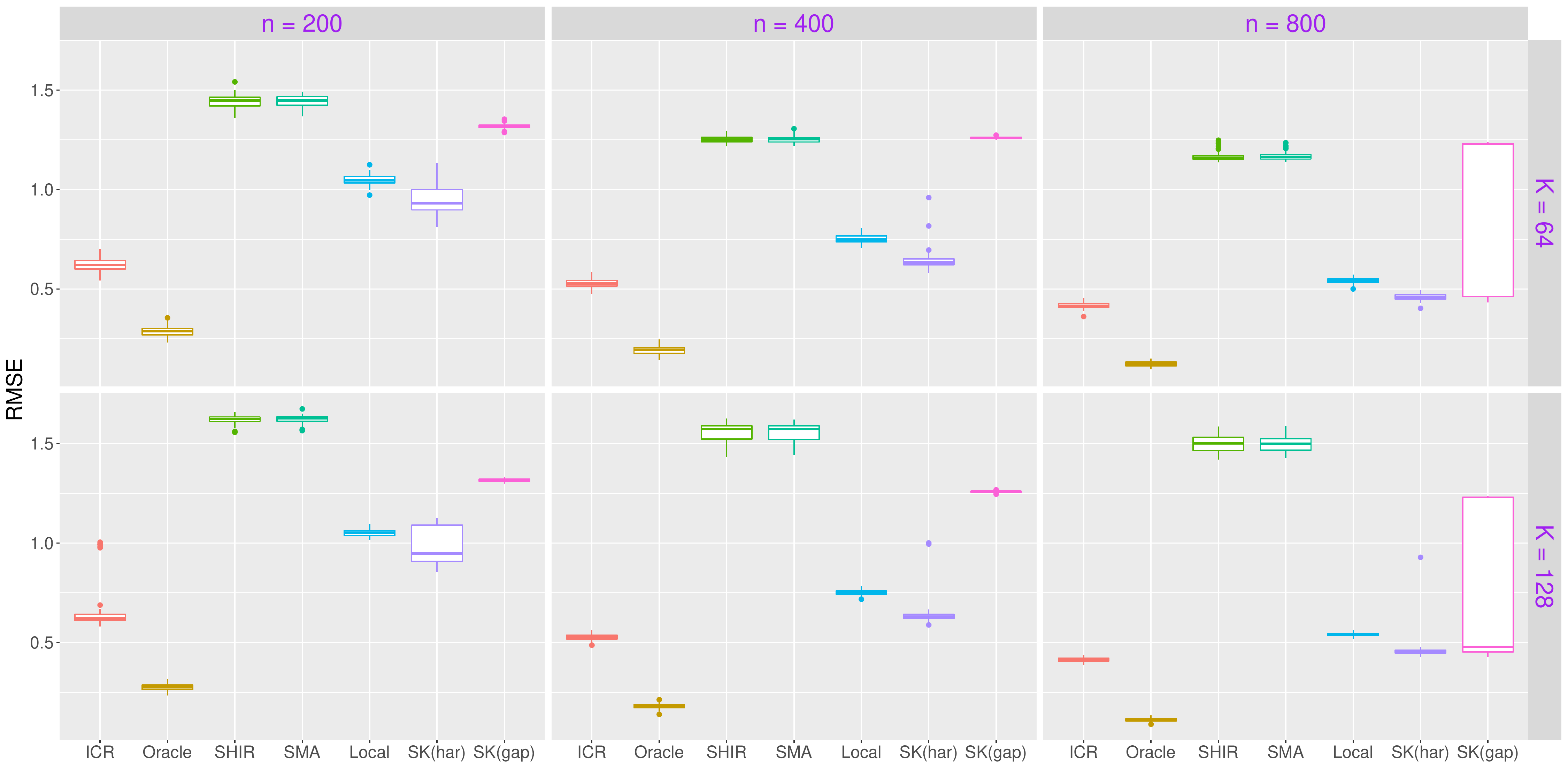}
\caption{ Simulation Example 2: boxplot of RMSE.}
\label{ex2_estimation}
\end{figure}

Results for Example 2 are summarized in Table \ref{ex2_variable}, Table \ref{ex2_cluster}, and Figure \ref{ex2_estimation} (Appendix B). It is observed that the proposed approach has higher variable selection and estimation accuracy, especially when $n$ is large. Its performance is much closer to Oracle. Compared to the Local approach, SHIR and SMA has worse estimation performance, which suggests that inappropriate data integration may not help. Table \ref{ex2_cluster} shows that, with the proposed approach and SK(har), the identified number of clusters is close to the true. Further, the RI and ARI values are close to 1, indicating satisfactory clustering performance. 
A larger $n$ leads to better performance. In comparison, SK(gap) usually underestimates the number of clusters and has much lower clustering accuracy.

\begin{figure}[hptb]
\centering
\includegraphics[height = 9cm, width = 18cm]{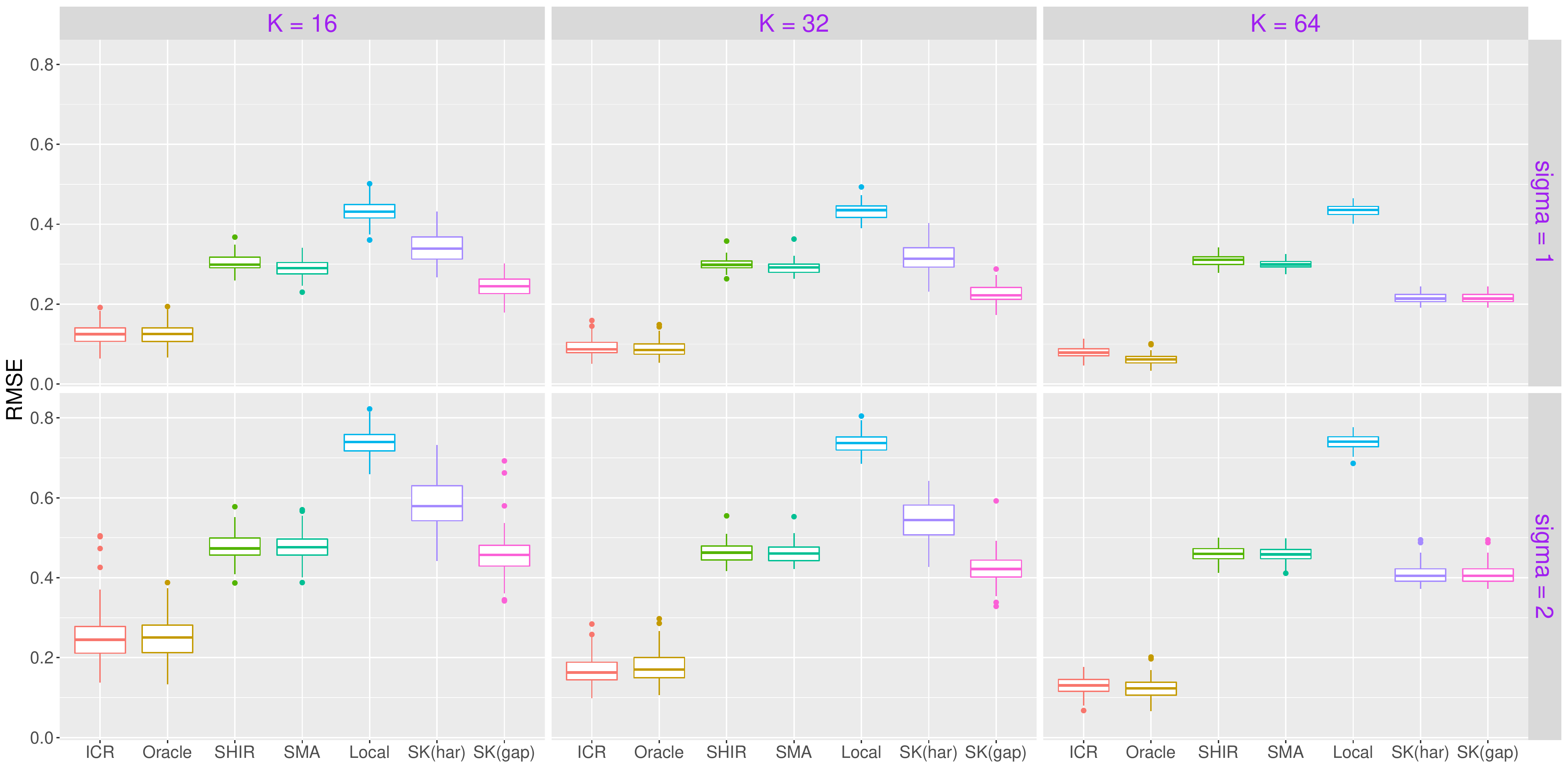}
\caption{ Simulation Example 3: boxplot of RMSE.
}
\label{ex3_estimation}
\end{figure}

\begin{figure}[hptb]
\centering
\includegraphics[height = 9cm, width = 18cm]{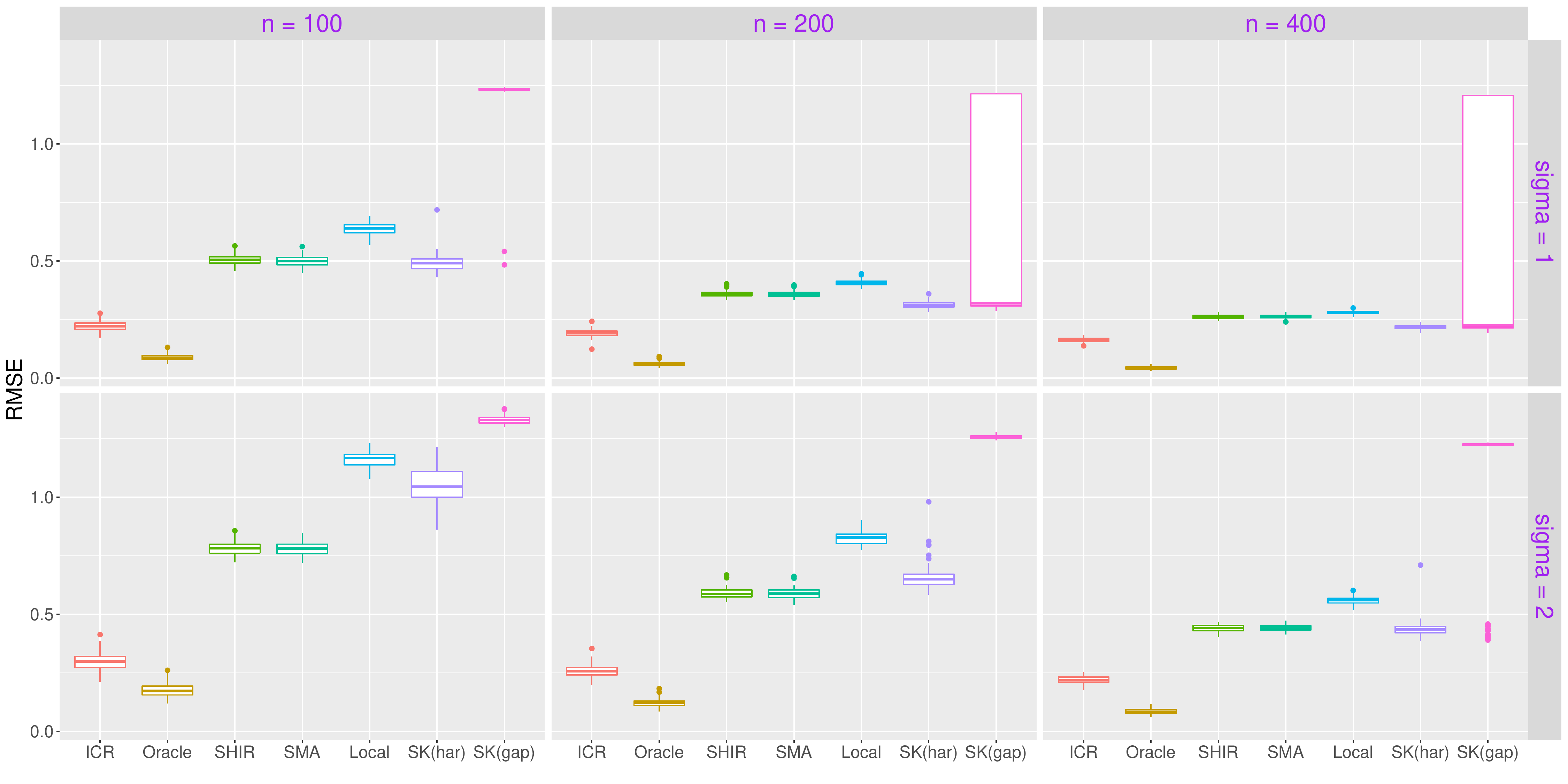}
\caption{ Simulation Example 4: boxplot of RMSE.
}
\label{ex4_estimation}
\end{figure}

Results for Examples 3 and 4 are summarized in Tables \ref{ex3_variable}-\ref{ex4_cluster} and Figures \ref{ex3_estimation}-\ref{ex4_estimation} (Appendix B). The overall findings are very similar to Examples 1-2.

\linespread{1.2}

\begin{table}[H]
	\centering{\scriptsize
		\caption{Simulation Example 1, variable selection accuracy: mean (sd) based on 100 replicates.
		}\label{ex1_variable}
		\setlength{\tabcolsep}{2mm}{
		\begin{tabular}{cccccccccccccc}
		
			\hline
                               \cline{1-11}
                           &&\multicolumn{3}{c}{$K=16$}&\multicolumn{3}{c}{$K=32$}&\multicolumn{3}{c}{$K=64$}\\
                           \cmidrule(lr){3-5}
                           \cmidrule(lr){6-8}
                           \cmidrule(lr){9-11}

		&  Method &TPR&FPR&MS &TPR&FPR&MS &TPR&FPR&MS    \\
		
			\hline

         $p=100$                        &      ICR	  & 0.990              &0.000       &15.840       &1.000     &0.000     &16.000  &1.000     &0.000     &16.000 \\
                               &&(0.058)&(0.000)&(0.929)&(0.000)&(0.000)&(0.000)&(0.000)&(0.000)&(0.000)\\                                     
                                            &      SHIR	  &1.000               &  0.010     &128.900       &1.000     &0.008     &256.730  &1.000     &0.007     &512.670 \\
                               &&(0.000)&(0.012)&(1.087)&(0.000)&(0.009)&(0.863)&(0.000)&(0.010)&(0.922)\\         
                                            &      SMA	  & 1.000              &0.008       &128.780       &1.000     &0.007     &256.66  &1.000     &0.003     &512.920 \\
                               &&(0.000)&(0.011)&(0.991)&(0.000)&(0.008)&(0.742)&(0.000)&(0.007)&(6.402)\\         
                                            &      Local	  & 0.889              & 0.102      &264.180       &0.893     &0.100     &524.470  &0.891     &0.102     &1056.990 \\
                               &&(0.026)&(0.020)&(29.586)&(0.018)&(0.013)&(37.805)&(0.013)&(0.009)&(54.532)\\         
                                            &      SK(har)	  & 0.985              &0.350       &173.69         &0.996     &0.516     &268.43  & 1.000    &0.728     &379.450\\
                               &&(0.051)&(0.161)&(57.414)&(0.021)&(0.210)&(87.220)&(0.000)&(0.183)&(104.023)\\       
                                            &      SK(gap)	  &1.000     &0.571     &121.000       & 1.000    &0.817     &166.400  &1.000     & 0.964    &193.300 \\
                               &&(0.000)&(0.109)&(20.013)&(0.000)&(0.075)&(13.775)&(0.000)&(0.025)&(4.613)\\       
                                                              
          $p=500$                        &      ICR	  & 0.921              &0.000       &14.740       &0.999     &0.000     &16.000  &1.000     &0.000     &16.340 \\
                               &&(0.100)&(0.000)&(1.599)&(0.013)&(0.000)&(0.284)&(0.000)&(0.001)&(0.945)\\                                     
                                            &      SHIR	  &0.999               &  0.004     &129.730       &0.999     &0.004     &257.730  &1.000     &0.002     &513.100 \\
                               &&(0.013)&(0.003)&(2.206)&(0.013)&(0.003)&(3.681)&(0.000)&(0.002)&(1.185)\\         
                                            &      SMA	  & 0.999              &0.004       &129.580       &1.000     &0.003     &257.230  &1.000     &0.001     &512.610 \\
                               &&(0.013)&(0.003)&(2.180)&(0.000)&(0.002)&(1.230)&(0.000)&(0.002)&(0.886)\\         
                                            &      Local	  & 0.849              & 0.034      &373.360       &0.846     &0.033     &741.420  &0.844     &0.034     &1493.910 \\
                               &&(0.029)&(0.006)&(49.342)&(0.021)&(0.004)&(71.624)&(0.016)&(0.003)&(107.543)\\        
                                            &      SK(har)	  &0.984               &0.119       & 281.850      &0.993     &0.181     &487.520  &1.000    &0.354     &865.340 \\
                               &&(0.052)&(0.067)&(121.622)&(0.043)&(0.106)&(211.896)&(0.000)&(0.180)&(357.825)\\           
                                            &      SK(gap)	  & 1.000              & 0.235      &248.190       &1.000     &0.418     &427.740  &1.000     &0.662     &667.620 \\
                               &&(0.000)&(0.061)&(59.641)&(0.000)&(0.061)&(59.943)&(0.000)&(0.050)&(49.399)\\                                                                                             
    \hline
		     \cline{1-11}	
		\end{tabular}
	}}
\end{table}

\begin{table}[H]
	\centering{\scriptsize
		\caption{Simulation Example 1, clustering accuracy: mean (sd) based on 100 replicates.
		}\label{ex1_cluster}
		\setlength{\tabcolsep}{1mm}{
		\begin{tabular}{cccccccccccccc}
		
			\hline
                               \cline{1-14}
                           &&\multicolumn{4}{c}{$K=16$}&\multicolumn{4}{c}{$K=32$}&\multicolumn{4}{c}{$K=64$}\\
                           \cmidrule(lr){3-6}
                           \cmidrule(lr){7-10}
                           \cmidrule(lr){11-14}

		&  Method &$\widehat M$&Per&RI&ARI &$\widehat M$&Per&RI&ARI &$\widehat M$&Per&RI&ARI     \\
		
			\hline

         $p=100$                        &      ICR	  &2.000               &1.000       &1.000       &1.000     &2.000               &1.000       &1.000       &1.000    &2.000               &1.000       &1.000       &1.000 \\
                               &&(0.000)&(-)&(0.000)&(0.000)&(0.000)&(-)&(0.000)&(0.000)&(0.000)&(-)&(0.000)&(0.000)\\                                        
                                            &      SK(har)	  &4.640               &0.000       &0.779         &0.539     & 5.230    &0.000  &0.759     &0.508     &5.220&0.000&0.734&0.462\\
                               &&(1.508)&(-)&(0.097)&(0.208)&(1.847)&(-)&(0.098)&(0.202)&(1.495)&(-)&(0.065)&(0.133)\\       
                                            &      SK(gap)	  &2.000               &1.000       &1.000         &1.000      &2.000               &1.000       &1.000         &1.000  &2.000               &1.000       &1.000         &1.000 \\
                               &&(0.000)&(-)&(0.000)&(0.000)&(0.000)&(-)&(0.000)&(0.000)&(0.000)&(-)&(0.000)&(0.000)\\       
         $p=500$                        &      ICR	  &2.000               &1.000       &1.000       &1.000     &2.000               &1.000       &1.000       &1.000    &2.000               &1.000       &1.000       &1.000 \\
                               &&(0.000)&(-)&(0.000)&(0.000)&(0.000)&(-)&(0.000)&(0.000)&(0.000)&(-)&(0.000)&(0.000)\\                                        
                                            &      SK(har)	  &4.580               &0.000       &0.793         &0.570     & 5.440    &0.000  &0.754     &0.499     &5.150&0.000&0.766&0.528\\
                               &&(1.505)&(-)&(0.101)&(0.216)&(1.684)&(-)&(0.097)&(0.199)&(1.720)&(-)&(0.098)&(0.199)\\       
                                            &      SK(gap)	  &2.010               &0.990       &0.999         &0.999      &2.000               &1.000       &1.000         &1.000  &2.000               &1.000       &1.000         &1.000 \\
                               &&(0.100)&(-)&(0.006)&(0.012)&(0.000)&(-)&(0.000)&(0.000)&(0.000)&(-)&(0.000)&(0.000)\\                                                                                              
    \hline
		     \cline{1-14}	
		\end{tabular}
	}}
\end{table}

\section{Data Application} 

With the emergence of technological innovations, cyberattacks (generally carried out by abnormal requests) are becoming increasingly serious, and may hinder enterprise operations or interrupt critical infrastructure systems. Web logs, which are generated by systems to record  detailed access information, have been widely used to detect abnormal requests in system monitoring and intrusion detection (also called anomaly detection) systems \citep{hu2017anomalous,guo2021anomaly,unal2022anomalyadapters}. Large-scale web logs are usually stored with distributed clients, and the transmission of raw logs from local clients to a central server is often infeasible. As discussed in \cite{guo2021anomaly}, on one hand, only a small part of raw logs contains useful information, and hence the full transmission of raw logs, which is very time-consuming, is not necessary. On the other hand, to facilitate log analysis, raw logs are often transferred to third-party analytic services, which increases the risks of privacy leakage. To tackle this problem, \cite{guo2021anomaly} resorted to federated learning for anomaly detection under distributed settings. A limitation of this study is that homogeneity among clients is assumed. 
\cite{hu2017anomalous} and \cite{unal2022anomalyadapters} pointed out the heterogeneity among clients and constructed client-specific models. This study can be limited with too many redundant parameters. In this section, we apply the proposed method, which takes into consideration both  multi-source heterogeneity and estimation efficiency, to a bank website logs data, which is stored in multiple interfaces (clients).

%In modern website operation, user interaction with a website relies on the parameter submission process in the http request. An increase in interaction can facilitate a user to obtain more information but can also increase the risk of network intrusion. To tackle the intrusion problem, many studies have examined detecting abnormal requests (or attacks) with the assistance of analytic tools \citep{8629941}. Here we conduct an anomaly detection analysis based on the http request logs extracted from a bank website. Specially, on the bank's website, there is an interface for personal online banking login and an interface corresponding to the search box. Users can use different interfaces when accessing different services. Each URL interface can generate a large number of request logs. In our analysis, each request log corresponds to one sample, and the URL interfaces correspond to the multiple clients. 

\begin{figure}[hptb]
\centering
\includegraphics[height = 6cm, width = 12cm]{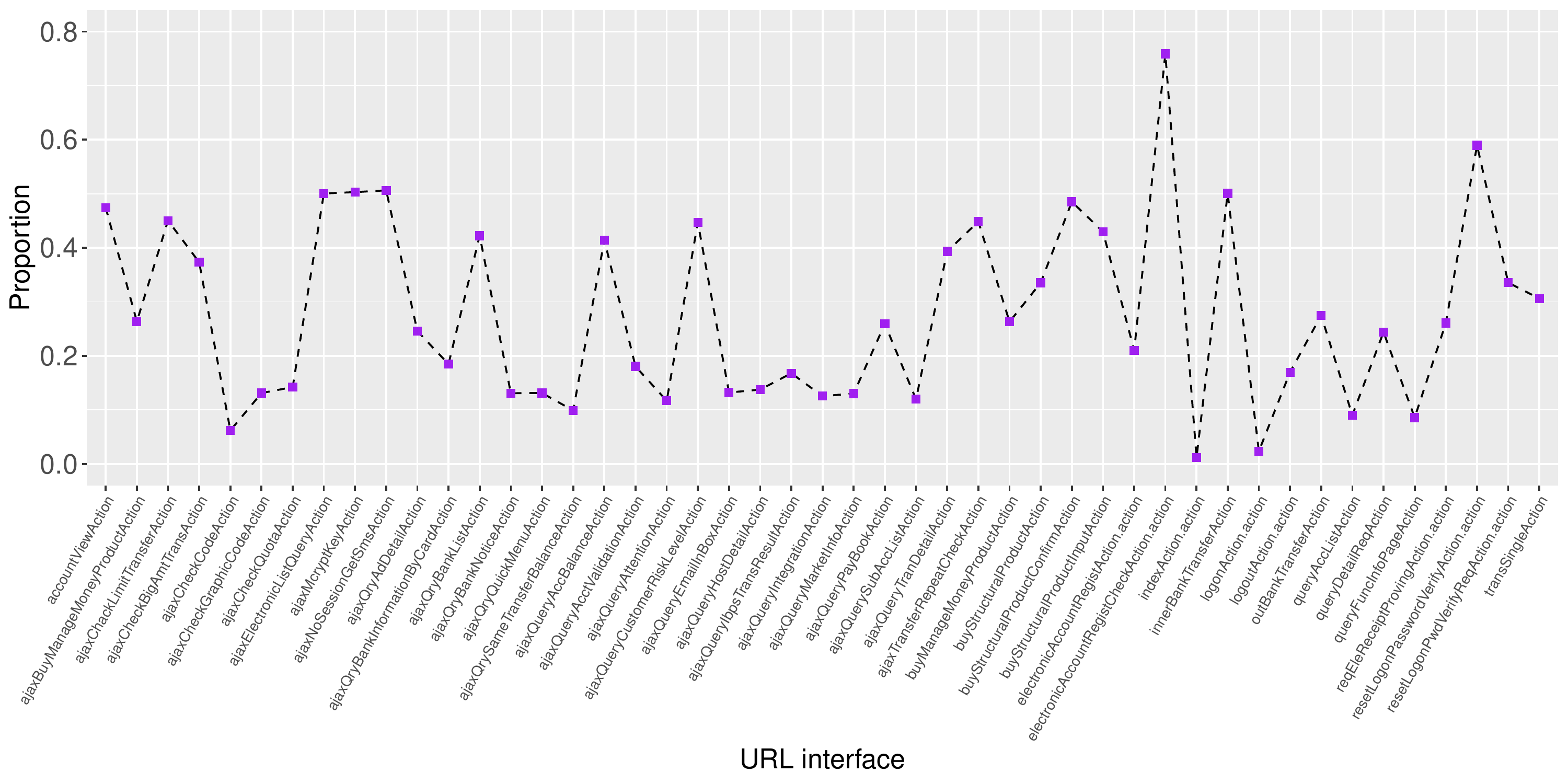}
\caption{Data analysis: abnormal proportions among the 47 URL interfaces.}
\label{prop}
\end{figure}

In this analysis, request type is the binary response and takes values ``normal'' and ``abnormal''. Our goal is to distinguish the abnormal requests from the normal ones based on the log contents, which poses a binary classification problem. There are a total of $K=123$ URL interfaces, and the sample sizes range from 60 to 25,552. The total sample size is $N=306,377$, and the overall percentage of abnormal request is 21.6\%. Among the 123 interfaces, 76 have the percentages of abnormal requests equal to 50\%, and for the rest 47 interfaces, the percentages range from 1.2\% to 75.9\%. In Figure \ref{prop} (Appendix B), we present the percentages of abnormal requests for these 47 URL interfaces.  The significant differences across interfaces suggest the possibility of heterogeneity.

The collected request logs can only be observed in the form of character strings. 
In particular, each request log contains the interface name and two submitted parameters from ``POST'' and ``GET'' access, respectively. In addition, each parameter from POST or GET access consists of a series of key-value pairs separated by ``\&''. For demonstration, in Table \ref{record}, we present one representative record of the initial request logs from URL interface ``ajaxNoSessionGetSmsAction''. The unstructured parameters can be difficult to model, and we extract statistical features from the submitted parameters as follows. First, we generate two features, namely Gnum and Pnum, which are defined as the number of GET and POST key-value pairs, respectively.  Second, we generate two features, namely Glen and Plen, which are defined as the total length of GET and POST parameter, respectively. Third, we generate a series of features to measure the lengths of some key-value pairs in the GET and POST parameters, denoted by $Glx$ and $Plx$, which are defined as the lengths of the (x+1)-th key-value pairs, respectively. Finally, we generate a series of features to measure the types of some key-value pairs in the GET and POST parameters, denoted by $Gtx$ and $Ptx$, which are defined as the types of the (x+1)-th key-value pairs, respectively. There are three types of key-value pairs, namely ``na'', ``str'' and ``num'', which indicate that the key-value pair is missing, string and numeric, respectively. Take the record in Table \ref{record} (Appendix B) as an example. We can obtain the following feature values: Gnum = 1, Glen = 9, Pnum = 4, Plen = 72,  $Gl0=9$, $Gl1=\cdots =Gl19=0$, $Pl0=19$, $Pl1=16$, $Pl2=10$, $Pl3=24$, $Pl4=\cdots =Pl19=0$, $Gt0=$``str'', $Gt1=\cdots =Gt19=$``na'', $Pt0=\cdots =Pt2=$``str'', $Pt3=$``num'', $Pt4=\cdots =Pt19=$``na''. Since the values of $Gl1,\dots, Gl19$ are 0 for all requests, we delete these features. 
To further utilize the character strings, we concatenate $Gt0$ -- $Gt19$ sequentially into a sequence of strings and train the Skip-gram model (which is a popular model of word2vec) to obtain a 40-dimensional continuous word vector features, denoted by $Gw1,\dots, Gw40$. Similarly, we obtain another 40-dimensional continuous word vector  features, denoted by $Pw1,\dots, Pw40$, from $Pt0$ -- $Pt19$. Overall, there are 105 features available for analysis.

Prior to analysis, we standardize the $p=105$ continuous variables to have means 0 and variances 1. The proposed method identifies seven nontrivial clusters (with sizes larger than one and denoted as ${\mathrm{ICR}^{(1)}},\dots, {\mathrm {ICR}^{(7)}}$), which have sizes 35, 34, 8, 7, 3, 2 and 2, respectively. Additionally, there are 32 interfaces forming their own individual clusters. In Table \ref{realdata}, we present important variables identified by the proposed method and/or the two integrative analysis alternatives SHIR and SMA. It is observed that different methods lead to quite different identification and selection results.
Note that, for the proposed method, we only present estimates for the nontrivial clusters. The results for the trivial clusters are omitted and available from the authors. 
From Table \ref{realdata}, we can see that, with the proposed method, Gnum has a negative effect on the probability of abnormal request, while $Gl0$, $Pl0$, $Pl1$ and $Pl2$ have positive effects. This suggests that requests with less GET key-value pairs, a longer length of the first key-value pair in GET, and longer lengths of the first three  key-value pairs in POST are more likely to be abnormal for most of the interfaces. This can potentially lead to a security rule for the initial screening of abnormal requests. Here we note that the traditional anomaly detection for logs is to extract security rules from samples \citep{el2018validation}. Additionally, the proposed analysis has reduced the number of models to 39 (7 clustered ones and 32 individual ones), which corresponds to a much lower cost of maintaining models than the alternative client-specific modeling methods (with 123 models).

\begin{figure}[hptb]
\centering
\subfloat[AUC]{
 \includegraphics[height=5.33cm, width =8cm]{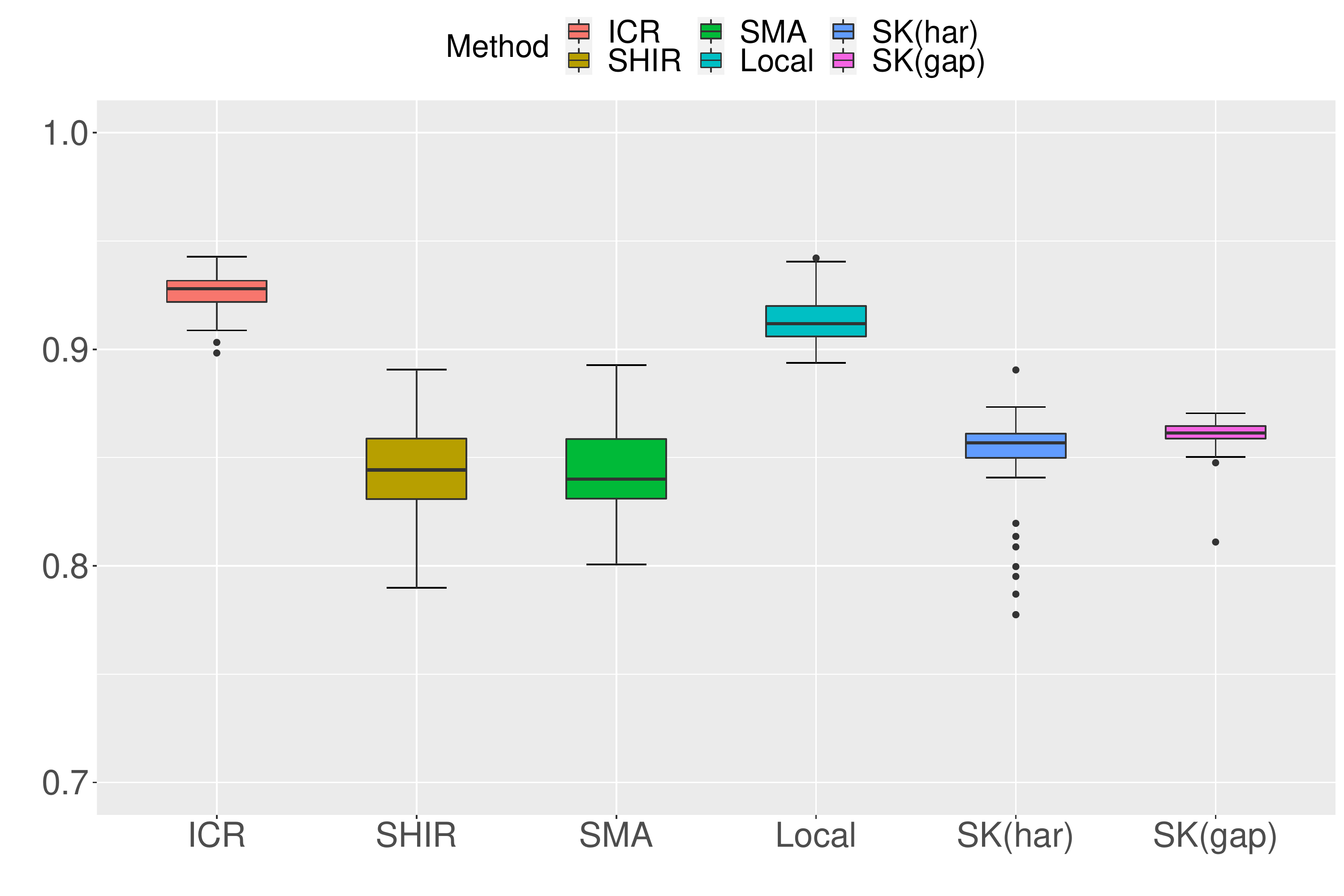}}
 \subfloat[Brier Score]{
 \includegraphics[height=5.33cm, width =8cm]{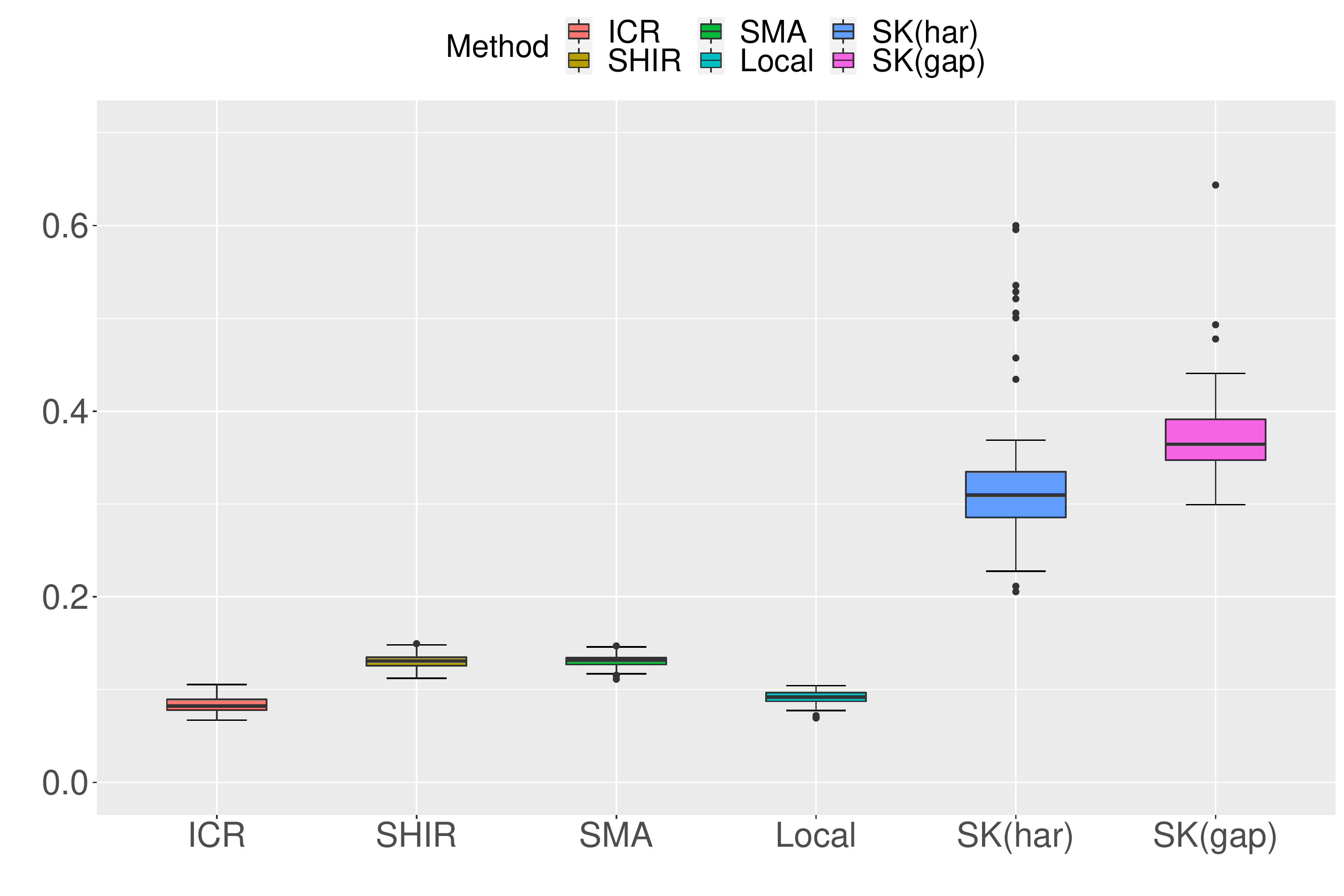}}\\
\centering
\caption{Data analysis: Boxplots of (a) AUC and (b) Brier Score based on 100 random splits.}\label{validation}
\end{figure}

With practical data, it is difficult to objectively evaluate identification and estimation results. To support our finding, we conduct a prediction evaluation. Specifically, we randomly select 3/4 of the samples and form the training data. In this selection, the normal:abnormal ratio is retained. The remaining samples forms the testing data. Estimation is conducted using the training data, and we evaluate prediction performance on the testing data using AUC and brier score. This process is repeated 100 times, and the results are summarized in Figure \ref{validation} (Appendix B). It is observed that the proposed method has better prediction performance than the alternatives. Additionally, the Local method also has competitive performance, which suggests that inappropriate integration may lead to inferior prediction. Compared to Local, the proposed method can have better interpretability.

\begin{table}[H]
	\centering{\small
		\caption{Data analysis: identified important variables and their estimates using the three integrative analysis methods. For the proposed method, only estimates for the nontrivial clusters are shown.
		}\label{realdata}
		\setlength{\tabcolsep}{2mm}{
		\begin{tabular}{cccccccccccc}
		
			\hline

		 Variable&${\mathrm {ICR}^{(1)}}$&${\mathrm {ICR}^{(2)}}$&${\mathrm {ICR}^{(3)}}$&${\mathrm {ICR}^{(4)}}$&${\mathrm {ICR}^{(5)}}$&${\mathrm {ICR}^{(6)}}$&${\mathrm {ICR}^{(7)}}$&SHIR&SMA  \\
		
			\hline

                       Intercept	             & -0.168              & 3.967                &-1.207           &9.173                    &4.863                  &4.330           &1.150                &-1.757       &-1.754           \\
                       Gnum	           & -0.027                & -0.522                &-0.002          &-0.430                     &-0.054              &-0.116         &-0.077                &--         &--          \\
                     Glen	                   & 0.011                    & -0.278             &0.096               &-0.709                    &0.126                &0.292           &-0.049               &-0.054        &-0.055                   \\
                       Pnum	         & 0.062                    & 0.036              &-0.052            &0.036                    & -0.079                 &0.736           &-0.228              &0.004         &--          \\
                   Plen	                  & 0.164                    & -0.066             &-0.740              &-0.754                       &0.465              & -1.374          &-0.109              &--         &--          \\
                   $Gl0$	                    & 0.142                    & 2.047               &0.603              &2.377                        &0.005             &1.564            &2.559               &0.152          &0.152          \\
                  $Pl0$	                    & 0.197                     & 1.841                &1.463                   &2.384                 &0.857                   &2.170      &0.930                &0.206         &0.212             \\
                   $Pl1$	                    & 0.523                     & 5.356                &4.939                  &8.852                 &3.925                   &8.744      &1.271                 &0.533         &0.542            \\
                    $Pl2$	                    & 0.396                     & 3.984                &5.607                   &9.839                 &8.197                   &8.885      &3.925                &0.437        &0.449              \\
                     $Pl4$	                    & 0.034                     & 0.011                &0.061                   &0.017                 &-0.016                   &-0.005      &-0.002            &0.068        &0.068                  \\
                      $Pl5$	                    & 0.006                     & 0.025                &0.133                   &0.011                 &-0.024                   &0.201      &-0.016               &0.026       &0.024                \\
                      $Pl6$	                    & 0.004                     & -0.002                &0.009                   &0.010                 &-0.021                   &0.002      &-0.018              &--       &--                  \\
                       $Pl9$	                    & --                     &--                 & --                  &--                 &--                   &--      &--                 &0.024         &0.025            \\
                       $Pl10$	          & --                     &--                 & --                  &--                 &--                   &--      &--                 &0.010         &0.010            \\
                      $Pl12$	             & 0.005                     & 0.005                &0.007                   &0.032                 &-0.007                   &-0.530      &0.005           &--        &--                    \\
                      $Pl17$	             & -0.003                    & 0.008                &0.014                   &-0.039                 &-0.005                   &-0.001      &0.001         &--         &--                     \\
                      $Gw11$	          & --                     &--                 & --                  &--                 &--                   &--      &--                &0.026          &0.026            \\
                       $Gw31$	          & --                     &--                 & --                  &--                 &--                   &--      &--                 &--          &-0.007           \\
                      $Gw32$	          & --                     &--                 & --                  &--                 &--                   &--      &--                 &-0.005       &--               \\
                      $Gw34$	          & --                     &--                 & --                  &--                 &--                   &--      &--                  &-0.006      &-0.006              \\
                      $Gw36$	          & --                     &--                 & --                  &--                 &--                   &--      &--                 &0.003         &0.004            \\
                       $Pw1$	             & 0.012                     & -0.062                &0.116                   &-0.043                 &-0.104             &-1.165      &-0.266            &--           &--                    \\
                       $Pw27$	          & --                     &--                 & --                  &--                 &--                   &--      &--                  &0.013        &0.018            \\
                       $Pw37$	          & --                     &--                 & --                  &--                 &--                   &--      &--                 &0.018          &0.015           \\

                                   \hline

		\end{tabular}
		
	}
	
}
\end{table}

\section{Conclusion}
\label{sec:conc}
In this article, we have developed a new integrative data analysis method that is based on summary statistics and hence can sufficiently protect privacy of individual clients’ data. The most significant advancement is that it allows for data/model heterogeneity and can automatically identify the underlying clustering structure. Our rigorous theoretical investigation has shown that the proposed method has multiple much desired consistency properties. Additionally, simulation and data analysis have shown its competitive numerical performance. 

This study can be potentially extended in multiple directions. The same as the existing one-shot methods, the proposed analysis only demands one communication between the local clients and central server. Our Theorem 1 suggests that the additional error due to the aggregation of summary statistics is asymptotically negligible when we properly restrict the divergence rate of $K$. If we allow multiple communications (which may lead to higher computational cost), this condition can be relaxed, and there is also a possibility of further improving numerical performance. The proposed method demands mild conditions on the lack-of-fit function. In numerical studies, we have investigated the logistic and linear regressions. The proposed strategy can be potentially applied to much broader models/loss functions. Another possible extension, as previously mentioned, is to apply two-level penalized selection and allow different sparsity structures for multiple clients. These aforementioned extensions will be postponed to future research.

%\bigskip
%\begin{center}
%{\large\bf SUPPLEMENTARY MATERIAL}
%\end{center}

%\begin{description}

%\item[Title:] Brief description. (file type)

%\item[R-package for  MYNEW routine:] R-package ÒMYNEWÓ containing code to perform the diagnostic methods described in the article. The package also contains all datasets used as examples in the article. (GNU zipped tar file)

%\item[HIV data set:] Data set used in the illustration of MYNEW method in Section~ 3.2. (.txt file)

%\end{description}

\section*{Acknowledgements}
This study has been partly supported by NSF 1916251 and 2209685.

%\clearpage
\bibliographystyle{ECA_jasa}
\bibliography{reference}

\clearpage
\section*{Appendix A}

\renewcommand{\theequation}{A.\arabic{equation}}
\setcounter{equation}{0}

This section includes two lemmas and the proof of Theorems 1 and 2.

\begin{lem}\label{lem1}
Suppose that $z_1,\dots,z_n\in\mathbb R$ are independent and centered sub-Gaussian random variables. Let $\mathbf z=(z_1,\dots, z_n)^{\top}$ and $\kappa=\max_{i\in[n]}||z_i||_{\psi_2}$. Then for any $\mathbf a=(a_1,\dots,a_n)^{\top}\in\mathbb R^n$ and  $t> 0$, there exists a constant $C_1>0$ such that
\[
P(|\mathbf a^{\top}\mathbf z|\ge t)\le 2\exp\bigg(-\frac{C_1t^2}{\kappa^2||\mathbf a||_2^2}\bigg).
\]
\end{lem}
\noindent \textbf{Proof.} Lemma \ref{lem1} follows directly from  Lemma 14.3, Chapter 14.2.2 of \cite{Buhlmann2011}. $\hfill \square$

\begin{lem}[]\label{lem2}
Assume that Conditions (C1), (C4), and (C5) hold.   If $K\ll p$ and $q\log p\ll n^*$, then with probability approaching 1, 
\[
\max_{k\in[K]}\Big|\Big|\widetilde{\mathbf V}^{(k)}-\mathbb E({\mathbf V}^{*(k)})\Big|\Big|_{\max}=O_p\big(\sqrt{{q\log p}/{n^*}}\big).
\]
\end{lem}

\noindent\textbf{Proof.} Note that,
\begin{equation}\label{I}
\begin{aligned}
\max_{k\in[K]}\Big|\Big|\widetilde{\mathbf V}^{(k)}-\mathbb E({\mathbf V}^{*(k)})\Big|\Big|_{\max}&\ \le \max_{k\in[K]}\Big|\Big|\widetilde{\mathbf V}^{(k)}-\mathbb E(\widetilde{\mathbf V}^{(k)})\Big|\Big|_{\max}+\max_{k\in[K]}\Big|\Big|\mathbb E(\widetilde{\mathbf V}^{(k)})-\mathbb E({\mathbf V}^{*(k)})\Big|\Big|_{\max}\\
&\ \triangleq I_1+I_2.\\
\end{aligned}
\end{equation}

First, we derive the bound of $I_1$.
Under Condition (C5), $\max_{k\in[K]}||\widetilde{\bm\theta}^{(k)}-\bm\theta^{*(k)}||_2=o_p(1)$. Then by Conditions (C1) and (C4), for all $k\in[K]$ and $j_1,j_2\in[p]$, with probability approaching 1,
\[
\Big|\widetilde{\mathbf V}^{(k)}_{j_1j_2}\Big|=\bigg|n_k^{-1}\sum_{i=1}^{n_k}f^{\prime\prime}(\widetilde{\bm\theta}^{(k)\top}\bm x_i^{(k)},y_i^{(k)})x_{ij_1}x_{ij_2}\bigg|\le C_LC_x^2.
\]
Then, for any given $t>0$, Hoeffding's inequality and the union bound yield that
\[
P(I_1>t)\le 2Kp^2\exp\bigg\{-\frac{\min_{k\in[K]}{n_k}t^2}{2C_x^4C_L^2}\bigg\},
\]
which leads to $I_1=O_p(\sqrt{\log p/n^*})$.

Next, we derive the bound of $I_2$. For all $k\in[K]$ and $j_1,j_2\in[p]$, under Conditions (C1) and (C4),
\[
\begin{aligned}
\Big|\mathbb E(\widetilde{\mathbf V}^{(k)}_{j_1j_2})-\mathbb E({\mathbf V}^{*(k)}_{j_1j_2})\Big|&\ \le\mathbb E\bigg(\Big|f^{\prime\prime}(\widetilde{\bm\theta}^{(k)\top}\bm x_i^{(k)},y_i^{(k)})x_{ij_1}x_{ij_2}-f^{\prime\prime}({\bm\theta}^{*(k)\top}\bm x_i^{(k)},y_i^{(k)})x_{ij_1}x_{ij_2}\Big|\bigg)\\
&\ \le C_LC_x^2\mathbb E\Big[\Big|\bm x_i^{(k)\top}(\widetilde{\bm\theta}^{(k)}-\bm\theta^{*(k)})\Big|\Big]\\
&\ \le C_LC_x^2\Big\{\max_{\bm v\in\mathcal B_1(\bm 0)}\mathbb E\big[(\bm v^{\top}\bm x_i)^2\big]\big|\big|\widetilde{\bm\theta}^{(k)}-\bm\theta^{*(k)}\big|\big|_2^2\Big\}^{1/2}\\
&\ \le C_LC_x^{5/2}\big|\big|\widetilde{\bm\theta}^{(k)}-\bm\theta^{*(k)}\big|\big|_2.
\end{aligned}
\]
Thus,
\[
I_2\le C_LC_x^{5/2}\max_{k\in[K]}\big|\big|\widetilde{\bm\theta}^{(k)}-\bm\theta^{*(k)}\big|\big|_2=O_p(\sqrt{q\log p/{n^*}}).
\]
Combining the bounds of $I_1$ and $I_2$ as well as (\ref{I}), we can prove the result. $\hfill \square$

\noindent\textbf{Proof of Theorem 1}

Recall the definitions of $\mathcal L(\bm\theta)$ in (\ref{cor_theta}) and $\mathcal L^{\mathcal G}(\bm\psi)$ in (\ref{cor_psi}). We further define two objective functions. If the true clustering structure and important covariate set are known, we can define the oracle estimator $\widehat{\bm\theta}^{or}=(\widehat{\bm\theta}_{\mathcal A}^{or\top},\bm 0_{(p-q)\times K}^{\top})^{\top}$ for $\bm\theta$ as
\begin{equation}\label{or_theta}
\mathop{\mathrm{arg\ min}}\limits_{\bm\theta\in\mathcal M_{\mathcal G},\bm\theta_{\mathcal A^c}=\bm 0}\mathcal L^{or}(\bm\theta)=\frac{1}{N}\sum_{k=1}^Kn_k\Big(\bm\theta^{(k)\top}\widetilde{\mathbf V}^{(k)}\bm\theta^{(k)}-2\bm\theta^{(k)\top}\widetilde{\bm\zeta}^{(k)}\Big).
\end{equation}

Accordingly, the oracle estimator $\widehat{\bm\psi}^{or}=(\widehat{\bm\psi}_{\mathcal A}^{or\top},\bm 0_{(p-q)\times M}^{\top})^{\top}$ of $\bm\psi$ can be defined as
\begin{equation}\label{or_obj}
\mathop{\mathrm{arg\ min}}\limits_{\bm\psi\in\mathbb R^{p\times M},\bm\psi_{\mathcal A^c}=\bm 0}\mathcal L^{or,\mathcal G}(\bm\psi)= \frac1{N}\sum_{m=1}^M\Bigg[\bm\psi^{(m)\top}\Bigg(\sum_{k\in\mathcal G^{(m)}}n_k\widetilde{\mathbf V}^{(k)}\Bigg)\bm\psi^{(m)}-2\bm\psi^{(m)\top}\Bigg(\sum_{k\in\mathcal G^{(m)}}n_k\widetilde{\bm\zeta}^{(k)}\Bigg)\Bigg].
\end{equation}

This theorem can be proved via two steps.

Step 1: Let $\bm\psi_{\mathcal A}=(\bm\psi_{\mathcal A}^{(1)},\dots,\bm\psi_{\mathcal A}^{(M)})$ and $\bm\psi_{\mathcal A}^{(m)}=(\psi_{\mathcal A,1}^{(m)},\dots,\psi_{\mathcal A,q}^{(m)})^{\top}$. Based on the definition of  $\mathcal L^{or,\mathcal G}(\bm\psi)$, we can further define
\begin{equation}\label{or_obj_A}
\mathcal L^{or,\mathcal G}_{\mathcal A}(\bm\psi_{\mathcal A})= \frac1{N}\sum_{m=1}^M\Bigg[\bm\psi_{\mathcal A}^{(m)\top}\Bigg(\sum_{k\in\mathcal G^{(m)}}n_k\widetilde{\mathbf V}^{(k)}_{\mathcal A\mathcal A}\Bigg)\bm\psi_{\mathcal A}^{(m)}-2\bm\psi_{\mathcal A}^{(m)\top}\Bigg(\sum_{k\in\mathcal G^{(m)}}n_k\widetilde{\bm\zeta}_{\mathcal A}^{(k)}\Bigg)\Bigg].
\end{equation}
The solution of (\ref{or_obj_A}) is denoted by $\widehat{\bm\psi}_{\mathcal A}^{{or}}=(\widehat{\bm\psi}_{\mathcal A}^{{or}(1)},\dots,\widehat{\bm\psi}_{\mathcal A}^{{or}(M)})$, and the corresponding true coefficient matrix is denoted by ${\bm\psi}_{\mathcal A}^*=({\bm\psi}_{\mathcal A}^{*(1)},\dots,{\bm\psi}_{\mathcal A}^{*(M)})$. Then we have $\mathcal L^{or,\mathcal G}_{\mathcal A}(\widehat{\bm\psi}_{\mathcal A}^{{or}})\le\mathcal L^{or,\mathcal G}_{\mathcal A}(\bm\psi_{\mathcal A}^*)$, and accordingly,
\begin{equation}\label{ineq:1}
\sum_{m=1}^M\Big(\widehat{\bm\psi}_{\mathcal A}^{{or}(m)\top}\widetilde{\mathbf V}_{\mathcal A\mathcal A}^{\mathcal G(m)}\widehat{\bm\psi}_{\mathcal A}^{{or}(m)}-2\widehat{\bm\psi}_{\mathcal A}^{{or}(m)\top}\widetilde{\bm\zeta}_{\mathcal A}^{\mathcal G(m)}\Big)\le \sum_{m=1}^M\Big({\bm\psi}_{\mathcal A}^{*(m)\top}\widetilde{\mathbf V}_{\mathcal A\mathcal A}^{\mathcal G(m)}{\bm\psi}_{\mathcal A}^{*(m)}-2{\bm\psi}_{\mathcal A}^{*(m)\top}\widetilde{\bm\zeta}_{\mathcal A}^{\mathcal G(m)}\Big),
\end{equation}
where 
\[
\widetilde{\mathbf V}^{\mathcal G(m)}=N^{-1}\sum_{k\in\mathcal G^{(m)}}n_k\widetilde{\mathbf V}^{(k)},\quad \widetilde{\bm\zeta}^{\mathcal G(m)}=N^{-1}\sum_{k\in\mathcal G^{(m)}}n_k\widetilde{\bm\zeta}^{(k)}.
\]

Motivated by \cite{cai2021individual}, for a vector or matrix $A(t)$ whose $(i,j)$th entry $A_{ij}(t)$ is a function of a scalar $t\in[0,1]$, we define $\int_0^1A(t)dt$ as the vector or matrix with its $(i,j)$th entry being $\int_0^1A_{ij}(t)dt$. Then, we can transform the term $-\widetilde{\bm\zeta}^{(k)}$ in (\ref{ineq:1}) into
\begin{equation}\label{expand_zeta}
\begin{aligned}
\widetilde{\mathbf g}^{(k)}-\widetilde{\mathbf V}^{(k)}\widetilde{\bm\theta}^{(k)}&\ =\mathbf g^{*(k)}-\widetilde{\mathbf V}^{(k)}{\bm\theta}^{*(k)}\\
&\ +\int_0^1\bigg\{\mathbf V^{(k)}\Big(\big[\bm\theta^{*(k)}+t(\widetilde{\bm\theta}^{(k)}-\bm\theta^{*(k)})\big]\Big)-\widetilde{\mathbf V}^{(k)}\bigg\}(\widetilde{\bm\theta}^{(k)}-\bm\theta^{*(k)})\ dt.
\end{aligned}
\end{equation}

Plugging (\ref{expand_zeta}) into (\ref{ineq:1}), we have
\begin{equation}\label{ineq:3}
\begin{aligned}
&\ {\sum_{m=1}^M\bigg[\Big(\widehat{\bm\psi}_{\mathcal A}^{{or}(m)}-\bm\psi_{\mathcal A}^{*(m)}\Big)^{\top}\widetilde{\mathbf V}_{\mathcal A\mathcal A}^{\mathcal G(m)}\Big(\widehat{\bm\psi}_{\mathcal A}^{{or}(m)}-\bm\psi_{\mathcal A}^{*(m)}\Big)\bigg]}\\
\le&\ -\sum_{m=1}^M\Big(\widehat{\bm\psi}_{\mathcal A}^{{or}(m)}-\bm\psi_{\mathcal A}^{*(m)}\Big)^{\top}\bigg[N^{-1}\sum_{k\in\mathcal G^{(m)}}n_k\mathbf g^{*(k)}\bigg]_{\mathcal A}-\sum_{m=1}^M\Big(\widehat{\bm\psi}_{\mathcal A}^{{or}(m)}-\bm\psi_{\mathcal A}^{*(m)}\Big)^{\top}\\
&\ \times\Bigg[N^{-1}\sum_{k\in\mathcal G^{(m)}}n_k\int_0^1\bigg\{\mathbf V^{(k)}\Big(\big[\bm\theta^{*(k)}+t(\widetilde{\bm\theta}^{(k)}-\bm\theta^{*(k)})\big]\Big)-\widetilde{\mathbf V}^{(k)}\bigg\}(\widetilde{\bm\theta}^{(k)}-\bm\theta^{*(k)})\ dt\Bigg]_{\mathcal A}.\\
\end{aligned}
\end{equation}

Let $\bm\alpha=\big[\text{vec}(\widehat{\bm\psi}_{\mathcal A}^{{or}})-\text{vec}({\bm\psi}_{\mathcal A}^{*})\big]$, $\widetilde{\mathbf V}^{(\mathcal G,\mathcal A)}=\text{bdiag}(\widetilde{\mathbf V}_{\mathcal A\mathcal A}^{\mathcal G(1)},\dots,\widetilde{\mathbf V}_{\mathcal A\mathcal A}^{\mathcal G(M)})$, $\bm\xi=\big(\bm\xi^{(1)},\dots,\bm\xi^{(M)}\big)$ and  $\bm\eta=\big(\bm\eta^{(1)},\dots,\bm\eta^{(M)}\big)$, where
\[
\bm\xi^{(m)}=\bigg[N^{-1}\sum_{k\in\mathcal G^{(m)}}n_k\mathbf g^{*(k)}\bigg],
\]
\[
\bm\eta^{(m)}=\Bigg[N^{-1}\sum_{k\in\mathcal G^{(m)}}n_k\int_0^1\bigg\{\mathbf V^{(k)}\Big(\big[\bm\theta^{*(k)}+t(\widetilde{\bm\theta}^{(k)}-\bm\theta^{*(k)})\big]\Big)-\widetilde{\mathbf V}^{(k)}\bigg\}(\widetilde{\bm\theta}^{(k)}-\bm\theta^{*(k)})\ dt\Bigg].
\]
Then by (\ref{ineq:3}), we have
\begin{equation}\label{main_ineq}
\bm\alpha^{\top}\widetilde{\mathbf V}^{(\mathcal G,\mathcal A)}\bm\alpha\le  \big|\bm\alpha^{\top}\mathrm{vec}(\bm\xi_{\mathcal A})\big|+\big|\bm\alpha^{\top}\mathrm{vec}(\bm\eta_{\mathcal A})\big|.
\end{equation}
Since $\bm\alpha^{\top}\widetilde{\mathbf V}^{(\mathcal G,\mathcal A)}\bm\alpha\ge\Lambda_{\min}(\widetilde{\mathbf V}^{(\mathcal G,\mathcal A)})||\bm\alpha||_2^2$,  (\ref{main_ineq}) turns to be
\begin{equation}\label{main_ineq2}
||\bm\alpha||_2^2\Lambda_{\min}(\widetilde{\mathbf V}^{(\mathcal G,\mathcal A)})\le  \big|\bm\alpha^{\top}\mathrm{vec}(\bm\xi_{\mathcal A})\big|+\big|\bm\alpha^{\top}\mathrm{vec}(\bm\eta_{\mathcal A})\big|.
\end{equation}

To bound $\Lambda_{\min}(\widetilde{\mathbf V}_{\mathcal A\mathcal A}^{(k)})$, noting that for each $k\in[K]$, we have
\begin{equation}\label{lambdamin}
\begin{aligned}
\Lambda_{\min}(\widetilde{\mathbf V}_{\mathcal A\mathcal A}^{(k)})\ge&\ \Lambda_{\min}\Big[\widetilde{\mathbf V}_{\mathcal A\mathcal A}^{(k)}-\mathbb E(\mathbf V_{\mathcal A\mathcal A}^{*(k)})\Big]+\Lambda_{\min}[\mathbb E(\mathbf V_{\mathcal A\mathcal A}^{*(k)})]\\
\ge&\  -\Big|\Big|\widetilde{\mathbf V}_{\mathcal A\mathcal A}^{(k)}-\mathbb E(\mathbf V_{\mathcal A\mathcal A}^{*(k)})\Big|\Big|_F+\Lambda_{\min}[\mathbb E(\mathbf V_{\mathcal A\mathcal A}^{*(k)})].
\end{aligned}
\end{equation}

By Lemma \ref{lem2}, since $q^3\log p\ll n^*$ by Condition (C7), for all $k\in[K]$, 
\begin{equation}\label{Fnorm}
\Big|\Big|\widetilde{\mathbf V}_{\mathcal A\mathcal A}^{(k)}-\mathbb E(\mathbf V_{\mathcal A\mathcal A}^{*(k)})\Big|\Big|_F\le \bigg(q^2\max_{k\in[K]}\Big|\Big|\widetilde{\mathbf V}^{(k)}-\mathbb E({\mathbf V}^{*(k)})\Big|\Big|_{\max}^2\bigg)^{1/2}=O_p(\sqrt{q^3\log p/n^*})=o_p(1).
\end{equation}
With (\ref{lambdamin}), (\ref{Fnorm}), and Condition (C3), for all $k\in[K]$, with probability approaching 1, $\Lambda_{\min}(\widetilde{\mathbf V}_{\mathcal A\mathcal A}^{(k)})\ge C_{\mathrm{min}}/2$. Consequently, from (\ref{main_ineq2}) and the Cauchy-Schwarz inequality, we have
\begin{equation}\label{main_ineq3}
\frac{C_{\min}N_{\min}}{2N}||\bm\alpha||_2^2\le ||\bm\alpha||_2\big|\big|\mathrm{vec}(\bm\xi_{\mathcal A})\big|\big|_2+||\bm\alpha||_2\big|\big|\mathrm{vec}(\bm\eta_{\mathcal A})\big|\big|_2.
\end{equation}

Now we bound $\big|\big|\mathrm{vec}(\bm\xi_{\mathcal A})\big|\big|_2$ and $\big|\big|\mathrm{vec}(\bm\eta_{\mathcal A})\big|\big|_2$. For $\big|\big|\mathrm{vec}(\bm\xi_{\mathcal A})\big|\big|_2$, since the product of  a sub-Gaussian random variable and a bounded random variable is also sub-Gaussian and $\mathbb E(n_k\mathbf g^{*(k)})=\mathbb E(\mathbf X^{(k)\top}\Phi^{(k)})$ with $\Phi^{(k)}=(f^{\prime}(\bm x_1^{(k)\top}\bm\theta^{*(k)}, y_1^{(k)}),\dots,f^{\prime}(\bm x_{n_k}^{(k)\top}\bm\theta^{*(k)}, y_{n_k}^{(k)}))^{\top}$, by Conditions (C1), (C2), and Lemma \ref{lem1}, for all $m\in[M], j\in[p]$ and any $t>0$,
\[
P\Bigg(\frac1{\sqrt{N_m}}\Bigg|\sum_{k\in\mathcal G^{(m)}}n_k\mathbf g_j^{*(k)}\Bigg|\ge t\Bigg)\le 2\exp\bigg(-\frac{C_1t^2}{C_x^2\kappa_x^2}\bigg).
\]
Accordingly, there exists a constant $C_2>0$ such that $\mathbb E[(\sum_{k\in\mathcal G^{(m)}}n_k\mathbf g_j^{*(k)})^2]\le C_2N_m$. Then,
\begin{equation}\label{xi}
P\Big(\big|\big|\mathrm{vec}(\bm\xi_{\mathcal A})\big|\big|_2\ge t\Big)\le\frac{\sum_{m=1}^m\mathbb E(||\bm\xi_{\mathcal A}^{(m)}||_2^2)}{t^2}\le\frac{\sum_{m=1}^M\sum_{j\in\mathcal A}(\sum_{k\in\mathcal G^{(m)}}n_k\mathbf g_j^{*(k)})^2}{N^2t^2}\le\frac{C_2q}{Nt^2},
\end{equation}
which leads to $\big|\big|\mathrm{vec}(\bm\xi_{\mathcal A})\big|\big|_2=O_p(\sqrt{q/N})$.

For $\big|\big|\mathrm{vec}(\bm\eta_{\mathcal A})\big|\big|_2$, note that, for all $k\in [K]$ and any $t\in[0,1]$, by Conditions (C1), (C4) and (C5), we have
\[
\begin{aligned}
&\ \Bigg|\Bigg|\bigg\{\mathbf V^{(k)}\Big(\big[\bm\theta^{*(k)}+t(\widetilde{\bm\theta}^{(k)}-\bm\theta^{*(k)})\big]\Big)-\widetilde{\mathbf V}^{(k)}\bigg\}(\widetilde{\bm\theta}^{(k)}-\bm\theta^{*(k)})\Bigg|\Bigg|_{\infty}\\
=&\ \Bigg|\Bigg|\frac{1}{n_k}\sum_{i=1}^{n_k}\bm x_{i}^{(k)}\bm x_{i}^{(k)\top}\Big\{f^{\prime\prime}\Big([\bm\theta^{*(k)}+t(\widetilde{\bm\theta}^{(k)}-\bm\theta^{*(k)})]^{\top}\bm x_{i}^{(k)},y_i^{(k)}\Big)-f^{\prime\prime}(\widetilde{\bm\theta}^{(k)\top}\bm x_{i}^{(k)},y_i^{(k)})\Big\}(\widetilde{\bm\theta}^{(k)}-\bm\theta^{*(k)})\Bigg|\Bigg|_{\infty}\\
\le&\ \frac{\max_{i,j,k}|x_{ij}^{(k)}|}{n_k}\sum_{i=1}^{n_k}\Big|(\widetilde{\bm\theta}^{(k)}-\bm\theta^{*(k)})^{\top}\bm x_{i}^{(k)}\Big|\cdot C_L\Big|(1-t)(\widetilde{\bm\theta}^{(k)}-\bm\theta^{*(k)})^{\top}\bm x_{i}^{(k)}\Big|\\
\le&\ \frac{C_LC_x}{n_k}\Big|\Big|\mathbf X^{(k)}(\widetilde{\bm\theta}^{(k)}-\bm\theta^{*(k)})\Big|\Big|_2^2= O_p({q\log p}/{n^*}).\\
\end{aligned}
\]
Thus, we have
\begin{equation}\label{eta}
\begin{aligned}
\big|\big|\mathrm{vec}(\bm\eta_{\mathcal A})\big|\big|_2=&\ \sqrt{\sum_{m=1}^M||\bm\eta_{\mathcal A}^{(m)}||_2^2}\le \sqrt{\sum_{m=1}^M(\sqrt q||\bm\eta^{(m)}||_{\infty})^2}\\
=&\ O_p\Bigg(\sqrt{\frac{\sum_{m=1}^M|\mathcal G^{(m)}|^2q^3(\log p)^2}{N^2}}\Bigg)= O_p\Bigg(\frac{M^{1/2}|\mathcal G_{\max}|q^{3/2}\log p}{N}\Bigg).\\
\end{aligned}
\end{equation}
Combining (\ref{main_ineq3})--(\ref{eta}),  with probability approaching 1, we have
\begin{equation}\label{alpha_bound}
||\bm\alpha||_2\le O_p(r_{1N}),
\end{equation}
where
\[
r_{1N}=\sqrt{\frac{(K/|\mathcal G_{\min}|)q}{N_{\min}}}+\frac{|\mathcal G_{\max}|M^{1/2}q^{3/2}\log p}{N_{\min}}.
\]

Step 2: Let $\widehat{\bm\psi}^{{or}}=(\widehat{\bm\psi}_{\mathcal A}^{{or}\top},\bm 0_{(p-q)\times M}^{\top})^{\top}$. We show that $\widehat{\bm\psi}^{{or}}$ is a local minimizer of $\mathcal L^{\mathcal G}(\bm\psi)$ in (\ref{cor_psi}) through verifying the  KKT conditions,
 \[
 \begin{aligned}
 &\sqrt{\sum_{m=1}^M\Big(|\mathcal G^{(m)}|^{1/2}\widehat\psi_j^{or(m)}\Big)^2}>\tau\lambda_1,  &j&\in\mathcal A,\\
 &\frac{\partial\mathcal L^{\mathcal G}(\widehat{\bm\psi}^{or})}{\partial\bm\psi_j}+\partial  p_{\tau}\Bigg(\sqrt{\sum_{m=1}^M\Big(|\mathcal G^{(m)}|^{1/2}\widehat{\psi}_j^{or(m)}\Big)^2},\lambda_1\Bigg)/\partial\bm\psi_j=\bm 0,\qquad  &j&\in\mathcal A^c,\\
 \end{aligned}
\]
 which are satisfied if the following conditions hold
 \begin{align}
&|\mathcal G_{\min}|^{1/2}\big|\big|\widehat{\bm\psi}_j^{or}\big|\big|_2>\tau\lambda_1,  &j&\in\mathcal A,\label{KKT1}\\
&\Bigg|\Bigg|\frac{\partial\mathcal L^{\mathcal G}(\widehat{\bm\psi}^{or})}{\partial\bm\psi_j}\Bigg|\Bigg|_2\le \lambda_1|\mathcal G_{\min}|^{1/2},\qquad\qquad  &j&\in\mathcal A^c.\label{KKT2}
 \end{align}

First, we show that Condition (\ref{KKT1}) is satisfied with probability approaching 1.  By the triangle inequality and (\ref{alpha_bound}), when $N$ is sufficiently large,
 \[
  \begin{aligned}
|\mathcal G_{\min}|^{1/2}\min_{j\in\mathcal A}||\widehat{\bm\psi}_j^{or}||_2&\ \ge|\mathcal G_{\min}|^{1/2}\bigg(\min_{j\in\mathcal A}||\bm\psi_j^*||_2-\max_{j\in\mathcal A}||\widehat{\bm\psi}_j^{or}-\bm\psi_j^*||_2\bigg)\\
&\ \ge|\mathcal G_{\min}|^{1/2}\bigg(\min_{j\in\mathcal A}||\bm\psi_j^*||_2-||\bm\alpha||_2\bigg)\ge |\mathcal G_{\min}|^{1/2}(d_1-Cr_{1N})>\tau\lambda_1,\\
 \end{aligned}
 \]
where $C$ is a constant. The last inequality is satisfied since  $|\mathcal G_{\min}|^{1/2}d_1> \tau\lambda_1$ and $\lambda_1\gg |\mathcal G_{\min}|^{1/2}r_{1N}$. Accordingly, (\ref{KKT1}) is satisfied with probability approaching 1 when $N\rightarrow\infty$. 

Second, we show that Condition (\ref{KKT2}) is satisfied with probability approaching 1. Note that $||\partial\mathcal L^{\mathcal G}(\widehat{\bm\psi}^{or})/\partial\bm\psi_j||_2=\sqrt{\sum_{m=1}^M(\partial\mathcal L^{\mathcal G}(\widehat{\bm\psi}^{or})/\partial\bm\psi_j^{(m)})^2}$. Then Condition (\ref{KKT2}) holds if 
\begin{equation}\label{KKT3}
\Bigg|\Bigg|\frac{\partial\mathcal L^{\mathcal G}(\widehat{\bm\psi}^{or})}{\partial\bm\psi_{\mathcal A^c}^{(m)}}\Bigg|\Bigg|_{\infty}\le\lambda_1(|\mathcal G_{\min}|/M)^{1/2},\quad m\in[M].
\end{equation}

Since for each $m\in[M]$, $\partial\mathcal L^{or,\mathcal G}(\widehat{\bm\psi}_{\mathcal A}^{or})/\partial\bm\psi_{\mathcal A}^{(m)}=\bm 0$, we have
\begin{equation}\label{e1}
\widehat{\bm\psi}_{\mathcal A}^{or(m)}-\bm\psi_{\mathcal A}^{*(m)}=\Big(\widetilde{\mathbf V}_{\mathcal A\mathcal A}^{\mathcal G(m)}\Big)^{-1}\big(\bm\xi_{\mathcal A}^{(m)}+\bm\eta_{\mathcal A}^{(m)}\big),\quad m\in[M].
\end{equation}
Thus, combining (\ref{KKT3}) and (\ref{e1}) leads to
\begin{equation}\label{I_all}
\begin{aligned}
\frac12\Bigg|\Bigg|\frac{\partial\mathcal L^{\mathcal G}(\widehat{\bm\psi}^{or})}{\partial\bm\psi_{\mathcal A^c}^{(m)}}\Bigg|\Bigg|_{\infty}=&\ \Big|\Big|\widetilde{\mathbf V}_{\mathcal A^c\mathcal A}^{\mathcal G(m)}\big(\widehat{\bm\psi}_{\mathcal A}^{or(m)}-\bm\psi_{\mathcal A}^{*(m)}\big)+(\bm\xi_{\mathcal A^c}^{(m)}+\bm\eta_{\mathcal A^c}^{(m)}\big)\Big|\Big|_{\infty}\\
\le&\ \Big|\Big|\widetilde{\mathbf V}_{\mathcal A^c\mathcal A}^{\mathcal G(m)}\Big(\widetilde{\mathbf V}_{\mathcal A\mathcal A}^{\mathcal G(m)}\Big)^{-1}\big(\bm\xi_{\mathcal A}^{(m)}+\bm\eta_{\mathcal A}^{(m)}\big)\Big|\Big|_{\infty}+\Big|\Big|\bm\xi_{\mathcal A^c}^{(m)}+\bm\eta_{\mathcal A^c}^{(m)}\Big|\Big|_{\infty}\\
\le&\ \Big(\Big|\Big|\widetilde{\mathbf V}_{\mathcal A^c\mathcal A}^{\mathcal G(m)}\Big(\widetilde{\mathbf V}_{\mathcal A\mathcal A}^{\mathcal G(m)}\Big)^{-1}\Big|\Big|_{\infty}+1\Big)\Big(\big|\big|\bm\xi^{(m)}\big|\big|_{\infty}+\big|\big|\bm\eta^{(m)}\big|\big|_{\infty}\Big).
\end{aligned}
\end{equation}

Following \cite{xue2012nonconcave}, we can derive the upper bound of $\Big|\Big|\widetilde{\mathbf V}_{\mathcal A^c\mathcal A}^{\mathcal G(m)}\Big(\widetilde{\mathbf V}_{\mathcal A\mathcal A}^{\mathcal G(m)}\Big)^{-1}\Big|\Big|_{\infty}, m\in[M]$. 
Following the definition of $\widetilde{\mathbf V}^{\mathcal G(m)} (m=1,\dots,M)$, we can also define ${\mathbf V}^{*\mathcal G(m)}(m=1,\dots,M)$ accordingly. By Condition (C3), we define
\begin{equation}\label{c}
c_m:=\Big|\Big|\big[\mathbb E\big({\mathbf V}_{\mathcal A\mathcal A}^{*\mathcal G(m)}\big)\big]^{-1}\Big|\Big|_{\infty}\le\sqrt q\Big|\Big|\big[\mathbb E\big({\mathbf V}_{\mathcal A\mathcal A}^{*\mathcal G(m)}\big)\big]^{-1}\Big|\Big|_{2}\le\frac{\sqrt qK}{|\mathcal G_{\min}|C_{\min}}.\\
\end{equation}
Furthermore, we define
\[
\begin{aligned}
\phi_{m}&=\Big|\Big|\widetilde{\mathbf V}_{\mathcal A^c\mathcal A}^{\mathcal G(m)}\Big(\widetilde{\mathbf V}_{\mathcal A\mathcal A}^{\mathcal G(m)}\Big)^{-1}-\mathbb E\big({\mathbf V}_{\mathcal A^c\mathcal A}^{*\mathcal G(m)}\big)\big[\mathbb E\big({\mathbf V}_{\mathcal A\mathcal A}^{*\mathcal G(m)}\big)\big]^{-1}\Big|\Big|_{\infty},\\
\phi_{1m}&= \Big|\Big|\big(\widetilde{\mathbf V}_{\mathcal A\mathcal A}^{\mathcal G(m)}\big)^{-1}-\big[\mathbb E\big({\mathbf V}_{\mathcal A\mathcal A}^{*\mathcal G(m)}\big)\big]^{-1}\Big|\Big|_{\infty},\\
\phi_{2m}&=\Big|\Big|\widetilde{\mathbf V}_{\mathcal A\mathcal A}^{\mathcal G(m)}-\mathbb E\big({\mathbf V}_{\mathcal A\mathcal A}^{*\mathcal G(m)}\big)\Big|\Big|_{\infty},\qquad\phi_{3m}=\Big|\Big|\widetilde{\mathbf V}_{\mathcal A^c\mathcal A}^{\mathcal G(m)}-\mathbb E\big({\mathbf V}_{\mathcal A^c\mathcal A}^{*\mathcal G(m)}\big)\Big|\Big|_{\infty}.\\
\end{aligned}
\]
Then by definition,
\[
\begin{aligned}
\phi_{m}=&\ \Big|\Big| \Big[\widetilde{\mathbf V}_{\mathcal A^c\mathcal A}^{\mathcal G(m)}-\mathbb E\big({\mathbf V}_{\mathcal A^c\mathcal A}^{*\mathcal G(m)}\big)\Big]\Big[\big(\widetilde{\mathbf V}_{\mathcal A\mathcal A}^{\mathcal G(m)}\big)^{-1}-\big[\mathbb E\big({\mathbf V}_{\mathcal A\mathcal A}^{*\mathcal G(m)}\big)\big]^{-1}\Big]\\
 &\qquad+\mathbb E\big({\mathbf V}_{\mathcal A^c\mathcal A}^{*\mathcal G(m)}\big)\big[\mathbb E\big({\mathbf V}_{\mathcal A\mathcal A}^{*\mathcal G(m)}\big)\big]^{-1}\Big[-\widetilde{\mathbf V}_{\mathcal A\mathcal A}^{\mathcal G(m)}+\mathbb E\big({\mathbf V}_{\mathcal A\mathcal A}^{*\mathcal G(m)}\big)\Big]\big(\widetilde{\mathbf V}_{\mathcal A\mathcal A}^{\mathcal G(m)}\big)^{-1}\\
&\qquad\qquad\qquad\qquad\qquad\qquad\quad+\Big[\widetilde{\mathbf V}_{\mathcal A^c\mathcal A}^{\mathcal G(m)}-\mathbb E\big({\mathbf V}_{\mathcal A^c\mathcal A}^{*\mathcal G(m)}\big)\Big]\big[\mathbb E\big({\mathbf V}_{\mathcal A\mathcal A}^{*\mathcal G(m)}\big)\big]^{-1}\Big|\Big|_{\infty}\\
\le&\ \phi_{3m}\phi_{1m}+\varphi^{\mathcal G(m)}\phi_{2m}\big|\big|\big(\widetilde{\mathbf V}_{\mathcal A\mathcal A}^{\mathcal G(m)}\big)^{-1}\big|\big|_{\infty}+\phi_{3m}c_m\\
\le&\ \phi_{3m}\phi_{1m}+\varphi^{\mathcal G(m)}\phi_{2m}(c_m+\phi_{1m})+\phi_{3m}c_m.\\
\end{aligned}
\]
Besides, $\phi_{1m}$ can be reformulated as
\[
\begin{aligned}
\phi_{1m}=&\ \Big|\Big|\big(\widetilde{\mathbf V}_{\mathcal A\mathcal A}^{\mathcal G(m)}\big)^{-1}\Big[\mathbb E\big({\mathbf V}_{\mathcal A\mathcal A}^{*\mathcal G(m)}\big)-\widetilde{\mathbf V}_{\mathcal A\mathcal A}^{\mathcal G(m)}\Big]\big[\mathbb E\big({\mathbf V}_{\mathcal A\mathcal A}^{*\mathcal G(m)}\big)\big]^{-1}\Big|\Big|_{\infty}\\
\le&\ \Big|\Big|\big(\widetilde{\mathbf V}_{\mathcal A\mathcal A}^{\mathcal G(m)}\big)^{-1}\Big|\Big|_{\infty}\cdot\Big|\Big|\mathbb E\big({\mathbf V}_{\mathcal A\mathcal A}^{*\mathcal G(m)}\big)-\widetilde{\mathbf V}_{\mathcal A\mathcal A}^{\mathcal G(m)}\Big|\Big|_{\infty}\cdot\Big|\Big|\big[\mathbb E\big({\mathbf V}_{\mathcal A\mathcal A}^{*\mathcal G(m)}\big)\big]^{-1}\Big|\Big|_{\infty}\\
\le&\ (c_m+\phi_{1m})\phi_{2m}c_m.
\end{aligned}
\]
Thus, as long as $\phi_{2m}c_m<1$, we have $\phi_{1m}\le\phi_{2m}c_m^2/(1-\phi_{2m}c_m)$, which yields
\begin{equation}\label{I1_m}
\phi_{m}\le (\phi_{3m}+\varphi^{\mathcal G(m)}\phi_{2m})\frac{c_m}{1-\phi_{2m}c_m}.
\end{equation}
Then, by Lemma \ref{lem2},
\begin{equation}\label{phi2}
\begin{aligned}
\phi_{2m}=\big|\big|\widetilde{\mathbf V}_{\mathcal A\mathcal A}^{\mathcal G(m)}-\mathbb E(\mathbf V_{\mathcal A\mathcal A}^{*\mathcal G(m)})\big|\big|_{\infty}\le&\  \frac{q\sum_{k\in\mathcal G^{(m)}}n_k\big|\big|\widetilde{\mathbf V}_{\mathcal A\mathcal A}^{(k)}-\mathbb E(\mathbf V_{\mathcal A\mathcal A}^{*(k)})\big|\big|_{\max}}{N}\\
=&\ O_p\Bigg(\frac{|\mathcal G_{\max}|}{K}\cdot\sqrt{\frac{q^3\log p}{n^*}}\Bigg).
\end{aligned}
\end{equation}
Combining (\ref{c}), (\ref{phi2}), and Condition (C7), we have
\begin{equation}\label{phi2c}
\phi_{2m}c_m=O_p\Bigg(\frac{|\mathcal G_{\max}|}{|\mathcal G_{\min}|}\cdot\sqrt{\frac{q^4\log p}{n^*}}\Bigg)=o_p(1).
\end{equation}
Following the proof of (\ref{phi2c}), we also have
\begin{equation}\label{phi3jc}
\phi_{3m}c_m=O_p\Bigg(\frac{|\mathcal G_{\max}|}{|\mathcal G_{\min}|}\cdot\sqrt{\frac{q^4\log p}{n^*}}\Bigg)=o_p(1).
\end{equation}
Combining (\ref{I1_m}), (\ref{phi2c}) and (\ref{phi3jc}), with probability approaching 1, for all $m\in[M]$ $\phi_{m}=o_p(\varphi_{\max})$,
\begin{equation}\label{I1}
\begin{aligned}
\max_{m\in[M]}\Big|\Big|\widetilde{\mathbf V}_{\mathcal A^c\mathcal A}^{\mathcal G(m)}\Big(\widetilde{\mathbf V}_{\mathcal A\mathcal A}^{\mathcal G(m)}\Big)^{-1}\Big|\Big|_{\infty}&\ \le \max_{m\in[M]}\ \phi_{m}+\max_{m\in[M]}\ \mathbb E\big({\mathbf V}_{\mathcal A^c\mathcal A}^{*\mathcal G(m)}\big)\big[\mathbb E\big({\mathbf V}_{\mathcal A\mathcal A}^{*\mathcal G(m)}\big)\big]^{-1}\\
&\ \le 2\varphi_{\max}.
\end{aligned}
\end{equation}

Now we focus on the term $||\bm\xi^{(m)}||_{\infty}, m\in[M]$. By Conditions (C1), (C2), Lemma \ref{lem1}, and the union bound
\[
P\Bigg(\max_{m\in[M]}\max_{j\in[p]}\ \Bigg|\sum_{k\in\mathcal G^{(m)}}n_k\mathbf g_j^{*(k)}\Bigg|\ge t\sqrt{N_{\max}\log p}\Bigg)\le 2pM\exp\Bigg(-\frac{C_1t^2\log p}{C_x^2\kappa_x^2}\Bigg)\le 2p^{2-C_3},
\]
where $C_3=C_1t^2/(C_x^2\kappa_x^2)$. When $t$ is sufficiently large and $2-C_3<0$, $2p^{2-C_3}\rightarrow 0$ when $p\rightarrow \infty$. Hence,
\begin{equation}\label{I2}
\max_{m\in[M]}\big|\big|\bm\xi^{(m)}\big|\big|_{\infty}=O_p\Bigg(\sqrt{\frac{|\mathcal G_{\max}|\log p}{KN}}\Bigg).
\end{equation}
Following the proof of (\ref{eta}), we have
\begin{equation}\label{I3}
\max_{m\in[M]}\big|\big|\bm\eta^{(m)}\big|\big|_{\infty}=O_p\Bigg({\frac{|\mathcal G_{\max}|q\log p}{N}}\Bigg).
\end{equation}
Combining (\ref{I_all}), (\ref{I1}), (\ref{I2}), (\ref{I3}), since $\varphi_{\max}r_{2N}\ll \lambda_1$, where
\[
r_{2N}=\sqrt{\frac{(|\mathcal G_{\max}|/|\mathcal G_{\min}|)M\log p}{KN}}+\frac{(|\mathcal G_{\max}|/|\mathcal G_{\min}|^{1/2})M^{1/2}q\log p}{N},
\]
we have verified that  (\ref{I_all}) is satisfied with probability approaching 1. Therefore, the KKT conditions have been verified, and the proof of Step 2 is completed.

\noindent\textbf{Proof of Theorem 2}

Define
\[
\mathcal Q(\bm\theta)=\mathcal L(\bm\theta)+\mathcal P_1(\bm\theta)+\mathcal P_2(\bm\theta),\qquad \mathcal Q^{\mathcal G}(\bm\psi)=\mathcal L^{\mathcal G}(\bm\psi)+\mathcal P_1^{\mathcal G}(\bm\psi)+\mathcal P_2^{\mathcal G}(\bm\psi),
\]
where 
\[
\begin{aligned}
&\mathcal L(\bm\theta)=\frac{1}{N}\sum_{k=1}^Kn_k\Big(\bm\theta^{(k)\top}\widetilde{\mathbf V}^{(k)}\bm\theta^{(k)}-2\bm\theta^{(k)\top}\widetilde{\bm\zeta}^{(k)}\Big)+\sum_{j=2}^{p}p_{\tau}\big(||\boldsymbol\theta_j||_2,\lambda_1\big),\\
&\mathcal P_2(\bm\theta)=\sum_{k<k^{\prime}}p_{\tau}\big(||\boldsymbol\theta^{(k)}-\boldsymbol\theta^{(k^{\prime})}||_2,\lambda_2\big),\\
&\mathcal L^{\mathcal G}(\bm\psi)=\frac{1}{N}\sum_{m=1}^M\Bigg[\bm\psi^{(m)\top}\Bigg(\sum_{k\in\mathcal G^{(m)}}n_k\widetilde{\mathbf V}^{(k)}\Bigg)\bm\psi^{(m)}-2\bm\psi^{(m)\top}\Bigg(\sum_{k\in\mathcal G^{(m)}}n_k\widetilde{\bm\zeta}^{(k)}\Bigg)\Bigg]\\
&\qquad\qquad\qquad\qquad+\sum_{j=2}^pp_{\tau}\Bigg(\bigg[\sum_{m=1}^M|\mathcal G^{(m)}|{\psi_{j}^{(m)}}^2\bigg]^{1/2},\lambda_1\Bigg),\\
&\mathcal P_2^{\mathcal G}(\bm\psi)=\sum_{m<m^{\prime}}\big|\mathcal G^{(m)}\big|\big|\mathcal G^{(m^{\prime})}\big|p_{\tau}\bigg(\Big|\Big|\bm\psi^{(m)}-\bm\psi^{(m^{\prime})}\Big|\Big|_2,\lambda_2\bigg).\\
\end{aligned}
\]

Define two mapping functions $\mathcal T^{\mathcal G}(\cdot)$ and $\mathcal T(\cdot)$. Specifically, let $\mathcal T^{\mathcal G}:\mathcal M_{\mathcal G}\rightarrow\mathbb R^{p\times M}$ be the function such that $\mathcal T^{\mathcal G}(\bm\theta)$ is the $p\times M$ matrix whose $m$th column equals the common coefficient vector of $\bm\theta^{(k)}$ for $k\in\mathcal G^{(m)}$. Additionally, let $\mathcal T: \mathbb R^{p\times K}\rightarrow\mathbb R^{p\times M}$ be the function such that $\mathcal T(\bm\theta)=\big\{|\mathcal G^{(m)}|^{-1}\sum_{k\in\mathcal G^{(m)}}\bm\theta^{(k)}\big\}_{m=1}^M$. Obviously, if $\bm\theta\in\mathcal M_{\mathcal G}$, $\mathcal T^{\mathcal G}(\bm\theta)=\mathcal T(\bm\theta)$.

For every $\bm\theta\in\mathcal M_{\mathcal G}$, we have $\mathcal P_2(\bm\theta)=\mathcal P_2^{\mathcal G}(\mathcal T^{\mathcal G}(\bm\theta))$. Similarly, for every $\bm\psi\in\mathbb R^{p\times M}$, we have $\mathcal P_2({\mathcal T^{\mathcal G}}^{-1}(\bm\psi))=\mathcal P_2^{\mathcal G}(\bm\psi)$. Hence, 
\begin{equation}\label{equal}
\mathcal Q(\bm\theta)=\mathcal Q^{\mathcal G}(\mathcal T^{\mathcal G}(\bm\theta)),\qquad \mathcal Q^{\mathcal G}(\bm\psi)=\mathcal Q({\mathcal T^{\mathcal G}}^{-1}(\bm\psi)).
\end{equation}

Consider the neighborhood of $\bm\theta^*$, denoted by $\Theta$, which is defined as
\[
\Theta_1=\Big\{\bm\theta\in\mathbb R^{p\times K}:  ||\bm\theta-\bm\theta^*||_F\le C|\mathcal G_{\max}|^{1/2}r_{1N}\Big\}.
\]
Define the event
\[
E_1=\Big\{||\widehat{\bm\theta}^{or}-\bm\theta^*||_F\le C|\mathcal G_{\max}|^{1/2}r_{1N}\Big\}.
\]
Then, by the result in Theorem 1, for any $\epsilon>0$, there exists a constant $C_{\epsilon}>0$ such that for any $C\ge C_{\epsilon}$, $P(E_1)\ge 1-\epsilon$. Accordingly, $\widehat{\bm\theta}^{or}\in\Theta_1$ with probability at least $1-\epsilon$. Furthermore, we define another neighborhood
\[
\Theta_2=\Big\{\bm\theta\in\mathbb R^{p\times K}:  ||\bm\theta-\widehat{\bm\theta}^{or}||_F\le t_N\Big\},
\]
where $t_N$ is a positive sequence.
For any $\bm\theta=(\bm\theta_{\mathcal A}^{\top},\bm\theta_{\mathcal A^c}^{\top})^{\top}\in\Theta_1$, let $\breve{\bm\theta}=(\bm\theta_{\mathcal A}^{\top},\bm 0_{(p-q)\times K}^{\top})^{\top}$ and $\breve{\bm\theta}^{\mathcal G}={\mathcal T^{\mathcal G}}^{-1}(\mathcal T(\breve{\bm\theta}))$. Then we show that $\widehat{\bm\theta}^{or}$ is a strictly local minimizer of objective function (\ref{obj3}) with probability approaching 1 through the following two steps
\begin{enumerate}[(a)]
\item On event $E_1$, $\mathcal Q(\breve{\bm\theta}^{\mathcal G})>\mathcal Q(\widehat{\bm\theta}^{or})$ for any $\bm\theta\in\Theta_1$ and $\bm\theta^{\mathcal G}\ne\widehat{\bm\theta}^{or}$;
\item On event $E_1$, $\mathcal Q(\bm\theta)\ge\mathcal Q(\breve{\bm\theta})\ge\mathcal Q(\breve{\bm\theta}^{\mathcal G})$ for any $\bm\theta\in\Theta_1\cap\Theta_2$ for a sufficiently large $N$.
\end{enumerate}

For any $\bm\theta\in\Theta_1$, let $\mathcal T(\breve{\bm\theta})=(\breve{\bm\psi}^{(1)},\dots,\breve{\bm\psi}^{(M)})$. Note that,
\[
\mathcal P_2^{\mathcal G}(\mathcal T(\breve{\bm\theta}))=\sum_{m<m^{\prime}}\big|\mathcal G^{(m)}\big|\big|\mathcal G^{(m^{\prime})}\big|p_{\tau}\bigg(\Big|\Big|\breve{\bm\psi}^{(m)}-\breve{\bm\psi}^{(m^{\prime})}\Big|\Big|_2,\lambda_2\bigg),
\]
and, for any $m, m^{\prime}\in[M]$ and $m\ne m^{\prime}$, we have
\begin{equation}\label{min_dis}
\begin{aligned}
\Big|\Big|\breve{\bm\psi}^{(m)}-\breve{\bm\psi}^{(m^{\prime})}\Big|\Big|_2\ge&\ \min_{m, m^{\prime}\in[M], m\ne m^{\prime}}\Big|\Big|\bm\psi^{*(m)}-\bm\psi^{*(m^{\prime})}\Big|\Big|_2-2\cdot\max_{m\in[M]}\Big|\Big|\breve{\bm\psi}^{(m)}-\bm\psi^{*(m)}\Big|\Big|_2\\
\ge&\ d_2-2\cdot\max_{m\in[M]}\Bigg|\Bigg|\big|\mathcal G^{(m)}\big|^{-1}\sum_{k\in\mathcal G^{(m)}}(\breve{\bm\theta}^{(k)}-\bm\theta^{*(k)})\Bigg|\Bigg|_2\\
\ge&\ d_2-2\cdot\max_{k\in[K]}\big|\big|\breve{\bm\theta}^{(k)}-\bm\theta^{*(k)}\big|\big|_2\ge d_2-2C|\mathcal G_{\max}|^{1/2}r_{1N}> \tau\lambda_2, \\
\end{aligned}
\end{equation}
where $C$ is a constant and the last  inequality follows from $d_2>\tau\lambda_2$ and $\lambda_2\gg |\mathcal G_{\max}|^{1/2}r_{1N}$. 
Consequently, for any $\bm\theta\in\Theta_1$, $\mathcal P_2^{\mathcal G}(\mathcal T(\breve{\bm\theta}))=C_N$, and $C_N>0$ is a constant. By the result of Theorem 1, $\widehat{\bm\psi}^{or}$ is the local minimizer of objective function $\mathcal L^{\mathcal G}(\bm\psi)$ with probability approaching 1. Thus we have $\mathcal L^{\mathcal G}(\widehat{\bm\psi}^{or})<\mathcal L^{\mathcal G}(\mathcal T(\breve{\bm\theta}))$ for any $\bm\theta\in\Theta_1$ and $\mathcal T(\breve{\bm\theta})\ne\widehat{\bm\psi}^{or}$. Combining this with $\mathcal P_2^{\mathcal G}(\mathcal T(\breve{\bm\theta}))=C_N$ for $\bm\theta\in\Theta_1$, we have $\mathcal Q^{\mathcal G}(\widehat{\bm\psi}^{or})<\mathcal Q^{\mathcal G}(\mathcal T(\breve{\bm\theta}))$. By (\ref{equal}), we have
\[
\mathcal Q^{\mathcal G}(\widehat{\bm\psi}^{or})=\mathcal Q({\mathcal T^{\mathcal G}}^{-1}(\widehat{\bm\psi}^{or}))=\mathcal Q(\widehat{\bm\theta}^{or}), \qquad \mathcal Q^{\mathcal G}(\mathcal T(\breve{\bm\theta}))=\mathcal Q({\mathcal T^{\mathcal G}}^{-1}(\mathcal T(\breve{\bm\theta})))=\mathcal Q(\breve{\bm\theta}^{\mathcal G}).
\]
Accordingly, we have $\mathcal Q(\widehat{\bm\theta}^{or})<\mathcal Q(\breve{\bm\theta}^{\mathcal G})$ for any $\bm\theta\in\Theta_1$ and $\breve{\bm\theta}^{\mathcal G}\ne\widehat{\bm\theta}^{or}$. This finishes the proof of the result in (a).

Next, we show that the result in (b) holds with probability approaching 1. First, we show that, on event $E_1$,  $\mathcal Q(\bm\theta)\ge\mathcal Q(\breve{\bm\theta})$ for any $\bm\theta\in\Theta_1\cap\Theta_2$.
By the Taylor series expansion, we have
\begin{equation}\label{omega1}
\mathcal Q(\bm\theta)-\mathcal Q(\breve{\bm\theta})=\bm\Omega_{11}+\bm\Omega_{12}+\bm\Omega_{13},
\end{equation}
where 
\[
\begin{aligned}
&\bm\Omega_{11}=\sum_{j=1}^p\frac{\partial\mathcal L(\overline{\bm\theta})}{\partial\bm\theta_j^{\top}}(\bm\theta_j-\breve{\bm\theta}_j),\qquad\bm\Omega_{12}=\sum_{j=2}^p\frac{\partial\mathcal P_1(\overline{\bm\theta})}{\partial\bm\theta_j^{\top}}(\bm\theta_j-\breve{\bm\theta}_j),\\
&\bm\Omega_{13}=\sum_{k<k^{\prime}}\Big[p_{\tau}(||\bm\theta^{(k)}-\bm\theta^{(k^{\prime})}||_2,\lambda_2)-p_{\tau}(||\breve{\bm\theta}^{(k)}-\breve{\bm\theta}^{(k^{\prime})}||_2,\lambda_2)\Big],
\end{aligned}
\]
in which $\overline{\bm\theta}=\delta_1\bm\theta+(1-\delta_1)\breve{\bm\theta}$ for some $\delta_1\in(0,1)$. Note that, for any $j\in\mathcal A$, $\bm\theta_j=\breve{\bm\theta}_j$, and for any $j\in\mathcal A^c$, $\breve{\bm\theta}_j=\bm 0$ and $\overline{\bm\theta}_j=\delta_1\bm\theta_j$. Since $p_{\tau}(t,\lambda_2)$ is a nondecreasing function of $t$ with $t\in[0,\infty)$, $\bm\Omega_{13}\ge 0$. Besides,
\begin{equation}\label{omega12}
\begin{aligned}
\bm\Omega_{12}=\sum_{j\in\mathcal A^c}p_{\tau}^{\prime}(||\overline{\bm\theta}_j||_2,\lambda_1)\frac{\overline{\bm\theta}_j^{\top}(\bm\theta_j-\breve{\bm\theta}_j)}{||\overline{\bm\theta}_j||_2}=&\ \sum_{j\in\mathcal A^c}p_{\tau}^{\prime}(\delta_1||{\bm\theta}_j||_2,\lambda_1)||\bm\theta_j||_2\ge \sum_{j\in\mathcal A^c}p_{\tau}^{\prime}(t_N,\lambda_1)||\bm\theta_j||_2,\\
\end{aligned}
\end{equation}
where the last inequality follows from the concavity of $p_{\tau}(t,\lambda_1)$. Furthermore,
\begin{equation}\label{omega11}
\begin{aligned}
|\bm\Omega_{11}|=\Bigg|\sum_{j\in\mathcal A^c}\frac{\partial\mathcal L(\overline{\bm\theta})}{\partial\bm\theta_j^{\top}}(\bm\theta_j-\breve{\bm\theta}_j)\Bigg|&\ \le\sum_{j\in\mathcal A^c}\Bigg|\Bigg|\frac{\partial\mathcal L(\overline{\bm\theta})}{\partial\bm\theta_j^{\top}}\Bigg|\Bigg|_2||\bm\theta_j||_2\\
&\ \le\sqrt K\max_{k\in[K]}\Bigg|\Bigg|\frac{\partial\mathcal L(\overline{\bm\theta})}{\partial\bm\theta_{\mathcal A^c}^{(k)}}\Bigg|\Bigg|_{\infty}\sum_{j\in\mathcal A^c}||\bm\theta_j||_2.\\
\end{aligned}
\end{equation}
Combining (\ref{omega1}), (\ref{omega12}), and (\ref{omega11}), we have
\begin{equation}\label{omegaminus}
\begin{aligned}
\mathcal Q(\bm\theta)-\mathcal Q(\breve{\bm\theta})=&\ \bm\Omega_{11}+\bm\Omega_{12}+\bm\Omega_{13}\\
\ge&\ \sum_{j\in\mathcal A^c}\Big[p_{\tau}^{\prime}(t_N,\lambda_1)-\sqrt K\max_{k\in[K]}\big|\big|{\partial\mathcal L(\overline{\bm\theta})}/{\partial\bm\theta_{\mathcal A^c}^{(k)}}\big|\big|_{\infty}\Big]||\bm\theta_j||_2.
\end{aligned}
\end{equation}

Following the proof from (\ref{e1}) to (\ref{I3}), we have
\[
\sqrt K\max_{k\in[K]}\big|\big|{\partial\mathcal L(\overline{\bm\theta})}/{\partial\bm\theta_{\mathcal A^c}^{(k)}}\big|\big|_{\infty}=O_p\Bigg(\varphi_{\max}\Big[\sqrt{{\log p}/{N}}+{K^{1/2}q\log p}/{N}\Big]+(Kq)^{1/2}t_{N}\Bigg).
\]
In addition, let $t_N=o(1)$, and then $p_{\tau}^{\prime}(t_N,\lambda_1)\to\lambda_1$. Furthermore, let $(Kq)^{1/2}t_N\ll \lambda_1$, and then
\[
\lambda_1\gg \varphi_{\max}\Big[r_{2N}+\sqrt{{\log p}/{N}}\Big]
\] 
leads to
\begin{equation}\label{order}
\lambda_1\gg \varphi_{\max}\Big[\sqrt{{\log p}/{N}}+{K^{1/2}q\log p}/{N}\Big]+(Kq)^{1/2}t_{N}.
\end{equation}
Then, by (\ref{omegaminus}) and (\ref{order}), when $N$ is sufficiently large, with probability approaching 1,
\[
\mathcal Q(\bm\theta)-\mathcal Q(\breve{\bm\theta})\ge 0.
\]

Next, we show that, on event $E_1$, $\mathcal Q(\breve{\bm\theta})\ge Q(\breve{\bm\theta}^{\mathcal G})$ for any $\bm\theta\in\Theta_1\cap\Theta_2$. By the Taylor series expansion, we have
\[
\mathcal Q(\breve{\bm\theta})-\mathcal Q(\breve{\bm\theta}^{\mathcal G})=\bm\Omega_{21}+\bm\Omega_{22}+\bm\Omega_{23},
\]
where
\[
\begin{aligned}
&\bm\Omega_{21}=\sum_{k=1}^K\frac{\partial\mathcal L(\breve{\bm\theta}^{\hbar})}{\partial\bm\theta^{(k)\top}}(\breve{\bm\theta}^{(k)}-\breve{\bm\theta}^{\mathcal G(k)}),\qquad\bm\Omega_{22}=\sum_{j=2}^p\frac{\partial\mathcal P_1(\breve{\bm\theta}^{\hbar})}{\partial\bm\theta_j^{\top}}(\breve{\bm\theta}_j-\breve{\bm\theta}_j^{\mathcal G}),\\
&\bm\Omega_{23}=\sum_{k=1}^K\frac{\partial\mathcal P_2(\breve{\bm\theta}^{\hbar})}{\partial\bm\theta^{(k)\top}}(\breve{\bm\theta}^{(k)}-\breve{\bm\theta}^{\mathcal G(k)}),
\end{aligned}
\]
in which $\breve{\bm\theta}^{\hbar}=\delta_2\breve{\bm\theta}+(1-\delta_2)\breve{\bm\theta}^{\mathcal G}$ for some $\delta_2\in(0,1)$. Next, we bound $\bm\Omega_{21}$, $\bm\Omega_{22}$, and $\bm\Omega_{23}$. Recall that, for any $j\in\mathcal A^c$, $\breve{\bm\theta}_{\mathcal A^c}^{\hbar}=\breve{\bm\theta}_{\mathcal A^c}=\breve{\bm\theta}_{\mathcal A^c}^{\mathcal G}=\bm 0$. Then,
\begin{equation}\label{omega22}
\bm\Omega_{22}=\sum_{j=2}^q\frac{\partial\mathcal P_1(\breve{\bm\theta}^{\hbar})}{\partial\bm\theta_j^{\top}}(\breve{\bm\theta}_j-\breve{\bm\theta}_j^{\mathcal G})=\sum_{j=2}^qp_{\tau}^{\prime}(||\breve{\bm\theta}_j^{\hbar}||_2,\lambda_1)\frac{\breve{\bm\theta}_j^{\hbar\top}(\breve{\bm\theta}_j-\breve{\bm\theta}_j^{\mathcal G})}{||\breve{\bm\theta}_j^{\hbar}||_2}.
\end{equation}

Note that,
\[
\begin{aligned}
||\breve{\bm\theta}_j^{\mathcal G}-\bm\theta_j^*||_2=&\ \sqrt{\sum_{m=1}^M|\mathcal G^{(m)}|\Bigg(\frac{\sum_{k\in\mathcal G^{(m)}}(\breve\theta_j^{(k)}-\psi_j^{*(m)})}{|\mathcal G^{(m)}|}\Bigg)^2}\\
\le&\ \sqrt{\sum_{m=1}^M|\mathcal G^{(m)}|\times\frac{|\mathcal G^{(m)}|\Big[\sum_{k\in\mathcal G^{(m)}}\big(\breve\theta_j^{(k)}-\psi_j^{*(m)}\big)^2\Big]}{|\mathcal G^{(m)}|^2}}=||\breve{\bm\theta}_j-\bm\theta_j^*||_2.
\end{aligned}
\]
Besides, since $\breve{\bm\theta}_j^{\hbar}=\delta_2\breve{\bm\theta}_j+(1-\delta_2)\breve{\bm\theta}_j^{\mathcal G}$, 
\[
||\breve{\bm\theta}_j^{\hbar}-\bm\theta_j^*||_2=\delta_2||\breve{\bm\theta}_j-\bm\theta_j^*||_2+(1-\delta_2)||\breve{\bm\theta}_j^{\mathcal G}-\bm\theta_j^*||_2\le||\breve{\bm\theta}_j-\bm\theta_j^*||_2.
\]
Hence, for any $j\in\mathcal A$, by the triangle inequality,
\begin{equation}\label{omega22_1}
\begin{aligned}
||\breve{\bm\theta}_j^{\hbar}||_2\ge&\ ||\bm\theta_j^*||_2-||\breve{\bm\theta}_j^{\hbar}-\bm\theta_j^*||_2\ge\min_{j\in\mathcal A}||\bm\theta_j^*||_2-\max_{j\in\mathcal A}||\breve{\bm\theta}_j-\bm\theta_j^*||_2\\
\ge&\ |\mathcal G_{\min}|^{1/2}d_1-C|\mathcal G_{\max}|^{1/2}r_{1N}>\tau\lambda_1,
\end{aligned}
\end{equation}
where the last inequality follows from $|\mathcal G_{\min}|^{1/2}d_1>\tau\lambda_1\gg|\mathcal G_{\max}|^{1/2}r_{1N}$. Combining (\ref{omega22}) and (\ref{omega22_1}), since $p_{\tau}(t,\lambda_1)$ is a constant for $t\ge \tau\lambda_1$, we have $\bm\Omega_{22}=0$.

Now consider $\bm\Omega_{23}$. Recall that, for any $j\in\mathcal A^c$, $\breve{\bm\theta}_{\mathcal A^c}^{\hbar}=\breve{\bm\theta}_{\mathcal A^c}=\breve{\bm\theta}_{\mathcal A^c}^{\mathcal G}=\bm 0$.  Then,
\begin{equation}\label{omega23}
\begin{aligned}
\bm\Omega_{23}=&\ \sum_{k<k^{\prime}}\bigg\{p_{\tau}^{\prime}\Big(\big|\big|\breve{\bm\theta}_{\mathcal A}^{\hbar(k)}-\breve{\bm\theta}_{\mathcal A}^{\hbar(k^{\prime})}\big|\big|_2,\lambda_2\Big)\big|\big|\breve{\bm\theta}_{\mathcal A}^{\hbar(k)}-\breve{\bm\theta}_{\mathcal A}^{\hbar(k^{\prime})}\big|\big|_2^{-1}\\
&\ \qquad\qquad\qquad\qquad\times\Big(\breve{\bm\theta}_{\mathcal A}^{\hbar(k)}-\breve{\bm\theta}_{\mathcal A}^{\hbar(k^{\prime})}\Big)^{\top}\Big[\big(\breve{\bm\theta}_{\mathcal A}^{(k)}-\breve{\bm\theta}_{\mathcal A}^{\mathcal G(k)}\big)-\big(\breve{\bm\theta}_{\mathcal A}^{(k^{\prime})}-\breve{\bm\theta}_{\mathcal A}^{\mathcal G(k^{\prime})}\big)\Big]\bigg\}\\
=&\ \sum_{m=1}^M\sum_{k,k^{\prime}\in\mathcal G^{(m)},k<k^{\prime}}p_{\tau}^{\prime}\Big(\big|\big|\breve{\bm\theta}_{\mathcal A}^{\hbar(k)}-\breve{\bm\theta}_{\mathcal A}^{\hbar(k^{\prime})}\big|\big|_2,\lambda_2\Big)\big|\big|\breve{\bm\theta}_{\mathcal A}^{(k)}-\breve{\bm\theta}_{\mathcal A}^{(k^{\prime})}\big|\big|_2\\
&\ +\sum_{m<m^{\prime}}\sum_{k\in\mathcal G^{(m)},k^{\prime}\in\mathcal G^{(m^{\prime})}}\bigg\{p_{\tau}^{\prime}\Big(\big|\big|\breve{\bm\theta}_{\mathcal A}^{\hbar(k)}-\breve{\bm\theta}_{\mathcal A}^{\hbar(k^{\prime})}\big|\big|_2,\lambda_2\Big)\big|\big|\breve{\bm\theta}_{\mathcal A}^{\hbar(k)}-\breve{\bm\theta}_{\mathcal A}^{\hbar(k^{\prime})}\big|\big|_2^{-1}\\
&\ \qquad\qquad\qquad\qquad\qquad \times\Big(\breve{\bm\theta}_{\mathcal A}^{\hbar(k)}-\breve{\bm\theta}_{\mathcal A}^{\hbar(k^{\prime})}\Big)^{\top}\Big[\big(\breve{\bm\theta}_{\mathcal A}^{(k)}-\breve{\bm\theta}_{\mathcal A}^{\mathcal G(k)}\big)-\big(\breve{\bm\theta}_{\mathcal A}^{(k^{\prime})}-\breve{\bm\theta}_{\mathcal A}^{\mathcal G(k^{\prime})}\big)\Big]\bigg\},\\
\end{aligned}
\end{equation}
where the first term of the second equality follows from the fact that, when $k,k^{\prime}\in\mathcal G^{(m)}$, $\breve{\bm\theta}_{\mathcal A}^{\mathcal G(k)}=\breve{\bm\theta}_{\mathcal A}^{\mathcal G(k^{\prime})}$ and $\breve{\bm\theta}_{\mathcal A}^{\hbar(k)}-\breve{\bm\theta}_{\mathcal A}^{\hbar(k^{\prime})}=\delta_2(\breve{\bm\theta}_{\mathcal A}^{(k)}-\breve{\bm\theta}_{\mathcal A}^{(k^{\prime})})$. Note that, for any $m\in[M]$, $k\in\mathcal G^{(m)}$,
\[
\big|\big|\breve{\bm\theta}_{\mathcal A}^{\mathcal G(k)}-\bm\theta_{\mathcal A}^{*(k)}\big|\big|_2=\Bigg|\Bigg|\frac{\sum_{k\in\mathcal G^{(m)}}\breve{\bm\theta}_{\mathcal A}^{(k)}}{|\mathcal G^{(m)}|}-\bm\psi_{\mathcal A}^{*(m)}\Bigg|\Bigg|_2\le\max_{k\in[K]}\big|\big|\breve{\bm\theta}_{\mathcal A}^{(k)}-\bm\theta_{\mathcal A}^{*(k)}\big|\big|_2.
\]
And then for any $k\in[K]$, we have $\big|\big|\breve{\bm\theta}_{\mathcal A}^{\hbar(k)}-\bm\theta_{\mathcal A}^{*(k)}\big|\big|_2\le\max_{k\in[K]}\big|\big|\breve{\bm\theta}_{\mathcal A}^{(k)}-\bm\theta_{\mathcal A}^{*(k)}\big|\big|_2$.
Hence, similar to (\ref{min_dis}), for any $m<m^{\prime}$, $k\in\mathcal G^{(m)}$, $k^{\prime}\in\mathcal G^{(m^{\prime})}$,
\begin{equation}\label{omega23_1}
\begin{aligned}
\big|\big|\breve{\bm\theta}_{\mathcal A}^{\hbar(k)}-\breve{\bm\theta}_{\mathcal A}^{\hbar(k^{\prime})}\big|\big|_2\ge&\ \min_{k\in\mathcal G^{(m)},k^{\prime}\in\mathcal G^{(m^{\prime})}}\big|\big|{\bm\theta}_{\mathcal A}^{*(k)}-{\bm\theta}_{\mathcal A}^{*(k^{\prime})}\big|\big|_2-2\max_{k\in[K]}\big|\big|\breve{\bm\theta}_{\mathcal A}^{\hbar(k)}-{\bm\theta}_{\mathcal A}^{*(k)}\big|\big|_2\\
\ge&\ \min_{k\in\mathcal G^{(m)},k^{\prime}\in\mathcal G^{(m^{\prime})}}\big|\big|{\bm\theta}_{\mathcal A}^{*(k)}-{\bm\theta}_{\mathcal A}^{*(k^{\prime})}\big|\big|_2-2\max_{k\in[K]}\big|\big|\breve{\bm\theta}_{\mathcal A}^{(k)}-{\bm\theta}_{\mathcal A}^{*(k)}\big|\big|_2\\
\ge&\ d_2-2C|\mathcal G_{\max}|^{1/2}r_{1N}>\tau\lambda_2.
\end{aligned}
\end{equation}
Combining (\ref{omega23}) and (\ref{omega23_1}), since  $\big|\big|\breve{\bm\theta}_{\mathcal A}^{\hbar(k)}-\breve{\bm\theta}_{\mathcal A}^{\hbar(k^{\prime})}\big|\big|_2\le2\max_{k\in[K]}\big|\big|\breve{\bm\theta}_{\mathcal A}^{(k)}-{\bm\theta}_{\mathcal A}^{*(k)}\big|\big|_2\le2C|\mathcal G_{\max}|^{1/2}r_{1N}$, we have
\begin{equation}\label{omega23_2}
\begin{aligned}
\bm\Omega_{23}=&\ \sum_{m=1}^M\sum_{k,k^{\prime}\in\mathcal G^{(m)},k<k^{\prime}}p_{\tau}^{\prime}\Big(\big|\big|\breve{\bm\theta}_{\mathcal A}^{\hbar(k)}-\breve{\bm\theta}_{\mathcal A}^{\hbar(k^{\prime})}\big|\big|_2,\lambda_2\Big)\big|\big|\breve{\bm\theta}_{\mathcal A}^{(k)}-\breve{\bm\theta}_{\mathcal A}^{(k^{\prime})}\big|\big|_2\\
\ge&\ \sum_{m=1}^M\sum_{k,k^{\prime}\in\mathcal G^{(m)},k<k^{\prime}}p_{\tau}^{\prime}\Big(2C|\mathcal G_{\max}|^{1/2}r_N,\lambda_2\Big)\big|\big|\breve{\bm\theta}_{\mathcal A}^{(k)}-\breve{\bm\theta}_{\mathcal A}^{(k^{\prime})}\big|\big|_2\\
\ge&\ \sum_{m=1}^M\sum_{k,k^{\prime}\in\mathcal G^{(m)},k<k^{\prime}}\frac{\lambda_2}{2}\big|\big|\breve{\bm\theta}_{\mathcal A}^{(k)}-\breve{\bm\theta}_{\mathcal A}^{(k^{\prime})}\big|\big|_2,\\
\end{aligned}
\end{equation}
where the last inequality follows from $2C|\mathcal G_{\max}|^{1/2}r_{1N}\to 0$ when $N\to\infty$.

We now consider the bound of $\bm\Omega_{21}$. For $k\in[K]$, we define
\[
\begin{aligned}
\bm\omega^{(k)}&\ :=\frac{\partial\mathcal L(\breve{\bm\theta}^{\hbar})}{\partial\bm\theta_{\mathcal A}^{(k)}} =2\Big\{\big[(n_k/N)\widetilde{\mathbf V}_{\mathcal A\mathcal A}^{(k)}\big]\breve{\bm\theta}_{\mathcal A}^{\hbar(k)}-(n_k/N)\widetilde{\bm\zeta}_{\mathcal A}^{(k)}\Big\}\\
&\ \ =2\bigg\{\big[(n_k/N)\widetilde{\mathbf V}_{\mathcal A\mathcal A}^{(k)}\big](\breve{\bm\theta}_{\mathcal A}^{\hbar(k)}-\bm\theta_{\mathcal A}^{*(k)})+(n_k/N)\mathbf g_{\mathcal A}^{*(k)}\\
&\ \qquad\qquad+(n_k/N)\int_0^1\bigg\{\mathbf V^{(k)}\Big(\big[\bm\theta^{*(k)}+t(\widetilde{\bm\theta}^{(k)}-\bm\theta^{*(k)})\big]\Big)-\widetilde{\mathbf V}^{(k)}\bigg\}(\widetilde{\bm\theta}^{(k)}-\bm\theta^{*(k)})\ dt\bigg\}\\
&\ \ :=\bm\omega_1^{(k)}+\bm\omega_2^{(k)}+\bm\omega_3^{(k)}.\\
\end{aligned}
\]
Then,
\[
\begin{aligned}
\bm\Omega_{21}&\ =\sum_{k=1}^K\bm\omega^{(k)\top}(\breve{\bm\theta}_{\mathcal A}^{(k)}-\breve{\bm\theta}_{\mathcal A}^{\mathcal G(k)})=\sum_{m=1}^M\sum_{k,k^{\prime}\in\mathcal G^{(m)}}\frac{\bm\omega^{(k)\top}(\breve{\bm\theta}_{\mathcal A}^{(k)}-\breve{\bm\theta}_{\mathcal A}^{(k^{\prime})})}{|\mathcal G^{(m)}|}\\
&\ =\sum_{m=1}^M\sum_{k,k^{\prime}\in\mathcal G^{(m)}}\frac{\bm\omega^{(k^{\prime})\top}(\breve{\bm\theta}_{\mathcal A}^{(k^{\prime})}-\breve{\bm\theta}_{\mathcal A}^{(k)})}{2|\mathcal G^{(m)}|}+\sum_{m=1}^M\sum_{k,k^{\prime}\in\mathcal G^{(m)}}\frac{\bm\omega^{(k)\top}(\breve{\bm\theta}_{\mathcal A}^{(k)}-\breve{\bm\theta}_{\mathcal A}^{(k^{\prime})})}{2|\mathcal G^{(m)}|}\\
&\ =\sum_{m=1}^M\sum_{k,k^{\prime}\in\mathcal G^{(m)}}\frac{(\bm\omega^{(k)}-\bm\omega^{(k^{\prime})})^{\top}(\breve{\bm\theta}_{\mathcal A}^{(k)}-\breve{\bm\theta}_{\mathcal A}^{(k^{\prime})})}{2|\mathcal G^{(m)}|}\\
&\ =\sum_{m=1}^M\sum_{k,k^{\prime}\in\mathcal G^{(m)},k<k^{\prime}}\frac{(\bm\omega^{(k)}-\bm\omega^{(k^{\prime})})^{\top}(\breve{\bm\theta}_{\mathcal A}^{(k)}-\breve{\bm\theta}_{\mathcal A}^{(k^{\prime})})}{|\mathcal G^{(m)}|}.\\
\end{aligned}
\]
Following the proof from (\ref{lambdamin}) to (\ref{eta}) in Theorem 1, we can show that
\[
\begin{aligned}
&\max_{k\in[K]}||\bm\omega_1^{(k)}||_2=O_p\big(|\mathcal G_{\max}|^{1/2}r_{1N}/K\big),\quad \max_{k\in[K]}||\bm\omega_2^{(k)}||_2=O_p\big(\sqrt{q/(KN)}\big),\\
&\max_{k\in[K]}||\bm\omega_1^{(k)}||_2=O_p\big(q^{3/2}\log p/N\big).
\end{aligned}
\]
Then,
\begin{equation}\label{omega21}
\begin{aligned}
|\bm\Omega_{21}|\le\sum_{m=1}^M\sum_{k,k^{\prime}\in\mathcal G^{(m)},k<k^{\prime}}\frac{2\max_{k\in[K]}||\bm\omega^{(k)}||_2}{|\mathcal G_{\min}|}\big|\big|\breve{\bm\theta}_{\mathcal A}^{(k)}-\breve{\bm\theta}_{\mathcal A}^{(k^{\prime})}\big|\big|_2.\\
\end{aligned}
\end{equation}
Combining (\ref{omega23_2}) and (\ref{omega21}), since $\lambda_2\gg |\mathcal G_{\max}|^{1/2}r_{1N}$, we have
\[
\lambda_2\gg \frac{|\mathcal G_{\max}|^{1/2}r_{1N}}{K|\mathcal G_{\min}|}+\sqrt{\frac{q}{KN|\mathcal G_{\min}|^2}}+\frac{q^{3/2}\log p}{N|\mathcal G_{\min}|},
\]
which leads to
\[
\mathcal Q(\breve{\bm\theta})-\mathcal Q(\breve{\bm\theta}^{\mathcal G})\ge\sum_{m=1}^M\sum_{k,k^{\prime}\in\mathcal G^{(m)},k<k^{\prime}}\bigg\{\frac{\lambda_2}{2}-\frac{2\max_{k\in[K]}||\bm\omega^{(k)}||_2}{|\mathcal G_{\min}|}\bigg\}\big|\big|\breve{\bm\theta}_{\mathcal A}^{(k)}-\breve{\bm\theta}_{\mathcal A}^{(k^{\prime})}\big|\big|_2\ge 0,
\]
for a sufficiently large $N$ with probability approaching 1. Thus, we have proved the result in (b). This finishes all the proof of Theorem 2.

\clearpage
\section*{Appendix B}

This section contains additional simulation results and data application results.

\begin{table}[H]
	\centering{\scriptsize
		\caption{ Simulation Example 2, variable selection accuracy: mean (sd) based on 100 replicates.		}\label{ex2_variable}
		\setlength{\tabcolsep}{2.3mm}{
		\begin{tabular}{ccccccccccccccccc}
		
			\hline
                               \cline{1-11}
                           &&\multicolumn{3}{c}{$n=200$}&\multicolumn{3}{c}{$n=400$}&\multicolumn{3}{c}{$n=800$}\\
                           \cmidrule(lr){3-5}
                           \cmidrule(lr){6-8}
                           \cmidrule(lr){9-11}

		&  Method &TPR&FPR&MS&TPR&FPR&MS &TPR&FPR&MS    \\
		
			\hline
    
        $K=64$                           &      ICR	             &1.000       & 0.000      &33.920     &1.000&0.000     &32.000   &1.000&0.000     &32.000\\
                                 &&(0.000)&(0.000)&(3.959)&(0.000)&(0.000)&(0.000)&(0.000)&(0.000)&(0.000)\\                                                                                               
                                               &      SHIR	             & 0.760      & 0.001      &389.220     &1.000&0.001     &512.700   &1.000&0.004     &533.770\\
                                 &&(0.042)&(0.004)&(21.718)&(0.000)&(0.003)&(6.400)&(0.000)&(0.005)&(30.463)\\                    
                                               &      SMA	             &0.763       &0.001       &390.500     &1.000&0.001     &513.330   &1.000&0.003     &528.000\\
                                 &&(0.042)&(0.004)&(21.367)&(0.000)&(0.003)&(9.002)&(0.000)&(0.005)&(27.852)\\                                            
                                               &      Local	             & 0.853      &0.148       &1309.750     &0.987&0.198     &1673.740   &1.000&0.218     &1794.750\\
                                 &&(0.020)&(0.010)&(63.847)&(0.005)&(0.010)&(56.069)&(0.001)&(0.011)&(63.438)\\              
                                              &      SK(har)	             &0.945       &0.526       &222.880     &0.996&0.934     &375.480   &1.000&0.967     &387.880\\
                                 &&(0.098)&(0.456)&(169.339)&(0.028)&(0.137)&(51.331)&(0.000)&(0.020)&(7.335)\\        
                                               &      SK(gap)	             &1.000       &0.995       & 199.000    &1.000&0.999     &199.840   &1.000&0.987     &273.200\\
                                 &&(0.000)&(0.008)&(1.518)&(0.000)&(0.003)&(0.615)&(0.000)&(0.020)&(92.285)\\

       $K=128$                           &      ICR	             & 1.000      &0.000       & 34.800    &1.000&0.000     &32.080   &1.000&0.000     &32.000\\
                                 &&(0.000)&(0.000)&(5.005)&(0.000)&(0.000)&(0.800)&(0.000)&(0.000)&(0.000)\\                                                                                               
                                               &      SHIR	             &0.784       &0.000       &802.560     &1.000&0.005     &1077.770   &1.000&0.011     &1152.000\\
                                 &&(0.068)&(0.000)&(70.120)&(0.000)&(0.005)&(63.485)&(0.000)&(0.000)&(0.000)\\                    
                                               &      SMA	             &0.781       & 0.000      &800.000     &1.000&0.004     &1067.530   &1.000&0.011     &1152.000\\
                                 &&(0.074)&(0.000)&(75.835)&(0.000)&(0.005)&(60.954)&(0.000)&(0.000)&(0.000)\\                                            
                                               &      Local	             & 0.851      & 0.147      & 2602.260    &0.987&0.199     &3351.110   &1.000&0.219     &3599.230\\
                                 &&(0.015)&(0.007)&(94.466)&(0.003)&(0.007)&(84.503)&(0.000)&(0.007)&(85.526)\\        
                                               &      SK(har)	             & 0.899      &0.505       &216.400     &1.000&0.988     &394.320   &1.000&0.980     &392.440\\
                                 &&(0.121)&(0.497)&(189.078)&(0.000)&(0.100)&(38.757)&(0.000)&(0.141)&(52.333)\\        
                                               &      SK(gap)	             &1.000       &1.000       &199.980     &1.000&1.000     &200.000   &1.000&0.990     &299.240\\
                                 &&(0.000)&(0.001)&(0.200)&(0.000)&(0.000)&(0.000)&(0.000)&(0.100)&(102.996)\\        
                                   \hline
		     \cline{1-11}	
		\end{tabular}
	}}
\end{table}

\begin{table}[H]
	\centering{\scriptsize
		\caption{Simulation Example 2, clustering accuracy: mean (sd) based on 100 replicates.
		}\label{ex2_cluster}
		\setlength{\tabcolsep}{1mm}{
		\begin{tabular}{cccccccccccccc}
		
			\hline
                               \cline{1-14}
                           &&\multicolumn{4}{c}{$n=200$}&\multicolumn{4}{c}{$n=400$}&\multicolumn{4}{c}{$n=800$}\\
                           \cmidrule(lr){3-6}
                           \cmidrule(lr){7-10}
                           \cmidrule(lr){11-14}

		&  Method &$\widehat M$&Per&RI&ARI &$\widehat M$&Per&RI&ARI &$\widehat M$&Per&RI&ARI     \\
		
			\hline

         $K=64$                        &      ICR	  &4.240               &0.790       &0.998       &0.995     &4.000               &1.000       &1.000       &1.000    &4.000               &1.000       &1.000       &1.000 \\
                               &&(0.495)&(-)&(0.004)&(0.010)&(0.000)&(-)&(0.000)&(0.000)&(0.000)&(-)&(0.000)&(0.000)\\                                        
                                             &      SK(har)	  &3.980               &0.980       &0.994       &0.984    &4.000               &1.000       &1.000       &1.000   &4.000               &1.000       &1.000       &1.000 \\
                               &&(0.141)&(-)&(0.020)&(0.048)&(0.000)&(-)&(0.000)&(0.000)&(0.000)&(-)&(0.000)&(0.000)\\                
                                             &      SK(gap)	  &2.000               &0.000       &0.746       & 0.487    &2.000               &0.000       &0.746       &0.488    & 2.780              &0.390       &0.845       &0.688 \\
                               &&(0.000)&(-)&(0.002)&(0.003)&(0.000)&(-)&(0.000)&(0.000)&(0.980)&(-)&(0.124)&(0.251)\\         
         $K=128$                        &      ICR	  & 4.350              &0.640       & 0.991      &0.978     & 4.010              &0.990       &1.000       &1.000    &4.000               &1.000       &1.00       &1.000 \\
                               &&(0.626)&(-)&(0.030)&(0.070)&(0.100)&(-)&(0.000)&(0.001)&(0.000)&(-)&(0.000)&(0.000)\\                                        
                                             &      SK(har)	  & 4.000              &0.960       &0.994       &0.984     &3.980               &0.980       &0.997       &0.994    &3.990               &0.990       &0.999       &0.997 \\
                               &&(0.201)&(-)&(0.019)&(0.044)&(0.141)&(-)&(0.018)&(0.041)&(0.100)&(-)&(0.013)&(0.029)\\                
                                             &      SK(gap)	  &2.000               &0.000       &0.748       &0.494     &2.000               & 0.000      &0.748       &0.494    & 3.030              & 0.510      & 0.878      & 0.754\\
                               &&(0.000)&(-)&(0.000)&(0.000)&(0.000)&(-)&(0.000)&(0.000)&(1.000)&(-)&(0.126)&(0.253)\\                                                                                                
    \hline
		     \cline{1-14}	
		\end{tabular}
	}}
\end{table}

\begin{table}[H]
	\centering{\scriptsize
		\caption{ Simulation Example 3, variable selection accuracy: mean (sd) based on 100 replicates.
		}\label{ex3_variable}
		\setlength{\tabcolsep}{2mm}{
		\begin{tabular}{cccccccccccccc}
		
			\hline
                               \cline{1-11}
                           &&\multicolumn{3}{c}{$K=16$}&\multicolumn{3}{c}{$K=32$}&\multicolumn{3}{c}{$K=64$}\\
                           \cmidrule(lr){3-5}
                           \cmidrule(lr){6-8}
                           \cmidrule(lr){9-11}

		&  Method &TPR&FPR&MS &TPR&FPR&MS &TPR&FPR&MS    \\
		
			\hline

         $\sigma=1$                        &      ICR	  & 1.000              &0.000       &16.000       & 1.000              &0.000       &16.000  &1.000     &0.000     &16.060 \\
                               &&(0.000)&(0.000)&(0.000)&(0.000)&(0.000)&(0.000)&(0.000)&(0.002)&(0.343)\\                                     
                                            &      SHIR	  & 1.000              &0.019       &129.920       &1.000     &0.022     &258.030  &1.000     &0.015     &513.410 \\
                               &&(0.000)&(0.016)&(2.347)&(0.000)&(0.015)&(1.374)&(0.000)&(0.014)&(1.248)\\         
                                            &      SMA	  & 1.000              & 0.001      &128.410       &1.000     &0.002     &256.190  &1.000     &0.003     &512.240 \\
                               &&(0.000)&(0.003)&(2.257)&(0.000)&(0.005)&(0.419)&(0.000)&(0.005)&(0.495)\\         
                                            &      Local	  &  0.990             &  0.116     &  297.450     &0.989     &0.117     &598.030  &0.990     & 0.115    &1184.390 \\
                               &&(0.008)&(0.022)&(32.006)&(0.007)&(0.015)&(43.364)&(0.004)&(0.011)&(62.616)\\    
                                            &      SK(har)	  &0.994               &0.382       &198.580       &0.998     &0.571     &300.420  & 1.000    &0.802     &383.890 \\
                               &&(0.033)&(0.183)&(74.091)&(0.018)&(0.201)&(95.568)&(0.000)&(0.142)&(90.112)\\                
                                             &      SK(gap)	  &1.000               & 0.646      & 134.820      &1.000     &0.861     &174.340  &1.000     &0.982     &196.700 \\
                               &&(0.000)&(0.104)&(19.080)&(0.000)&(0.062)&(11.353)&(0.000)&(0.019)&(3.416)\\              
                               
         $\sigma=2$                        &      ICR	  & 0.996              &0.000       &16.190       & 1.000              &0.000       &16.000  &1.000     &0.000     &16.060 \\
                               &&(0.021)&(0.003)&(1.522)&(0.000)&(0.000)&(0.000)&(0.000)&(0.002)&(0.343)\\                                     
                                            &      SHIR	  & 1.000              &0.135       &228.320       &1.000     &0.144     & 436.990 &1.000     &0.139     &762.890 \\
                               &&(0.000)&(0.064)&(81.849)&(0.000)&(0.054)&(116.607)&(0.000)&(0.034)&(134.836)\\         
                                            &      SMA	  & 1.000              &0.111       &  283.240     &1.000     &0.110     & 515.970 &1.000     &0.088     &772.130 \\
                               &&(0.000)&(0.036)&(55.266)&(0.000)&(0.057)&(182.428)&(0.000)&(0.048)&(266.909)\\         
                                            &      Local	  &0.816               &0.099       &250.050       & 0.818    &0.100     &504.280  &0.819     &0.099     &1000.380 \\
                               &&(0.034)&(0.020)&(31.474)&(0.022)&(0.015)&(44.000)&(0.016)&(0.011)&(65.717)\\       
                                            &      SK(har)	  & 0.986              & 0.319      &168.58       &0.996     &0.485     &255.770  &1.000     &0.701     &365.340 \\
                               &&(0.053)&(0.166)&(71.974)&(0.038)&(0.203)&(97.201)&(0.000)&(0.204)&(118.276)\\                
                                             &      SK(gap)	  &1.000               &0.555       &118.370       &1.000     &0.809     &165.630  &1.000     &0.967     &193.840 \\
                               &&(0.000)&(0.150)&(26.845)&(0.000)&(0.084)&(15.860)&(0.000)&(0.026)&(4.745)\\                                                                                              
    \hline
		     \cline{1-11}	
		\end{tabular}
	}}
\end{table}

\begin{table}[H]
	\centering{\scriptsize
		\caption{Simulation Example 3, clustering accuracy: mean (sd) based on 100 replicates.
		}\label{ex3_cluster}
		\setlength{\tabcolsep}{1mm}{
		\begin{tabular}{cccccccccccccc}
		
			\hline
                               \cline{1-14}
                           &&\multicolumn{4}{c}{$K=16$}&\multicolumn{4}{c}{$K=32$}&\multicolumn{4}{c}{$K=64$}\\
                           \cmidrule(lr){3-6}
                           \cmidrule(lr){7-10}
                           \cmidrule(lr){11-14}

		&  Method &$\widehat M$&Per&RI&ARI &$\widehat M$&Per&RI&ARI &$\widehat M$&Per&RI&ARI     \\
		
			\hline

         $\sigma=1$                        &      ICR	  &2.000               &1.000       &1.000       &1.000     &2.000               &1.000       &1.000       &1.000    &2.000               &1.000       &1.000       &1.000 \\
                               &&(0.000)&(-)&(0.000)&(0.000)&(0.000)&(-)&(0.000)&(0.000)&(0.000)&(-)&(0.000)&(0.000)\\                                        
                                             &      SK(har)	  &4.900               &0.000       &0.752       &0.483    &5.340               &0.000       & 0.739      &0.468   &4.790               &0.000       &0.744       &0.484 \\
                               &&(1.418)&(-)&(0.085)&(0.182)&(1.821)&(-)&(0.088)&(0.182)&(1.274)&(-)&(0.065)&(0.132)\\                
                                             &      SK(gap)	   &2.000               &1.000       &1.000       &1.000     &2.000               &1.000       &1.000       &1.000     &2.000               &1.000       &1.000       &1.000 \\
                               &&(0.000)&(-)&(0.000)&(0.000)&(0.000)&(-)&(0.000)&(0.000)&(0.000)&(-)&(0.000)&(0.000)\\         
         $\sigma=2$                        &      ICR	  & 2.020              &0.980       & 0.999      &0.998     &2.000               &1.000       &1.000       &1.000    &2.000               &1.000       &1.000       &1.000 \\
                               &&(0.141)&(-)&(0.008)&(0.017)&(0.000)&(-)&(0.000)&(0.000)&(0.000)&(-)&(0.000)&(0.000)\\                                        
                                             &      SK(har)	  & 4.800              &0.000       & 0.764      &0.507     &5.180               &0.000       &0.754       &0.497    &5.200               &0.000       & 0.736      & 0.466\\
                               &&(1.589)&(-)&(0.097)&(0.207)&(1.850)&(-)&(0.095)&(0.195)&(1.589)&(-)&(0.077)&(0.156)\\                
                                             &      SK(gap)	  &2.020               &0.980       &0.996       &0.993     &2.010               & 0.990      &0.998       &0.997    & 2.000              & 1.000      & 0.999      & 0.999\\
                               &&(0.141)&(-)&(0.019)&(0.039)&(0.100)&(-)&(0.013)&(0.026)&(0.000)&(-)&(0.004)&(0.009)\\                                                                                                
    \hline
		     \cline{1-14}	
		\end{tabular}
	}}
\end{table}

\begin{table}[H]
	\centering{\scriptsize
		\caption{ Simulation Example 4, variable selection accuracy: mean (sd) based on 100 replicates.
		}\label{ex4_variable}
		\setlength{\tabcolsep}{2mm}{
		\begin{tabular}{cccccccccccccc}
		
			\hline
                               \cline{1-11}
                           &&\multicolumn{3}{c}{$n=100$}&\multicolumn{3}{c}{$n=200$}&\multicolumn{3}{c}{$n=400$}\\
                           \cmidrule(lr){3-5}
                           \cmidrule(lr){6-8}
                           \cmidrule(lr){9-11}

		&  Method &TPR&FPR&MS &TPR&FPR&MS &TPR&FPR&MS    \\
		
			\hline

         $\sigma=1$                        &      ICR	  &1.000               &0.000       &32.040       &1.000               &0.000       &32.000  &1.000               &0.000       &32.000 \\
                               &&(0.000)&(0.001)&(0.400)&(0.000)&(0.000)&(0.000)&(0.000)&(0.000)&(0.000)\\                                     
                                                     &      SHIR	  &1.000               & 0.077      &  531.640     &1.000               &0.081       &524.450  &1.000     &0.072     &523.670 \\
                               &&(0.000)&(0.029)&(31.566)&(0.000)&(0.030)&(17.815)&(0.000)&(0.030)&(17.975)\\            
                                                     &      SMA	  & 1.000              &0.060       &539.570       & 1.000              &0.078       &524.870  &1.000     &0.073     &525.040 \\
                               &&(0.000)&(0.026)&(38.983)&(0.000)&(0.030)&(19.013)&(0.000)&(0.028)&(21.384)\\            
                                                     &      Local	  & 0.989              & 0.204      &1709.120       & 1.000              & 0.199      &1685.360  &1.000     &0.195     &1658.000 \\
                               &&(0.005)&(0.012)&(70.651)&(0.000)&(0.011)&(63.342)&(0.000)&(0.011)&(64.948)\\       
                                                      &      SK(har)	  & 0.999              &0.932       &375.080       &1.000               &0.933       &375.160  &1.000     &0.937     &376.680 \\
                               &&(0.013)&(0.137)&(50.832)&(0.000)&(0.098)&(36.120)&(0.000)&(0.036)&(13.237)\\     
                                                      &      SK(gap)	  &1.000               &0.998       &203.380       &1.000               &0.967       & 301.720 & 1.000    &0.959     &320.040 \\
                               &&(0.000)&(0.010)&(24.503)&(0.000)&(0.036)&(89.220)&(0.000)&(0.039)&(83.909)\\     
                               
         $\sigma=2$                        &      ICR	 &1.000               &0.000       &32.400       &1.000               &0.000       &32.000  &1.000               &0.000       &32.040 \\
                               &&(0.000)&(0.000)&(1.752)&(0.000)&(0.000)&(0.000)&(0.000)&(0.001)&(0.400)\\                                     
                                                     &      SHIR	  & 1.000              &0.270       &1542.320       &1.000               & 0.229      &1246.210  &1.000     &0.197     &1084.500 \\
                               &&(0.000)&(0.080)&(486.815)&(0.000)&(0.076)&(367.035)&(0.000)&(0.068)&(350.924)\\            
                                                     &      SMA	  &  1.000             &0.263       &1575.730       &1.000               & 0.230      &1246.960  & 1.000    & 0.188    &1058.500 \\
                               &&(0.000)&(0.0840)&(469.041)&(0.000)&(0.076)&(391.310)&(0.000)&(0.068)&(374.399)\\            
                                                     &      Local	  &0.719               & 0.127      &1116.360       &0.944               &0.176       &1519.300  &0.998     &0.194     &1650.780 \\
                               &&(0.030)&(0.013)&(85.809)&(0.014)&(0.012)&(75.027)&(0.002)&(0.011)&(64.969)\\     
                                                      &      SK(har)	  &  0.908             &0.285       &140.730       &0.994               &0.868       &351.280  & 0.999    &0.928     &373.640 \\
                               &&(0.106)&(0.410)&(161.999)&(0.033)&(0.223)&(83.034)&(0.013)&(0.101)&(37.494)\\     
                                                      &      SK(gap)	  &0.998               &0.980       &196.360       &1.000               &0.998       &199.580  &1.000     & 0.991    &230.660 \\
                               &&(0.025)&(0.100)&(18.750)&(0.000)&(0.005)&(0.955)&(0.000)&(0.022)&(68.584)\\      
                                                                                                                            
    \hline
		     \cline{1-11}	
		\end{tabular}
	}}
\end{table}

\begin{table}[H]
	\centering{\scriptsize
		\caption{Simulation Example 4, clustering accuracy: mean (sd) based on 100 replicates.
		}\label{ex4_cluster}
		\setlength{\tabcolsep}{1mm}{
		\begin{tabular}{cccccccccccccc}
		
			\hline
                               \cline{1-14}
                           &&\multicolumn{4}{c}{$n=100$}&\multicolumn{4}{c}{$n=200$}&\multicolumn{4}{c}{$n=400$}\\
                           \cmidrule(lr){3-6}
                           \cmidrule(lr){7-10}
                           \cmidrule(lr){11-14}

		&  Method &$\widehat M$&Per&RI&ARI &$\widehat M$&Per&RI&ARI &$\widehat M$&Per&RI&ARI     \\
		
			\hline

         $\sigma=1$                        &      ICR	  &4.000               &1.000       &1.000       &1.000     &4.000               &1.000       &1.000       &1.000    &4.000               &1.000       &1.000       &1.000 \\
                               &&(0.000)&(-)&(0.000)&(0.000)&(0.000)&(-)&(0.000)&(0.000)&(0.000)&(-)&(0.000)&(0.000)\\                                        
                                             &      SK(har)	   &4.000               &1.000       &1.000       &1.000     &4.000               &1.000       &1.000       &1.000  &4.000               &1.000       &1.000       &1.000  \\
                               &&(0.000)&(-)&(0.000)&(0.000)&(0.000)&(-)&(0.000)&(0.000)&(0.000)&(-)&(0.000)&(0.000)\\                
                                             &      SK(gap)	  & 2.040              &0.020       & 0.751      &0.498    &3.140               &0.570       & 0.891      & 0.780  &  3.350             & 0.670      &0.917       &0.833 \\
                               &&(0.281)&(-)&(0.036)&(0.072)&(0.995)&(-)&(0.126)&(0.255)&(0.936)&(-)&(0.119)&(0.240)\\        
         $\sigma=2$                        &      ICR	  & 4.050              &0.950       & 1.000      &0.999     &4.000               &1.000       &1.000       &1.000    &4.000               &1.000       &1.00       &1.000 \\
                               &&(0.219)&(-)&(0.002)&(0.005)&(0.000)&(-)&(0.000)&(0.000)&(0.000)&(-)&(0.000)&(0.000)\\                                        
                                             &      SK(har)	  &4.150               &0.860       &0.977       &0.935   &4.000               &1.000       &1.000       &1.000    &4.000               &1.000       &1.000       &1.000  \\
                               &&(0.386)&(-)&(0.021)&(0.058)&(0.000)&(-)&(0.000)&(0.000)&(0.000)&(-)&(0.000)&(0.000)\\                
                                             &      SK(gap)	  &2.000               &0.000       &0.745       &0.485    &2.000               &0.000       &0.746       &0.488   &2.340               & 0.170      &0.789       &0.575\\
                               &&(0.000)&(-)&(0.005)&(0.010)&(0.000)&(-)&(0.000)&(0.000)&(0.755)&(-)&(0.096)&(0.193)\\                                                                                                 
    \hline
		     \cline{1-14}	
		\end{tabular}
	}}
\end{table}

\begin{table}[H]
	\centering{\scriptsize
		\caption{Data analysis: one record of the initial request logs.
		}\label{record}
		\setlength{\tabcolsep}{4mm}{
		\begin{tabular}{ccc}
		
			\hline

		 URL interface&GET Parameter & POST Parameter\\

			\hline

                  	     		 \multirow{2}{*}{ ajaxNoSessionGetSmsAction}& \multirow{2}{*}{s=captcha} &\_method=\_\_construct\&filter[]=phpinfo\&\\
			 &&method=get\&server[REQUEST\_METHOD]=1\\

                                   \hline

		\end{tabular}
		
	}
	
	}

\end{table}

\end{document}